\documentclass[aps,prl,twocolumn,amsmath,amssymb,longbibliography,superscriptaddress,floatfix]{revtex4-2}
\usepackage{amsmath,amssymb,amsfonts,bm}
\usepackage{ifthen}
\usepackage{graphicx}
\usepackage[colorlinks=true,breaklinks=true,citecolor=teal,linkcolor=teal,urlcolor=teal]{hyperref}
\usepackage{mathtools}
\usepackage{braket}
\usepackage{mathrsfs}
\usepackage{xcolor}
\usepackage{xparse}
\usepackage{orcidlink}
\usepackage{bbm}

\newcommand{\be}{\begin{equation}}
\newcommand{\ee}{\end{equation}}
\newcommand{\bea}{\begin{aligned}}
\newcommand{\eea}{\end{aligned}}
\newcommand{\average}[1]{\langle #1 \rangle}
\newcommand{\e}{\mathrm{e}}
\newcommand{\dd}{\mathrm{d}}
\newcommand{\Tr}{\operatorname{Tr}}
\newcommand{\id}{\mathbbm{1}}
\newcommand{\urho}{\check{\rho}}
\newcommand{\Tpass}{\mathcal{T}}
\newcommand{\tp}{T_{\mathrm{P}}}
\newcommand{\Wsp}{\mathcal{W}}

\newcommand{\exv}{\mathbb{E}}

\begin{document}

\title{Universal purification dynamics of monitored Clifford circuits}

\author{Beatrice Magni~\orcidlink{0009-0009-8577-0525}}
\email{bmagni@uni-koeln.de}
\affiliation{Institut f\"ur Theoretische Physik, Universit\"at zu K\"oln, Z\"ulpicher Strasse 77, 50937 K\"oln, Germany}

\author{Federico Gerbino~\orcidlink{0009-0008-1485-3764}}
\email{federico.gerbino@universite-paris-saclay.fr}
\affiliation{Laboratoire de Physique Th\'eorique et Mod\`eles Statistiques, Universit\'e Paris-Saclay, CNRS, 91405 Orsay, France}

\author{Xhek Turkeshi~\orcidlink{0000-0003-1093-3771}}
\email{xturkesh@uni-koeln.de}
\affiliation{Institut f\"ur Theoretische Physik, Universit\"at zu K\"oln, Z\"ulpicher Strasse 77, 50937 K\"oln, Germany}

\author{Andrea De Luca~\orcidlink{0000-0003-0272-5083}}
\email{andrea.de-luca@ens.fr}
\affiliation{Laboratoire de Physique de l'\'Ecole Normale Sup\'erieure, CNRS, ENS \& PSL University, Sorbonne Universit\'e, Universit\'e Paris Cit\'e, 75005 Paris, France}

\date{\today}

\begin{abstract}
Quantum circuits under sufficiently weak monitoring purify on a timescale $T_P$ exponentially long in the system size. This slowness underlies a universal purification dynamics, whose quantitative description has so far required the replica trick, with a delicate analytic continuation. We show that monitored Clifford circuits on $L$ qudits of prime dimension $q$ bypass this construction entirely: in the scaling limit at fixed $x = t/T_P(L)$, purification reduces to the Markovian decay of the density-matrix rank, an exactly solvable death process descending from infinity. We compute the full scaling functions in compact form: all R\'enyi entropies collapse onto a universal curve $\average{S(x)}$. Exact stabilizer simulations at $q=2,3,5$ confirm the predictions, with no fitting parameter for the global model and $T_P$ as the only fitted scale for local brick-wall circuits. Also, the replica problem amounts to a tilted version of the same Markov process, in agreement with exact computations from the Clifford commutant. Finally, the quantization of the rank leaves two hallmarks that distinguish Clifford dynamics from generic monitored circuits: the entropy fluctuations saturate at short scaled times $x\to0$ to an $O(1)$ variance, instead of vanishing, and observables develop a temporal modulation periodic in $\log_q x$, which cannot be captured by the replica approach.
\end{abstract}

\maketitle

\textit{Introduction.}---Monitored quantum many-body dynamics realizes a genuinely nonequilibrium form of competition: unitary evolution scrambles and entangles~\cite{nahum2017quantum,nahum2018operator,von_Keyserlingk_2018}, while local measurements extract information and tend to disentangle~\cite{fisher2023,potter2022entanglementdynamicsin,lunt2022quantumsimulationusing,sierant2023, ippoliti2021entanglement, sierant2022measurement}.  As the measurement rate is increased, monitored circuits undergo a measurement-induced phase transition (MIPT) between a weak-measurement phase, where the steady state obeys a volume law for the entanglement entropy, and a strong-measurement phase, where it obeys an area law~\cite{li2018quantum,li2019measurementdrivenentanglement,skinner2019measurementinducedphase,chan2019unitaryprojective,szyniszewski2019entanglementtransitionfrom,zabalo2020criticalpropertiesof,turkeshi2020measurementinducedcriticality,turkeshi2022measurementinducedcriticality,sierant2022universalbehaviorbeyond, block2022measurement, PhysRevResearch.4.L022066, PhysRevB.107.014306, SciPostPhysCore.5.2.023, PRXQuantum.6.010351, nambi2026postselectedcriticalitymeasurementinducedphase}; the transition has by now been probed on superconducting and trapped-ion hardware~\cite{Noel_2022,google2021,koh2022experimentalrealizationof,feng2025postselectionfreeexperimentalobservationmeasurementinduced, Sierant2022dissipativefloquet}. Due to the intrinsic randomness of the measurement outcomes, the transition falls into the category of disordered systems.
The average over disorder is commonly treated through the ``replica trick'': $N$ identical copies of the system give access to the integer moments of the quantities of interest, and the correct probabilistic weighting is restored in the formal limit $N \to 1$. This approach allows monitored circuits to be mapped onto an effective statistical mechanics model with a
symmetry consisting of permutations of the $N$ replicas: MIPTs are thus framed within the symmetry-breaking framework, between a broken phase (weak measurements) and a disordered phase (strong measurements)~\cite{bao2020theoryofthe,choi2020quantumerrorcorrection,jian2020measurementinducedcriticality,nahum2021measurement,bao2021symmetryenrichedphases,li2021conformal,li2021entanglementdomainwalls,li2021statisticalclifford,jian2022criticality,jian2023measurement,nahum2023renormalizationgroupfor,zhang2021emergentreplica,giachetti2023elusive}. However, the fate of this analysis as $N \to 1$ sometimes remains unclear~\cite{giachetti2023elusive}.

A complementary and operationally sharp diagnostic of the transition is \emph{purification}~\cite{gullans2020dynamicalpurificationphase,gullans2020scalable,bentsen2021measurementinducedpurification,gopalakrishnan2021entanglementandpurification,ippoliti2022dynamicalpurificationand,Fidkowski_2021, ippoliti2021postselection, leontica2023purification, anzai2026disordered, han2022measurement, kawabata2023entanglement, claeys2022exact, bompais2026rate, PhysRevB.102.224311, PhysRevLett.127.235701, yu2022measurementinducedentanglementphasetransition, PhysRevB.107.L220204, wang2025stabilityquantumchaosweak, PhysRevB.106.214304}. One starts from a maximally mixed state -- equivalently, a system maximally entangled with an ancillary reference -- and asks how long the monitored dynamics needs to drive it (close) to a pure state. In the strong-measurement phase the purification time scales logarithmically with the system size $L$; in the weak-measurement phase, it is exponentially long, $\tp(L)\sim \e^{\alpha L}$. In the replica/statistical-mechanics picture, this exponential separation reflects an extensive free-energy barrier
present in the broken phase connecting distinct symmetry sectors: purification requires a rare domain wall (an instanton in replica space) sweeping the entire system~\cite{bao2020theoryofthe,giachetti2023elusive,choi2020quantumerrorcorrection}.

Because purification is the slowest relevant process in the volume-law phase, all microscopic relaxation channels have time to act before a single measurement succeeds in extracting global information. This suggests that the slow purification dynamics is \emph{universal}: upon rescaling time $x = t/\tp(L)$, suitable spectral probes of the density matrix collapse onto scaling functions that no longer remember the lattice geometry, the dimensionality, or the precise local gates~\cite{deluca2025universality,gerbino2026universal,ippoliti2022dynamicalpurificationand, PhysRevLett.134.140401, patel2026universalmonitoreddynamicsmultimode}. The intuitive mechanism is that, between two informative measurements, the unitary dynamics has ample time to scramble completely, so that the effective propagator converges to a Haar-random matrix on the whole Hilbert space; the many-body problem reduces to a $0{+}1$-dimensional random-matrix process~\cite{bulchandani_random-matrix_2023,gerbino2024dyson,mehta2004random}. Recent works extracted the corresponding universal R\'enyi-entropy scaling functions and identified distinct universality classes labeled by the emergent symmetry group -- unitary or orthogonal~\cite{deluca2025universality,gerbino2026universal}. However, these analyses are technical, mostly based on perturbative small-$x$ expansions and a careful combinatorial control of the replica limit $N \to 1$.

This Letter focuses on monitored \emph{Clifford} circuits on qudits of arbitrary prime dimension $q$ -- the paradigmatic efficiently simulable platform for the MIPT through the stabilizer states formalism~\cite{turkeshi2020measurementinducedcriticality,lavasani2021measurementinducedtopological,li2021statisticalclifford,bittel2025completetheorycliffordcommutant,magni2025anticoncentrationstatedesigndoped, PhysRevB.104.155111, PhysRevB.108.184204}. Crucially, the corresponding density matrix is always a multiple of a projector and purification reduces to the \emph{decay of a single parameter}, its rank. Introducing a minimal model, we isolate the universal content via a map of the purification process to a pure-death Markov process. We obtain the complete distribution of the rank in the scaling limit, which fully characterizes the spectral statistics of the system's density matrix. In addition, we obtain computable expressions for the von Neumann entropy and purity as a function of $x$. We support these predictions with exact stabilizer simulations at $q=2,3,5$, both for the global model (with parameter-free agreement) and for local brick-wall circuits, using purification time as the only (non-universal) parameter.
Finally, in the End Matter we show that $N \neq 1$ replicas amount to a Feynman--Kac tilt of the same death process, and match it against the algebraic computation based on the Clifford commutant at small $N$. While for $q=2$ qubits the Clifford dynamics coincides with the Haar-unitary one for $N \leq 3$ (the Clifford group is a unitary 3-design), the identification fails in the Born limit $N \to 1$, which is sensitive to all $N$, even for low-replica observables such as the purity~\cite{webb2016clifford,zhu2017multiqubit,gross2021schur,bittel2025completetheorycliffordcommutant}.

\textit{Symplectic structure and mixed stabilizer states.}---We consider $L$ qudits of prime dimension $q$, with total Hilbert-space dimension $D=q^L$. We briefly recall the symplectic formalism at the origin of our calculation~\cite{gottesman1997stabilizer,aaronson2004improvedsimulationof}. Let $\omega=\e^{2\pi i/q}$ and let $X\ket{m}=\ket{m+1\bmod q}$, $Z\ket{m}=\omega^m\ket{m}$ be the clock and shift operators.
Consider the finite field $\mathbb F_q$; 
to a vector $\mathbf v=(\mathbf a|\mathbf b)\in\mathbb F_q^{2L}$,  one associates the $L$-qudit Pauli (Weyl) string $P=X^{\mathbf a}Z^{\mathbf b}=\bigotimes_{j=1}^L X^{a_j}_j Z^{b_j}_j$, where $X_j,Z_j$ act on site $j$.
One obtains a $1$-to-$1$ mapping $\mathbf{v} \leftrightarrow P$, where multiplication of strings becomes addition of vectors. Two strings obey $PP'=\omega^{\langle\mathbf v,\mathbf v'\rangle}P'P$, where $\mathbf{v}, \mathbf{v}'$ are the corresponding vectors in $\mathbb{F}_q^{2L}$ and the \emph{symplectic product}
\be
\label{eq:symplprod}
\langle\mathbf v,\mathbf v'\rangle=\sum_{j=1}^{L}\big(a_j b_j'-b_j a_j'\big)\bmod q.
\ee
Thus, they commute if and only if the symplectic product vanishes.
A linear subspace of dimension $k$ of $\mathbb{F}_q^{2L}$ corresponds to a set of $q^k$ Pauli strings closed under multiplication -- the finite cardinality on which all probabilities below rest.
The Clifford group contains all unitary maps of Pauli strings to Pauli strings~\cite{gottesman1997stabilizer,aaronson2004improvedsimulationof,gidney2021stimfaststabilizer}. Thanks to the vector formalism, they can be identified with automorphisms of $\mathbb{F}_q^{2L}$ that preserve the symplectic structure \eqref{eq:symplprod}.

Consider an Abelian group \(\mathcal{W}\) of mutually commuting Pauli operators, called the \emph{stabilizer group}.
In symplectic language, $\Wsp$ is an \emph{isotropic} subspace of $\mathbb F_q^{2L}$, i.e. all its elements mutually commute. A maximal \(\mathcal{W}\) has \(L\) independent generators $\{P_k\}_{k=1}^L$. Each Pauli string $P$ has eigenvalues $\omega^{\varphi}$, with $\varphi \in \{0,\ldots, q-1\}$ and the choice of $\{\varphi_k\}_{k=1}^L$, for each generator $P_k$, identifies  a unique pure quantum state \(|\psi\rangle\), \emph{a stabilizer state}, at the intersection of all the corresponding eigenspaces.
A smaller stabilizer group $\mathcal{W}$ with $L-r$ commuting generators, $0 \leq r \leq L$, together with the choice $\pmb{ \varphi} := \{\varphi_{k}\}_{k=1}^{L-r}$, no longer identifies a single pure state, but rather a subspace of the Hilbert space of dimension \(q^r\), known as the \emph{codespace} associated with \(\mathcal{W}\). A \emph{mixed stabilizer state} $\rho$ is defined as the maximally mixed state supported on this codespace,
\begin{equation}
\rho = \frac{P_{\Wsp,\pmb{\varphi}}}{q^r} =
q^{-L}\prod_{k=1}^{L-r} \left(\sum_{m=0}^{q-1} \omega^{-m \varphi_k} P_k^m\right)
\label{eq:rhofromWphi}
\end{equation}
where \(P_{\Wsp,\pmb{\varphi}}\) denotes the projector onto the codespace. The second equality follows from the fact that $\mathcal{W}$ is a group. 
The rank of $\rho$ equals $q^{r}$ and a larger $\Wsp$ ($|\Wsp|=q^{\,L-r}$) means more constraints and a smaller rank: the integer $r$ counts the unconstrained qudits and is independent of the choice of $\pmb{\varphi}$.

We quantify how mixed $\rho$ is by the R\'enyi entropies
\be
S_n(\rho)=\frac{1}{1-n}\log_q\Tr\rho^{\,n}\qquad(n>0),
\label{eq:renyi}
\ee
which reduces to the von Neumann entropy for $n\to1$. The purity is related to the $n=2$ entropy by $\mathcal P=\Tr\rho^2=q^{-S_2}$. A mixed stabilizer density matrix has $q^r$ equal eigenvalues $q^{-r}$ and all R\'enyi entropies coincide,
\be
S_n(\rho)=r\ \ \text{(in units of $\log q$, for every $n$)}.
\label{eq:obs}
\ee

The key object accompanying $\Wsp$ is its \emph{symplectic complement} (or centralizer)
\be
\Wsp^{\perp}=\big\{\mathbf v\in\mathbb F_q^{2L}:\ \langle\mathbf v,\mathbf w\rangle=0\ \ \forall\,\mathbf w\in\Wsp\big\},
\label{eq:Wperp}
\ee
or equivalently, the set of Pauli strings commuting with \emph{every} stabilizer in $\mathcal{W}$. As $\mathcal{W}$ is abelian, $\mathcal{W}^{\perp} \supset \mathcal{W}$. Also, each independent generator of $\Wsp$ imposes one linear constraint, so $\Wsp^\perp$ is itself a subspace, of dimension $2L-(L-r)=L+r$ and cardinality $|\Wsp^{\perp}| =q^{\,L+r}$.

\textit{The effective global model.}---
To study the purification dynamics, we assume that the initial state is maximally mixed $\rho_0 = \id/q^L$. Equivalently, the system shares $L$ maximally entangled pairs with an $L$-qudit reference $E$, $\rho_0=\Tr_E\ket{\Phi}\!\bra{\Phi}$ with $\ket{\Phi}=\bigotimes_{i=1}^L q^{-1/2}\sum_{m=0}^{q-1}\ket{m_i m_{E_i}}$.
In the stabilizer formalism, $\rho_0$ is a completely mixed state characterized by a trivial stabilizer group $\Wsp = \{ \id \}$ and $r = L$.

We consider monitored dynamics combining Clifford unitaries and projective measurements of Pauli strings; time is intrinsically discrete and counts the circuit depth. In the weak-measurement phase, the effectiveness of the measurements is highly diluted, and a long stretch of unitary dynamics elapses between two measurements that effectively extract information. We therefore consider the minimal model that repeats two operations: 1) a unitary step, drawing a uniformly random element of the global Clifford group; 2) the projective measurement of a Pauli string $P$ with outcome $\omega^m$ (the unitary step makes the specific choice of $P$ irrelevant). We denote by $t$ the number of combined steps.
At any time $t$, the state is described by a pair $(\mathcal{W}, \pmb{\varphi})$ according to Eq.~\eqref{eq:rhofromWphi}.
Measurements slowly decrease $r$, and purification is precisely the process $r \searrow 0$. Assume the dynamics has reached a stabilizer group $\Wsp$ with $L-r$ generators. The effect of the measurement on $r$ is dictated entirely by where the Pauli string $P$ sits relative to $\Wsp$ and $\Wsp^{\perp}$:
(i) if $P\in\Wsp$, its value is already fixed and nothing happens;
(ii) if $P\notin\Wsp^{\perp}$, it fails to commute with some stabilizer, and the measurement swaps a generator, changing the state but not the rank;
(iii) if $P\in\Wsp^{\perp}\!\setminus\Wsp$, it commutes with all stabilizers yet is not itself determined, so it adds one independent constraint and \emph{divides the rank by $q$}, $r\to r-1$.
Only the strings in $\Wsp^{\perp}\!\setminus\Wsp$ purify the state decreasing $r$: this geometric fact is what controls purification.
Furthermore, in cases (ii) and (iii) all $q$ outcomes $\omega^m$ are equally likely under the Born rule, and the post-measurement rank depends on the class of $P$ but not on $m$: the sum over outcomes cancels the Born weights exactly, so that the trajectory average reduces to the statistics of the three classes and $r$ remains the only dynamical variable [see Sec.~A in the End Matter; for $N\neq1$ replicas a residual Born weight survives (Sec.~D)]. The class probabilities amount to ratios of cardinalities. There are $q^{2L}-1$ nontrivial strings (all Pauli vectors except the identity); the rank-reducing ones of class (iii) are the elements of $\Wsp^{\perp}\!\setminus\Wsp$, in number $|\Wsp^{\perp}|-|\Wsp|=q^{L+r}-q^{L-r}$. Hence, the probability that the rank drops, $r\to r-1$, is
\be
p_L(r)=\frac{|\Wsp^{\perp}\!\setminus\Wsp|}{q^{2L}-1}=\frac{q^{L+r}-q^{L-r}}{q^{2L}-1}.
\label{eq:pL}
\ee
For completeness, class (i) comprises the $|\Wsp|-1=q^{L-r}-1$ already-determined strings and class (ii) the remaining
$q^{2L}-q^{L+r}$ rank-preserving ones; the three probabilities sum to one (see Sec.~A in the End Matter). The induced chain for $r$ is a discrete-time pure-death process, $r\to r-1$ with probability $p_L(r)$ and $r\to r$ otherwise. For fixed $r$ and large $L$, the $-1$ in the denominator is negligible, and
\be
p_L(r)\;\xrightarrow{L\to\infty}\; \gamma_r\,q^{-L},\qquad \gamma_r=q^{r}-q^{-r}.
\label{eq:rates}
\ee
The prefactor $q^{-L}$ identifies the purification timescale. In general, $\tp(L)$ is a \emph{nonuniversal} quantity, fixed by the model-dependent free-energy cost of a replica domain wall; in the global toy model, this quantity equals the Hilbert space dimension $\tp=q^{L}$
exactly as for the Haar random-matrix ensembles studied previously~\cite{deluca2025universality,gerbino2026universal}. We now consider the scaling limit of a large number of steps $t$ and a large number of qudits $L$, fixing the scaled time $x=t/\tp=q^{-L}t$.

\textit{Scaling limit: a death process from infinity.}---In the limit $L\to\infty$ at fixed $x$, the discrete chain converges to a continuous-time pure-death process $R(x)$ on $r=0,1,2,\dots$ with rates $\gamma_r$ from \eqref{eq:rates}. The master equation for the marginal $P(r, x) := \operatorname{Prob}(R(x) = r)$ reads
\be
\partial_x P(r,x)=\gamma_{r+1}P(r+1,x)-\gamma_r P(r,x),\quad \gamma_0=0,
\label{eq:master}
\ee
where $x$ plays the role of a continuous time. We denote by $\exv[\ldots]$ the expectation with respect to the $R(x)$-process, and by $\average{\ldots}$ the Born average of observables: as shown in Sec.~A of the End Matter, once the measurement outcomes are summed with their Born weights, the former is the only average left, e.g.\ $\average{S(x)}=\exv[R(x)]$.
Because the rates grow as $\gamma_r\simeq q^r$, the process \emph{descends from infinity} in finite rescaled time: starting from $r=L\to\infty$, the front reaches any fixed level almost instantaneously and then slows down. We define the waiting times $\tau_r$ as the time spent at $R=r$. They follow the exponential distribution with rate $\gamma_r$, i.e. $\tau_r \sim \mathrm{Exp}(\gamma_r)$. Then, with the entrance (first-passage) times $\Tpass_r=\sum_{k>r}\tau_k$ at which the rank first reaches level $r$, one has $R(x)=r\Leftrightarrow \Tpass_r\le x<\Tpass_{r-1}$, and the Laplace transform of $P$ is (see Sec.~B in the End Matter)
\be
\hat P(r,u)=\int_0^\infty\!\dd x\, \e^{-ux}P(r,x)=\frac1{\gamma_r}\prod_{k\ge r}\Big(1+\frac{u}{\gamma_k}\Big)^{-1}.
\label{eq:laplace}
\ee
All universal observables follow from \eqref{eq:master}--\eqref{eq:laplace}.

\textit{Universal entropy and purity.}---By \eqref{eq:obs} the averaged R\'enyi entropies are all equal to the mean rank, $\average{S(x)}=\exv[R(x)] = \sum_r P(r,x) r$, while the averaged purity is the first negative moment, $\average{\mathcal P(x)}=\exv[q^{-R(x)}]$. Whereas $\average{S}$ starts large and decreases, the purity starts small and grows to one. To explicitly compute them, we introduce the moments
\be
m_s(x):=\exv[q^{-sR(x)}]=\sum_{r=0}^\infty q^{-sr}P(r,x),
\ee
so that $\average{\mathcal P}=m_1$ and, more generally, $\exv[\Tr\rho^{s}]=m_{s-1}$. From \eqref{eq:master} we obtain the hierarchy
\be
\tfrac{\dd}{\dd x}m_s=(q^s-1)\,(m_{s-1}-m_{s+1}),\qquad m_0=1 \;.
\label{eq:momhier}
\ee
This equation is exact for any $s \in \mathbb{R}^+$, but for $s = n \in \mathbb{N}^+$, it provides an effective recursion.
We insert the ansatz expansion $m_n(x)=x^{n}\sum_{k\ge0}a_k(n)x^{2k}$ into
\eqref{eq:momhier} and obtain the recursion
\be
(n+2k)\,a_k(n)=(q^n-1)\big[a_k(n-1)-a_{k-1}(n+1)\big],
\label{eq:akrec}
\ee
with $a_{-1}\equiv0$. The solution of the above equation for $k=0$ gives the leading coefficient in closed form
\begin{equation}
    a_0(n) = (-1)^n\frac{(q;q)_n}{\Gamma(n+1)}\,,
    \label{eq:a0}
\end{equation}
with $(a;p)_n = \prod_{j=0}^{n-1} (1 - a p^j)$ the $q$-Pochhammer symbol. Hence, we obtain $\exv[\Tr\rho^{n}] =  a_0(n-1)\,x^{\,n-1} + O(x^{n+1})$ at small $x$.
The $a_1(n)$ can also be obtained in closed form, see Eq.~\eqref{eq:b1}. In particular, for the purity, we obtain
\be
\average{\mathcal P(x)}=(q-1)\,x-(q-1)^2(q^2-1)\,\frac{x^3}{6}+O(x^5).
\label{eq:purity}
\ee
It is tempting to use these analytical expressions for the integer moments to derive the behavior of entropy, using $\average{S(x)}=-(\ln q)^{-1}\partial_s m_s|_{s=0}$, by applying the so-called “replica trick”. A more thorough analysis based on Eq.~\eqref{eq:laplace} (Sec.~G in End Matter) shows, however, that the polynomial expansion of $m_n(x)$ does not capture a non-analytic behavior at small $x$ that emerges only for $s \notin \mathbb{N}$. In fact, the correct small-$x$ expansion of entropy takes the form
\be
\average{S(x)}=\log_q\!\frac1x+C_q+\phi\!\big(\log_q\tfrac1x \big)+c_2(q)\,x^2+O(x^4) \;.
\label{eq:Ssmall}
\ee
The constant $C_q$ and the coefficient $c_2(q)$ of the quadratic term can be extracted from the analytic continuation of $a_k(n)$ around $n \to 1$ and read
\begin{align}
C_q&=\sum_{k=1}^{\infty}\frac{1}{q^{k}-1}-\frac12-\frac{\gamma_{\rm E}}{\ln q} \;,
\label{eq:C} \\
c_2(q)&=\frac{1}{2\ln q}+\frac{q\,(q^2-2q-1)}{2\,(q^2-1)} \;.
\label{eq:c2}
\end{align}
with $\gamma_{\rm E}$ the Euler-Mascheroni constant. For qubits $C_2=0.2739\ldots$ and $c_2(2)=\tfrac{1}{2\ln2}-\tfrac13$ (see End Matter). In contrast, $\phi$ is a zero-mean function, periodic of unit period, such that $\phi(\log_q 1/x) = \phi(\{\log_q 1/x\})$, where $\{\log_q(1/x)\} \in[0, 1)$ is the fractional part. By construction, it cannot be determined from the integer moments, and it requires expanding the distribution $P(r,x)$ at small $x$ (see Sec.~G in End Matter).
Comparing Eq.~\eqref{eq:Ssmall} with the expansions obtained for the dynamics of generic monitored circuits~\cite{deluca2025universality, gerbino2026universal}, we see that log-periodicity is a trait of Clifford dynamics and of the quantized rank of the stabilizer states, which persists in the form of discrete invariance $r \to r-1, x \to q x$.
Its amplitude is very small for qubits ($\sim 10^{-6}$) but grows steeply with the local size, and in general, its presence highlights the sometimes hidden subtleties of the replica trick.
The rank quantization has a second, much larger imprint on the fluctuations: whereas in generic monitored circuits the entropy becomes deterministic at small $x$, with variance of order $x^2$~\cite{deluca2025universality}, here the rank distribution converges at small $x$ to a limit shape of $O(1)$ width centered on $\log_q(1/x)$, so that the entropy variance saturates as $x\to0$ to a $q$-dependent constant, e.g.\ $\sigma_q^2\simeq0.763$ for $q=2$ (see Eq.~\eqref{eq:varR}, Sec.~G in the End Matter). The limit shape retains a residual periodic dependence on the fractional part $\theta=\{\log_q(1/x)\}$, negligible at $q=2$ but increasingly important as $q$ grows; $\sigma_q^2$ refers to its average over $\theta$.

In contrast, at late times, the process is controlled by  the last decay $r = 1 \to 0$
with the slowest rate $\gamma_1=q-q^{-1}$. The survival probability at $r=1$ controls the late-time behavior of all quantities. For the entropy
\be
\average{S(x)}\simeq A_q\,\e^{-(q-q^{-1})\,x},\qquad A_q=\frac{q^2+1}{q\,(q-1)} \;.
\label{eq:Slarge}
\ee
While the decay rate $\gamma_1$ depends only on the slowest mode, the amplitude $A_q = \prod_{k\ge2}\gamma_k/(\gamma_k-\gamma_1)$ sees all of them as it is determined by the residue of \eqref{eq:laplace} at $u=-\gamma_1$ (see Sec.~B in the End Matter). For qubits, $A_2=\tfrac52$.
Analogously, the purity large-$x$ expansion is
\be
1-\average{\mathcal P(x)}\simeq \frac{q^2+1}{q^2}\,\e^{-(q-q^{-1})x},
\label{eq:Plarge}
\ee
governed by the same spectral gap $\gamma_1$ that controls the late-time entropy \eqref{eq:Slarge}.

\begin{figure*}[t]
\centering
\includegraphics[width=0.99\textwidth]{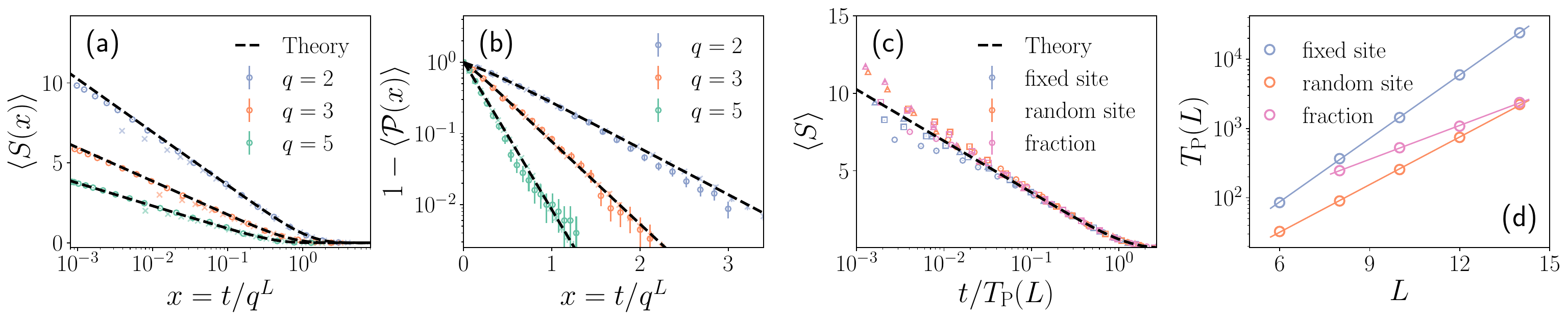}
\caption{Exact stabilizer numerics against the theory prediction (dashed), i.e.\ the death process \eqref{eq:master} evaluated through its first-passage representation \eqref{eq:laplace}. (a)~Global protocol: average entropy $\average{S(x)}$, in units of $\log q$, versus $x=t/q^L$, for $q=2$ (blue) at $L\in\{8,12\}$, $q=3$ (orange) at $L\in\{4,8\}$, and $q=5$ (green) at $L\in\{3,6\}$, the bigger size represented by circles, and the smaller one represented by crosses; the data collapse with no fitting parameter, the smaller sizes departing only in the entrance regime. (b)~Purity deficit $1-\average{\mathcal P(x)}$, same data; the late-time slopes are the spectral gaps $\gamma_1=q-q^{-1}$ and the amplitudes $(q^2+1)/q^2$, cf.~\eqref{eq:Plarge}. Error bars are smaller at the smaller sizes and grow at late times for the bigger ones. (c)~Local brick-wall circuits ($q=2$): entropy versus the renormalized time $t/\tp(L)$ for three measurement protocols -- a single fixed site, a single uniformly random site, and a small fixed fraction $p=0.06$ of the sites per step (volume phase) -- and several $L$ each  ($L = 8, 10, 14$ represented by circles, squares and triangles respectively). After the initial times, all collapse onto the same universal curve. $\tp(L)$ is fixed model by model from the single crossing $\average{S(x_0 \tp(L),L)}=\average{S(x_0)}$ at a reference level (here $\average{S}_{\rm ref}=3/2$ with $x_0 \approx 0.45$), with no fit. (d)~$\tp(L)$ versus $L$ (semilog) for the three local protocols, with exponential fits: all grow exponentially (volume law); the fixed site reproduces the global scaling $\tp\propto q^L$ ($\simeq2^{1.01L}$, prefactor $\simeq1.4$), while the random-site ($\simeq2^{0.76L}$) and fraction ($\simeq2^{0.54L}$) protocols purify faster.}
\label{fig:numerics}
\end{figure*}

\textit{Numerical checks.}---We test the theory with exact stabilizer simulations over $\mathbb F_q$. In practice, it is convenient to purify the state by treating each trajectory as the evolution of $L+L$ qudits of the system plus reference $E$, initially maximally entangled~\cite{aaronson2004improvedsimulationof}. Global Cliffords are drawn uniformly through the Bravyi--Maslov canonical form~\cite{bravyi2021hadamard} extended to odd primes~\cite{turkeshi2026sampler} and applied directly to the stabilizer generators, and the rank $r$ is read off as a symplectic rank over $\mathbb F_q$. For the global model, we simulate $q=2, 3, 5$ at two system sizes each, averaging over independent trajectories on horizons of several purification times. The data are indistinguishable from the exact chain \eqref{eq:pL}, with residuals within statistical accuracy over the entire time range. Upon rescaling $x=t/q^L$, different system sizes per $q$ collapse directly onto the universal curves [Fig.~\ref{fig:numerics}(a,b)]. The smaller sizes show discrepancy at small $x$: larger $L$ are required to approach the initially large value of $\average{S(x)} \sim -\log_q x$ in Eq.~\eqref{eq:Ssmall}.

We then test universality on a local $(1{+}1)$-dimensional circuit: each time step applies a brick-wall layer of uniformly random two-qudit Cliffords (one even and one odd sublayer with open boundary conditions)  followed by a computational-basis measurement. We consider three measurement protocols: a single fixed site, a single uniformly random site, and a small fixed fraction $p$ of the sites per step, which extract information at different rates and through different geometries. We keep the fraction of sites small enough to remain in the volume-law phase ($p = 0.06 < p_c \approx 0.16$) \cite{li2019measurementdrivenentanglement}. 
For each protocol and size, we fix $\tp(L)$ by a single crossing condition, $\average{S(x_0 \tp(L),L)}=\average{S(x_0)}$ at one reference level $x_0$; universality then shows up as the convergence of the rescaled curves to the theory for all $x \neq x_0$ as $L$ increases. Remarkably, all protocols show quick collapse in system size onto the same universal curve [Fig.~\ref{fig:numerics}(c)], confirming that locality and the choice of measurement leave the scaling function untouched.
The model-dependent timescale is shown in Fig.~\ref{fig:numerics}(d): $\tp(L)$ grows exponentially with $L$ for every protocol, consistently with the volume-law phase. The fixed-site dynamics retains a very slow purification, reproducing the global scaling $\tp\propto q^L$ up to an $O(1)$ prefactor ($\simeq1.4$ at $q=2$). Instead, the random-site and fraction protocols purify faster, with a smaller exponent set by the less efficient scrambling properties of the local brick-wall circuit. Thus, the universal content -- the scaling functions and constants of Eqs.~\eqref{eq:purity}--\eqref{eq:Ssmall} -- is entirely decoupled from the nonuniversal $\tp(L)$.

Our results establish that, after a single renormalization of the timescale, global and local Clifford dynamics follow the same exactly computed universal curve, with no further fitting parameters. 

\textit{Discussion and outlook.}---We have shown that the universal slow-purification
dynamics of monitored circuits, normally accessed through a replica
statistical-mechanics mapping, is captured \emph{exactly} for Clifford dynamics at any
prime local dimension by the Markovian decay of the stabilizer rank. The interest of
this result is twofold. On the one hand, Clifford circuits are the paradigmatic
efficiently simulable many-body dynamics, at the core of quantum error correction and
of the monitored physics probed on present-day quantum
platforms~\cite{Noel_2022,google2021,koh2022experimentalrealizationof,feng2025postselectionfreeexperimentalobservationmeasurementinduced}.
On the other hand, they provide a tractable arena for the purification problem: the
full complexity of the monitored quantum dynamics is encoded in a purely classical
stochastic process -- the pure-death decay of the stabilizer rank --
from which all the universal scaling functions follow in compact form.

Our exact solution also settles the fate of purification universality for Clifford
circuits, a question raised in Ref.~\cite{deluca2025universality}: the Clifford process
defines a dynamical universality class of its own, distinct from those of generic
monitored dynamics, whether unitary~\cite{deluca2025universality} or
orthogonal~\cite{gerbino2026universal}. The agreement with the generic classes is confined to very short times, where the entropy displays the
characteristic $-\log_q x+\mathrm{const}$ behavior;
even there, unlike its generic counterpart, the constant $C_q$ depends explicitly on $q$.
The hallmarks anticipated in Ref.~\cite{deluca2025universality} are confirmed and
sharpened here: all R\'enyi entropies coincide and are quantized in units of $\log q$,
and the quantization survives the scaling limit both as the log-periodic modulation
$\phi$ of Eq.~\eqref{eq:Ssmall} and as the $O(1)$ saturation of the entropy
fluctuations, Eq.~\eqref{eq:varR}, in sharp contrast with the vanishing $O(x^2)$
fluctuations of the generic classes.

Turning to perspectives, real-valued Clifford ensembles should provide an exactly solvable stabilizer counterpart of the orthogonal
universality class recently identified for real monitored
dynamics~\cite{gerbino2026universal,magni2025anticoncentrationstatedesigndoped}.
Conversely, these results delimit the regime where slow-purification universality can be expected, its mechanism being the free-energy barrier from the breaking of replica permutation symmetry. Monitored systems lacking this structure -- charge-sharpening transitions, which are not driven by permutation-symmetry breaking~\cite{agrawal2022entanglmentandchargesharpening,bao2021symmetryenrichedphases,nahum2025bayesian}, or Gaussian fermions, governed by a continuous replica symmetry and nonlinear sigma models~\cite{fava2023nonlinear,poboiko2023measurementinduced, loio2023, fava2024monitored, buchhold2022revealingmeasurementinduced, PhysRevB.110.054313, PhysRevResearch.4.033001, PhysRevB.110.L140301, PhysRevB.111.224301, PhysRevB.108.094304, zhu2023qubitfractionalizationemergentmajorana} -- purify on far shorter timescales, and whether they support an analogous universal regime is an open question. A sharper direction is to dope
the Clifford dynamics with magic $T$ ~\cite{Leone2024learningtdoped,bejan2024dynamical,lami2024quantum,haug2024probingquantumcomplexityuniversal,fux2025disentanglingunitarydynamicsclassically,nakhl2025stabilizer,liu2024classicalsimulabilitycliffordtcircuits,Leone_2026,turkeshi2025magicspreadingrandomquantum, aditya2025mpembaeffectsquantumcomplexity,paviglianiti2025emergencegenericentanglementstructure,loio2025quantumstatedesignsmagic,aditya2025growthspreadingquantumresources,varikuti2025impactcliffordoperationsnonstabilizing,haferkamp2022, Heinrich2019robustnessofmagic, True2022transitionsin, Magni_2025, p8dn-glcw}, interpolating between the
Clifford and unitary classes: in the absence of measurements, a (poly)logarithmic
number of $T$ states suffices to restore Haar-like anticoncentration~\cite{p8dn-glcw};
how much magic is needed to erase these Clifford fingerprints and recover the
unitary scaling functions is a well-posed problem. More broadly, the stabilizer rank
stands out as a replica-free laboratory for the universal physics
of monitored many-body systems.

\begin{acknowledgements}
\textit{Acknowledgements.---}
    This work is dedicated to Ada.
    B.M. and X.T. acknowledge support from DFG under Germany's Excellence Strategy – Cluster of Excellence Matter and Light for Quantum Computing (ML4Q) EXC 2004/2 – 390534769, and DFG Collaborative Research Center (CRC) 183 Project No. 277101999 - project B01, and DFG Emmy Noether Programme proposal ``Digital Quantum Matter Out-of-Equilibrium'' No. 560726973.

\textit{Code and Data Availability.---}
Code and Data will be publicly shared in Zenodo at publication.

\end{acknowledgements}

\bibliography{arxiv0707/biblio}

\begin{thebibliography}{114}%
\makeatletter
\providecommand \@ifxundefined [1]{%
 \@ifx{#1\undefined}
}%
\providecommand \@ifnum [1]{%
 \ifnum #1\expandafter \@firstoftwo
 \else \expandafter \@secondoftwo
 \fi
}%
\providecommand \@ifx [1]{%
 \ifx #1\expandafter \@firstoftwo
 \else \expandafter \@secondoftwo
 \fi
}%
\providecommand \natexlab [1]{#1}%
\providecommand \enquote  [1]{``#1''}%
\providecommand \bibnamefont  [1]{#1}%
\providecommand \bibfnamefont [1]{#1}%
\providecommand \citenamefont [1]{#1}%
\providecommand \href@noop [0]{\@secondoftwo}%
\providecommand \href [0]{\begingroup \@sanitize@url \@href}%
\providecommand \@href[1]{\@@startlink{#1}\@@href}%
\providecommand \@@href[1]{\endgroup#1\@@endlink}%
\providecommand \@sanitize@url [0]{\catcode `\\12\catcode `\$12\catcode `\&12\catcode `\#12\catcode `\^12\catcode `\_12\catcode `\%12\relax}%
\providecommand \@@startlink[1]{}%
\providecommand \@@endlink[0]{}%
\providecommand \url  [0]{\begingroup\@sanitize@url \@url }%
\providecommand \@url [1]{\endgroup\@href {#1}{\urlprefix }}%
\providecommand \urlprefix  [0]{URL }%
\providecommand \Eprint [0]{\href }%
\providecommand \doibase [0]{https://doi.org/}%
\providecommand \selectlanguage [0]{\@gobble}%
\providecommand \bibinfo  [0]{\@secondoftwo}%
\providecommand \bibfield  [0]{\@secondoftwo}%
\providecommand \translation [1]{[#1]}%
\providecommand \BibitemOpen [0]{}%
\providecommand \bibitemStop [0]{}%
\providecommand \bibitemNoStop [0]{.\EOS\space}%
\providecommand \EOS [0]{\spacefactor3000\relax}%
\providecommand \BibitemShut  [1]{\csname bibitem#1\endcsname}%
\let\auto@bib@innerbib\@empty
%</preamble>
\bibitem [{\citenamefont {Nahum}\ \emph {et~al.}(2017)\citenamefont {Nahum}, \citenamefont {Ruhman}, \citenamefont {Vijay},\ and\ \citenamefont {Haah}}]{nahum2017quantum}%
  \BibitemOpen
  \bibfield  {author} {\bibinfo {author} {\bibfnamefont {A.}~\bibnamefont {Nahum}}, \bibinfo {author} {\bibfnamefont {J.}~\bibnamefont {Ruhman}}, \bibinfo {author} {\bibfnamefont {S.}~\bibnamefont {Vijay}},\ and\ \bibinfo {author} {\bibfnamefont {J.}~\bibnamefont {Haah}},\ }\bibfield  {title} {\bibinfo {title} {Quantum entanglement growth under random unitary dynamics},\ }\href {https://doi.org/10.1103/PhysRevX.7.031016} {\bibfield  {journal} {\bibinfo  {journal} {Phys. Rev. X}\ }\textbf {\bibinfo {volume} {7}},\ \bibinfo {pages} {031016} (\bibinfo {year} {2017})}\BibitemShut {NoStop}%
\bibitem [{\citenamefont {Nahum}\ \emph {et~al.}(2018)\citenamefont {Nahum}, \citenamefont {Vijay},\ and\ \citenamefont {Haah}}]{nahum2018operator}%
  \BibitemOpen
  \bibfield  {author} {\bibinfo {author} {\bibfnamefont {A.}~\bibnamefont {Nahum}}, \bibinfo {author} {\bibfnamefont {S.}~\bibnamefont {Vijay}},\ and\ \bibinfo {author} {\bibfnamefont {J.}~\bibnamefont {Haah}},\ }\bibfield  {title} {\bibinfo {title} {Operator spreading in random unitary circuits},\ }\href {https://doi.org/10.1103/PhysRevX.8.021014} {\bibfield  {journal} {\bibinfo  {journal} {Phys. Rev. X}\ }\textbf {\bibinfo {volume} {8}},\ \bibinfo {pages} {021014} (\bibinfo {year} {2018})}\BibitemShut {NoStop}%
\bibitem [{\citenamefont {von Keyserlingk}\ \emph {et~al.}(2018)\citenamefont {von Keyserlingk}, \citenamefont {Rakovszky}, \citenamefont {Pollmann},\ and\ \citenamefont {Sondhi}}]{von_Keyserlingk_2018}%
  \BibitemOpen
  \bibfield  {author} {\bibinfo {author} {\bibfnamefont {C.~W.}\ \bibnamefont {von Keyserlingk}}, \bibinfo {author} {\bibfnamefont {T.}~\bibnamefont {Rakovszky}}, \bibinfo {author} {\bibfnamefont {F.}~\bibnamefont {Pollmann}},\ and\ \bibinfo {author} {\bibfnamefont {S.~L.}\ \bibnamefont {Sondhi}},\ }\bibfield  {title} {\bibinfo {title} {Operator hydrodynamics, otocs, and entanglement growth in systems without conservation laws},\ }\href {https://doi.org/10.1103/PhysRevX.8.021013} {\bibfield  {journal} {\bibinfo  {journal} {Phys. Rev. X}\ }\textbf {\bibinfo {volume} {8}},\ \bibinfo {pages} {021013} (\bibinfo {year} {2018})}\BibitemShut {NoStop}%
\bibitem [{\citenamefont {Fisher}\ \emph {et~al.}(2023)\citenamefont {Fisher}, \citenamefont {Khemani}, \citenamefont {Nahum},\ and\ \citenamefont {Vijay}}]{fisher2023}%
  \BibitemOpen
  \bibfield  {author} {\bibinfo {author} {\bibfnamefont {M.~P.}\ \bibnamefont {Fisher}}, \bibinfo {author} {\bibfnamefont {V.}~\bibnamefont {Khemani}}, \bibinfo {author} {\bibfnamefont {A.}~\bibnamefont {Nahum}},\ and\ \bibinfo {author} {\bibfnamefont {S.}~\bibnamefont {Vijay}},\ }\bibfield  {title} {\bibinfo {title} {Random quantum circuits},\ }\href {https://doi.org/10.1146/annurev-conmatphys-031720-030658} {\bibfield  {journal} {\bibinfo  {journal} {Ann. Rev. Cond. Matter Phys.}\ }\textbf {\bibinfo {volume} {14}},\ \bibinfo {pages} {335} (\bibinfo {year} {2023})}\BibitemShut {NoStop}%
\bibitem [{\citenamefont {Potter}\ and\ \citenamefont {Vasseur}(2022)}]{potter2022entanglementdynamicsin}%
  \BibitemOpen
  \bibfield  {author} {\bibinfo {author} {\bibfnamefont {A.~C.}\ \bibnamefont {Potter}}\ and\ \bibinfo {author} {\bibfnamefont {R.}~\bibnamefont {Vasseur}},\ }\bibinfo {title} {Entanglement dynamics in hybrid quantum circuits},\ in\ \href {https://doi.org/10.1007/978-3-031-03998-0_9} {\emph {\bibinfo {booktitle} {Entanglement in Spin Chains: From Theory to Quantum Technology Applications}}},\ \bibinfo {editor} {edited by\ \bibinfo {editor} {\bibfnamefont {A.}~\bibnamefont {Bayat}}, \bibinfo {editor} {\bibfnamefont {S.}~\bibnamefont {Bose}},\ and\ \bibinfo {editor} {\bibfnamefont {H.}~\bibnamefont {Johannesson}}}\ (\bibinfo  {publisher} {Springer},\ \bibinfo {address} {Cham},\ \bibinfo {year} {2022})\ pp.\ \bibinfo {pages} {211--249}\BibitemShut {NoStop}%
\bibitem [{\citenamefont {Lunt}\ \emph {et~al.}(2022)\citenamefont {Lunt}, \citenamefont {Richter},\ and\ \citenamefont {Pal}}]{lunt2022quantumsimulationusing}%
  \BibitemOpen
  \bibfield  {author} {\bibinfo {author} {\bibfnamefont {O.}~\bibnamefont {Lunt}}, \bibinfo {author} {\bibfnamefont {J.}~\bibnamefont {Richter}},\ and\ \bibinfo {author} {\bibfnamefont {A.}~\bibnamefont {Pal}},\ }\bibinfo {title} {Quantum simulation using noisy unitary circuits and measurements},\ in\ \href {https://doi.org/10.1007/978-3-031-03998-0_10} {\emph {\bibinfo {booktitle} {Entanglement in Spin Chains: From Theory to Quantum Technology Applications}}},\ \bibinfo {editor} {edited by\ \bibinfo {editor} {\bibfnamefont {A.}~\bibnamefont {Bayat}}, \bibinfo {editor} {\bibfnamefont {S.}~\bibnamefont {Bose}},\ and\ \bibinfo {editor} {\bibfnamefont {H.}~\bibnamefont {Johannesson}}}\ (\bibinfo  {publisher} {Springer},\ \bibinfo {address} {Cham},\ \bibinfo {year} {2022})\ pp.\ \bibinfo {pages} {251--284}\BibitemShut {NoStop}%
\bibitem [{\citenamefont {Sierant}\ \emph {et~al.}(2023)\citenamefont {Sierant}, \citenamefont {Schir\`o}, \citenamefont {Lewenstein},\ and\ \citenamefont {Turkeshi}}]{sierant2023}%
  \BibitemOpen
  \bibfield  {author} {\bibinfo {author} {\bibfnamefont {P.}~\bibnamefont {Sierant}}, \bibinfo {author} {\bibfnamefont {M.}~\bibnamefont {Schir\`o}}, \bibinfo {author} {\bibfnamefont {M.}~\bibnamefont {Lewenstein}},\ and\ \bibinfo {author} {\bibfnamefont {X.}~\bibnamefont {Turkeshi}},\ }\bibfield  {title} {\bibinfo {title} {Entanglement growth and minimal membranes in ($d+1$) random unitary circuits},\ }\href {https://doi.org/10.1103/PhysRevLett.131.230403} {\bibfield  {journal} {\bibinfo  {journal} {Phys. Rev. Lett.}\ }\textbf {\bibinfo {volume} {131}},\ \bibinfo {pages} {230403} (\bibinfo {year} {2023})}\BibitemShut {NoStop}%
\bibitem [{\citenamefont {Ippoliti}\ \emph {et~al.}(2021)\citenamefont {Ippoliti}, \citenamefont {Gullans}, \citenamefont {Gopalakrishnan}, \citenamefont {Huse},\ and\ \citenamefont {Khemani}}]{ippoliti2021entanglement}%
  \BibitemOpen
  \bibfield  {author} {\bibinfo {author} {\bibfnamefont {M.}~\bibnamefont {Ippoliti}}, \bibinfo {author} {\bibfnamefont {M.~J.}\ \bibnamefont {Gullans}}, \bibinfo {author} {\bibfnamefont {S.}~\bibnamefont {Gopalakrishnan}}, \bibinfo {author} {\bibfnamefont {D.~A.}\ \bibnamefont {Huse}},\ and\ \bibinfo {author} {\bibfnamefont {V.}~\bibnamefont {Khemani}},\ }\bibfield  {title} {\bibinfo {title} {Entanglement phase transitions in measurement-only dynamics},\ }\href {https://doi.org/10.1103/PhysRevX.11.011030} {\bibfield  {journal} {\bibinfo  {journal} {Phys. Rev. X}\ }\textbf {\bibinfo {volume} {11}},\ \bibinfo {pages} {011030} (\bibinfo {year} {2021})}\BibitemShut {NoStop}%
\bibitem [{\citenamefont {Sierant}\ \emph {et~al.}(2022{\natexlab{a}})\citenamefont {Sierant}, \citenamefont {Schir\`o}, \citenamefont {Lewenstein},\ and\ \citenamefont {Turkeshi}}]{sierant2022measurement}%
  \BibitemOpen
  \bibfield  {author} {\bibinfo {author} {\bibfnamefont {P.}~\bibnamefont {Sierant}}, \bibinfo {author} {\bibfnamefont {M.}~\bibnamefont {Schir\`o}}, \bibinfo {author} {\bibfnamefont {M.}~\bibnamefont {Lewenstein}},\ and\ \bibinfo {author} {\bibfnamefont {X.}~\bibnamefont {Turkeshi}},\ }\bibfield  {title} {\bibinfo {title} {Measurement-induced phase transitions in $(d+1)$-dimensional stabilizer circuits},\ }\href {https://doi.org/10.1103/PhysRevB.106.214316} {\bibfield  {journal} {\bibinfo  {journal} {Phys. Rev. B}\ }\textbf {\bibinfo {volume} {106}},\ \bibinfo {pages} {214316} (\bibinfo {year} {2022}{\natexlab{a}})}\BibitemShut {NoStop}%
\bibitem [{\citenamefont {Li}\ \emph {et~al.}(2018)\citenamefont {Li}, \citenamefont {Chen},\ and\ \citenamefont {Fisher}}]{li2018quantum}%
  \BibitemOpen
  \bibfield  {author} {\bibinfo {author} {\bibfnamefont {Y.}~\bibnamefont {Li}}, \bibinfo {author} {\bibfnamefont {X.}~\bibnamefont {Chen}},\ and\ \bibinfo {author} {\bibfnamefont {M.~P.~A.}\ \bibnamefont {Fisher}},\ }\bibfield  {title} {\bibinfo {title} {Quantum zeno effect and the many-body entanglement transition},\ }\href {https://doi.org/10.1103/PhysRevB.98.205136} {\bibfield  {journal} {\bibinfo  {journal} {Phys. Rev. B}\ }\textbf {\bibinfo {volume} {98}},\ \bibinfo {pages} {205136} (\bibinfo {year} {2018})}\BibitemShut {NoStop}%
\bibitem [{\citenamefont {Li}\ \emph {et~al.}(2019)\citenamefont {Li}, \citenamefont {Chen},\ and\ \citenamefont {Fisher}}]{li2019measurementdrivenentanglement}%
  \BibitemOpen
  \bibfield  {author} {\bibinfo {author} {\bibfnamefont {Y.}~\bibnamefont {Li}}, \bibinfo {author} {\bibfnamefont {X.}~\bibnamefont {Chen}},\ and\ \bibinfo {author} {\bibfnamefont {M.~P.~A.}\ \bibnamefont {Fisher}},\ }\bibfield  {title} {\bibinfo {title} {Measurement-driven entanglement transition in hybrid quantum circuits},\ }\href {https://doi.org/10.1103/PhysRevB.100.134306} {\bibfield  {journal} {\bibinfo  {journal} {Phys. Rev. B}\ }\textbf {\bibinfo {volume} {100}},\ \bibinfo {pages} {134306} (\bibinfo {year} {2019})}\BibitemShut {NoStop}%
\bibitem [{\citenamefont {Skinner}\ \emph {et~al.}(2019)\citenamefont {Skinner}, \citenamefont {Ruhman},\ and\ \citenamefont {Nahum}}]{skinner2019measurementinducedphase}%
  \BibitemOpen
  \bibfield  {author} {\bibinfo {author} {\bibfnamefont {B.}~\bibnamefont {Skinner}}, \bibinfo {author} {\bibfnamefont {J.}~\bibnamefont {Ruhman}},\ and\ \bibinfo {author} {\bibfnamefont {A.}~\bibnamefont {Nahum}},\ }\bibfield  {title} {\bibinfo {title} {Measurement-induced phase transitions in the dynamics of entanglement},\ }\href {https://doi.org/10.1103/PhysRevX.9.031009} {\bibfield  {journal} {\bibinfo  {journal} {Phys. Rev. X}\ }\textbf {\bibinfo {volume} {9}},\ \bibinfo {pages} {031009} (\bibinfo {year} {2019})}\BibitemShut {NoStop}%
\bibitem [{\citenamefont {Chan}\ \emph {et~al.}(2019)\citenamefont {Chan}, \citenamefont {Nandkishore}, \citenamefont {Pretko},\ and\ \citenamefont {Smith}}]{chan2019unitaryprojective}%
  \BibitemOpen
  \bibfield  {author} {\bibinfo {author} {\bibfnamefont {A.}~\bibnamefont {Chan}}, \bibinfo {author} {\bibfnamefont {R.~M.}\ \bibnamefont {Nandkishore}}, \bibinfo {author} {\bibfnamefont {M.}~\bibnamefont {Pretko}},\ and\ \bibinfo {author} {\bibfnamefont {G.}~\bibnamefont {Smith}},\ }\bibfield  {title} {\bibinfo {title} {Unitary-projective entanglement dynamics},\ }\href {https://doi.org/10.1103/PhysRevB.99.224307} {\bibfield  {journal} {\bibinfo  {journal} {Phys. Rev. B}\ }\textbf {\bibinfo {volume} {99}},\ \bibinfo {pages} {224307} (\bibinfo {year} {2019})}\BibitemShut {NoStop}%
\bibitem [{\citenamefont {Szyniszewski}\ \emph {et~al.}(2019)\citenamefont {Szyniszewski}, \citenamefont {Romito},\ and\ \citenamefont {Schomerus}}]{szyniszewski2019entanglementtransitionfrom}%
  \BibitemOpen
  \bibfield  {author} {\bibinfo {author} {\bibfnamefont {M.}~\bibnamefont {Szyniszewski}}, \bibinfo {author} {\bibfnamefont {A.}~\bibnamefont {Romito}},\ and\ \bibinfo {author} {\bibfnamefont {H.}~\bibnamefont {Schomerus}},\ }\bibfield  {title} {\bibinfo {title} {Entanglement transition from variable-strength weak measurements},\ }\href {https://doi.org/10.1103/PhysRevB.100.064204} {\bibfield  {journal} {\bibinfo  {journal} {Phys. Rev. B}\ }\textbf {\bibinfo {volume} {100}},\ \bibinfo {pages} {064204} (\bibinfo {year} {2019})}\BibitemShut {NoStop}%
\bibitem [{\citenamefont {Zabalo}\ \emph {et~al.}(2020)\citenamefont {Zabalo}, \citenamefont {Gullans}, \citenamefont {Wilson}, \citenamefont {Gopalakrishnan}, \citenamefont {Huse},\ and\ \citenamefont {Pixley}}]{zabalo2020criticalpropertiesof}%
  \BibitemOpen
  \bibfield  {author} {\bibinfo {author} {\bibfnamefont {A.}~\bibnamefont {Zabalo}}, \bibinfo {author} {\bibfnamefont {M.~J.}\ \bibnamefont {Gullans}}, \bibinfo {author} {\bibfnamefont {J.~H.}\ \bibnamefont {Wilson}}, \bibinfo {author} {\bibfnamefont {S.}~\bibnamefont {Gopalakrishnan}}, \bibinfo {author} {\bibfnamefont {D.~A.}\ \bibnamefont {Huse}},\ and\ \bibinfo {author} {\bibfnamefont {J.~H.}\ \bibnamefont {Pixley}},\ }\bibfield  {title} {\bibinfo {title} {Critical properties of the measurement-induced transition in random quantum circuits},\ }\href {https://doi.org/10.1103/PhysRevB.101.060301} {\bibfield  {journal} {\bibinfo  {journal} {Phys. Rev. B}\ }\textbf {\bibinfo {volume} {101}},\ \bibinfo {pages} {060301} (\bibinfo {year} {2020})}\BibitemShut {NoStop}%
\bibitem [{\citenamefont {Turkeshi}\ \emph {et~al.}(2020)\citenamefont {Turkeshi}, \citenamefont {Fazio},\ and\ \citenamefont {Dalmonte}}]{turkeshi2020measurementinducedcriticality}%
  \BibitemOpen
  \bibfield  {author} {\bibinfo {author} {\bibfnamefont {X.}~\bibnamefont {Turkeshi}}, \bibinfo {author} {\bibfnamefont {R.}~\bibnamefont {Fazio}},\ and\ \bibinfo {author} {\bibfnamefont {M.}~\bibnamefont {Dalmonte}},\ }\bibfield  {title} {\bibinfo {title} {Measurement-induced criticality in $(2+1)$-dimensional hybrid quantum circuits},\ }\href {https://doi.org/10.1103/PhysRevB.102.014315} {\bibfield  {journal} {\bibinfo  {journal} {Phys. Rev. B}\ }\textbf {\bibinfo {volume} {102}},\ \bibinfo {pages} {014315} (\bibinfo {year} {2020})}\BibitemShut {NoStop}%
\bibitem [{\citenamefont {Turkeshi}(2022)}]{turkeshi2022measurementinducedcriticality}%
  \BibitemOpen
  \bibfield  {author} {\bibinfo {author} {\bibfnamefont {X.}~\bibnamefont {Turkeshi}},\ }\bibfield  {title} {\bibinfo {title} {Measurement-induced criticality as a data-structure transition},\ }\href {https://doi.org/10.1103/PhysRevB.106.144313} {\bibfield  {journal} {\bibinfo  {journal} {Phys. Rev. B}\ }\textbf {\bibinfo {volume} {106}},\ \bibinfo {pages} {144313} (\bibinfo {year} {2022})}\BibitemShut {NoStop}%
\bibitem [{\citenamefont {Sierant}\ and\ \citenamefont {Turkeshi}(2022)}]{sierant2022universalbehaviorbeyond}%
  \BibitemOpen
  \bibfield  {author} {\bibinfo {author} {\bibfnamefont {P.}~\bibnamefont {Sierant}}\ and\ \bibinfo {author} {\bibfnamefont {X.}~\bibnamefont {Turkeshi}},\ }\bibfield  {title} {\bibinfo {title} {Universal behavior beyond multifractality of wave functions at measurement-induced phase transitions},\ }\href {https://doi.org/10.1103/PhysRevLett.128.130605} {\bibfield  {journal} {\bibinfo  {journal} {Phys. Rev. Lett.}\ }\textbf {\bibinfo {volume} {128}},\ \bibinfo {pages} {130605} (\bibinfo {year} {2022})}\BibitemShut {NoStop}%
\bibitem [{\citenamefont {Block}\ \emph {et~al.}(2022)\citenamefont {Block}, \citenamefont {Bao}, \citenamefont {Choi}, \citenamefont {Altman},\ and\ \citenamefont {Yao}}]{block2022measurement}%
  \BibitemOpen
  \bibfield  {author} {\bibinfo {author} {\bibfnamefont {M.}~\bibnamefont {Block}}, \bibinfo {author} {\bibfnamefont {Y.}~\bibnamefont {Bao}}, \bibinfo {author} {\bibfnamefont {S.}~\bibnamefont {Choi}}, \bibinfo {author} {\bibfnamefont {E.}~\bibnamefont {Altman}},\ and\ \bibinfo {author} {\bibfnamefont {N.~Y.}\ \bibnamefont {Yao}},\ }\bibfield  {title} {\bibinfo {title} {Measurement-induced transition in long-range interacting quantum circuits},\ }\href {https://doi.org/10.1103/PhysRevLett.128.010604} {\bibfield  {journal} {\bibinfo  {journal} {Phys. Rev. Lett.}\ }\textbf {\bibinfo {volume} {128}},\ \bibinfo {pages} {010604} (\bibinfo {year} {2022})}\BibitemShut {NoStop}%
\bibitem [{\citenamefont {Altland}\ \emph {et~al.}(2022)\citenamefont {Altland}, \citenamefont {Buchhold}, \citenamefont {Diehl},\ and\ \citenamefont {Micklitz}}]{PhysRevResearch.4.L022066}%
  \BibitemOpen
  \bibfield  {author} {\bibinfo {author} {\bibfnamefont {A.}~\bibnamefont {Altland}}, \bibinfo {author} {\bibfnamefont {M.}~\bibnamefont {Buchhold}}, \bibinfo {author} {\bibfnamefont {S.}~\bibnamefont {Diehl}},\ and\ \bibinfo {author} {\bibfnamefont {T.}~\bibnamefont {Micklitz}},\ }\bibfield  {title} {\bibinfo {title} {Dynamics of measured many-body quantum chaotic systems},\ }\href {https://doi.org/10.1103/PhysRevResearch.4.L022066} {\bibfield  {journal} {\bibinfo  {journal} {Phys. Rev. Res.}\ }\textbf {\bibinfo {volume} {4}},\ \bibinfo {pages} {L022066} (\bibinfo {year} {2022})}\BibitemShut {NoStop}%
\bibitem [{\citenamefont {Han}\ and\ \citenamefont {Chen}(2023)}]{PhysRevB.107.014306}%
  \BibitemOpen
  \bibfield  {author} {\bibinfo {author} {\bibfnamefont {Y.}~\bibnamefont {Han}}\ and\ \bibinfo {author} {\bibfnamefont {X.}~\bibnamefont {Chen}},\ }\bibfield  {title} {\bibinfo {title} {Entanglement structure in the volume-law phase of hybrid quantum automaton circuits},\ }\href {https://doi.org/10.1103/PhysRevB.107.014306} {\bibfield  {journal} {\bibinfo  {journal} {Phys. Rev. B}\ }\textbf {\bibinfo {volume} {107}},\ \bibinfo {pages} {014306} (\bibinfo {year} {2023})}\BibitemShut {NoStop}%
\bibitem [{\citenamefont {Sharma}\ \emph {et~al.}(2022)\citenamefont {Sharma}, \citenamefont {Turkeshi}, \citenamefont {Fazio},\ and\ \citenamefont {Dalmonte}}]{SciPostPhysCore.5.2.023}%
  \BibitemOpen
  \bibfield  {author} {\bibinfo {author} {\bibfnamefont {S.}~\bibnamefont {Sharma}}, \bibinfo {author} {\bibfnamefont {X.}~\bibnamefont {Turkeshi}}, \bibinfo {author} {\bibfnamefont {R.}~\bibnamefont {Fazio}},\ and\ \bibinfo {author} {\bibfnamefont {M.}~\bibnamefont {Dalmonte}},\ }\bibfield  {title} {\bibinfo {title} {Measurement-induced criticality in extended and long-range unitary circuits},\ }\href {https://doi.org/10.21468/SciPostPhysCore.5.2.023} {\bibfield  {journal} {\bibinfo  {journal} {SciPost Phys. Core}\ }\textbf {\bibinfo {volume} {5}},\ \bibinfo {pages} {023} (\bibinfo {year} {2022})}\BibitemShut {NoStop}%
\bibitem [{\citenamefont {Iadecola}\ \emph {et~al.}(2025)\citenamefont {Iadecola}, \citenamefont {Wilson},\ and\ \citenamefont {Pixley}}]{PRXQuantum.6.010351}%
  \BibitemOpen
  \bibfield  {author} {\bibinfo {author} {\bibfnamefont {T.}~\bibnamefont {Iadecola}}, \bibinfo {author} {\bibfnamefont {J.~H.}\ \bibnamefont {Wilson}},\ and\ \bibinfo {author} {\bibfnamefont {J.}~\bibnamefont {Pixley}},\ }\bibfield  {title} {\bibinfo {title} {Concomitant entanglement and control criticality driven by collective measurements},\ }\href {https://doi.org/10.1103/PRXQuantum.6.010351} {\bibfield  {journal} {\bibinfo  {journal} {PRX Quantum}\ }\textbf {\bibinfo {volume} {6}},\ \bibinfo {pages} {010351} (\bibinfo {year} {2025})}\BibitemShut {NoStop}%
\bibitem [{\citenamefont {Nambi}\ \emph {et~al.}(2026)\citenamefont {Nambi}, \citenamefont {Khanna}, \citenamefont {Allocca}, \citenamefont {Iadecola}, \citenamefont {Hickey}, \citenamefont {Vasseur},\ and\ \citenamefont {Wilson}}]{nambi2026postselectedcriticalitymeasurementinducedphase}%
  \BibitemOpen
  \bibfield  {author} {\bibinfo {author} {\bibfnamefont {D.}~\bibnamefont {Nambi}}, \bibinfo {author} {\bibfnamefont {K.}~\bibnamefont {Khanna}}, \bibinfo {author} {\bibfnamefont {A.}~\bibnamefont {Allocca}}, \bibinfo {author} {\bibfnamefont {T.}~\bibnamefont {Iadecola}}, \bibinfo {author} {\bibfnamefont {C.}~\bibnamefont {Hickey}}, \bibinfo {author} {\bibfnamefont {R.}~\bibnamefont {Vasseur}},\ and\ \bibinfo {author} {\bibfnamefont {J.~H.}\ \bibnamefont {Wilson}},\ }\href {https://arxiv.org/abs/2603.15744} {\bibinfo {title} {Post-selected criticality in measurement-induced phase transitions}} (\bibinfo {year} {2026}),\ \Eprint {https://arxiv.org/abs/2603.15744} {arXiv:2603.15744 [quant-ph]} \BibitemShut {NoStop}%
\bibitem [{\citenamefont {Noel}\ \emph {et~al.}(2022)\citenamefont {Noel}, \citenamefont {Niroula}, \citenamefont {Zhu}, \citenamefont {Risinger}, \citenamefont {Egan}, \citenamefont {Biswas}, \citenamefont {Cetina}, \citenamefont {Gorshkov}, \citenamefont {Gullans}, \citenamefont {Huse},\ and\ \citenamefont {Monroe}}]{Noel_2022}%
  \BibitemOpen
  \bibfield  {author} {\bibinfo {author} {\bibfnamefont {C.}~\bibnamefont {Noel}}, \bibinfo {author} {\bibfnamefont {P.}~\bibnamefont {Niroula}}, \bibinfo {author} {\bibfnamefont {D.}~\bibnamefont {Zhu}}, \bibinfo {author} {\bibfnamefont {A.}~\bibnamefont {Risinger}}, \bibinfo {author} {\bibfnamefont {L.}~\bibnamefont {Egan}}, \bibinfo {author} {\bibfnamefont {D.}~\bibnamefont {Biswas}}, \bibinfo {author} {\bibfnamefont {M.}~\bibnamefont {Cetina}}, \bibinfo {author} {\bibfnamefont {A.~V.}\ \bibnamefont {Gorshkov}}, \bibinfo {author} {\bibfnamefont {M.~J.}\ \bibnamefont {Gullans}}, \bibinfo {author} {\bibfnamefont {D.~A.}\ \bibnamefont {Huse}},\ and\ \bibinfo {author} {\bibfnamefont {C.}~\bibnamefont {Monroe}},\ }\bibfield  {title} {\bibinfo {title} {Measurement-induced quantum phases realized in a trapped-ion quantum computer},\ }\href {https://doi.org/10.1038/s41567-022-01619-7} {\bibfield  {journal} {\bibinfo  {journal} {Nat. Phys.}\ }\textbf {\bibinfo {volume} {18}},\ \bibinfo {pages} {760–764} (\bibinfo
  {year} {2022})}\BibitemShut {NoStop}%
\bibitem [{\citenamefont {Chen}\ \emph {et~al.}(2021)\citenamefont {Chen}, \citenamefont {Satzinger}, \citenamefont {Atalaya}, \citenamefont {Korotkov}, \citenamefont {Dunsworth}, \citenamefont {Sank}, \citenamefont {Quintana}, \citenamefont {McEwen}, \citenamefont {Barends}, \citenamefont {Klimov}, \citenamefont {Hong}, \citenamefont {Jones}, \citenamefont {Petukhov}, \citenamefont {Kafri}, \citenamefont {Demura}, \citenamefont {Burkett}, \citenamefont {Gidney}, \citenamefont {Fowler}, \citenamefont {Paler}, \citenamefont {Putterman}, \citenamefont {Aleiner}, \citenamefont {Arute}, \citenamefont {Arya}, \citenamefont {Babbush}, \citenamefont {Bardin}, \citenamefont {Bengtsson}, \citenamefont {Bourassa}, \citenamefont {Broughton}, \citenamefont {Buckley}, \citenamefont {Buell}, \citenamefont {Bushnell}, \citenamefont {Chiaro}, \citenamefont {Collins}, \citenamefont {Courtney}, \citenamefont {Derk}, \citenamefont {Eppens}, \citenamefont {Erickson}, \citenamefont {Farhi}, \citenamefont {Foxen}, \citenamefont
  {Giustina}, \citenamefont {Greene}, \citenamefont {Gross}, \citenamefont {Harrigan}, \citenamefont {Harrington}, \citenamefont {Hilton}, \citenamefont {Ho}, \citenamefont {Huang}, \citenamefont {Huggins}, \citenamefont {Ioffe}, \citenamefont {Isakov}, \citenamefont {Jeffrey}, \citenamefont {Jiang}, \citenamefont {Kechedzhi}, \citenamefont {Kim}, \citenamefont {Kitaev}, \citenamefont {Kostritsa}, \citenamefont {Landhuis}, \citenamefont {Laptev}, \citenamefont {Lucero}, \citenamefont {Martin}, \citenamefont {McClean}, \citenamefont {McCourt}, \citenamefont {Mi}, \citenamefont {Miao}, \citenamefont {Mohseni}, \citenamefont {Montazeri}, \citenamefont {Mruczkiewicz}, \citenamefont {Mutus}, \citenamefont {Naaman}, \citenamefont {Neeley}, \citenamefont {Neill}, \citenamefont {Newman}, \citenamefont {Niu}, \citenamefont {O’Brien}, \citenamefont {Opremcak}, \citenamefont {Ostby}, \citenamefont {Pató}, \citenamefont {Redd}, \citenamefont {Roushan}, \citenamefont {Rubin}, \citenamefont {Shvarts}, \citenamefont
  {Strain}, \citenamefont {Szalay}, \citenamefont {Trevithick}, \citenamefont {Villalonga}, \citenamefont {White}, \citenamefont {Yao}, \citenamefont {Yeh}, \citenamefont {Yoo}, \citenamefont {Zalcman}, \citenamefont {Neven}, \citenamefont {Boixo}, \citenamefont {Smelyanskiy}, \citenamefont {Chen}, \citenamefont {Megrant},\ and\ \citenamefont {Kelly}}]{google2021}%
  \BibitemOpen
  \bibfield  {author} {\bibinfo {author} {\bibfnamefont {Z.}~\bibnamefont {Chen}}, \bibinfo {author} {\bibfnamefont {K.~J.}\ \bibnamefont {Satzinger}}, \bibinfo {author} {\bibfnamefont {J.}~\bibnamefont {Atalaya}}, \bibinfo {author} {\bibfnamefont {A.~N.}\ \bibnamefont {Korotkov}}, \bibinfo {author} {\bibfnamefont {A.}~\bibnamefont {Dunsworth}}, \bibinfo {author} {\bibfnamefont {D.}~\bibnamefont {Sank}}, \bibinfo {author} {\bibfnamefont {C.}~\bibnamefont {Quintana}}, \bibinfo {author} {\bibfnamefont {M.}~\bibnamefont {McEwen}}, \bibinfo {author} {\bibfnamefont {R.}~\bibnamefont {Barends}}, \bibinfo {author} {\bibfnamefont {P.~V.}\ \bibnamefont {Klimov}}, \bibinfo {author} {\bibfnamefont {S.}~\bibnamefont {Hong}}, \bibinfo {author} {\bibfnamefont {C.}~\bibnamefont {Jones}}, \bibinfo {author} {\bibfnamefont {A.}~\bibnamefont {Petukhov}}, \bibinfo {author} {\bibfnamefont {D.}~\bibnamefont {Kafri}}, \bibinfo {author} {\bibfnamefont {S.}~\bibnamefont {Demura}}, \bibinfo {author} {\bibfnamefont {B.}~\bibnamefont
  {Burkett}}, \bibinfo {author} {\bibfnamefont {C.}~\bibnamefont {Gidney}}, \bibinfo {author} {\bibfnamefont {A.~G.}\ \bibnamefont {Fowler}}, \bibinfo {author} {\bibfnamefont {A.}~\bibnamefont {Paler}}, \bibinfo {author} {\bibfnamefont {H.}~\bibnamefont {Putterman}}, \bibinfo {author} {\bibfnamefont {I.}~\bibnamefont {Aleiner}}, \bibinfo {author} {\bibfnamefont {F.}~\bibnamefont {Arute}}, \bibinfo {author} {\bibfnamefont {K.}~\bibnamefont {Arya}}, \bibinfo {author} {\bibfnamefont {R.}~\bibnamefont {Babbush}}, \bibinfo {author} {\bibfnamefont {J.~C.}\ \bibnamefont {Bardin}}, \bibinfo {author} {\bibfnamefont {A.}~\bibnamefont {Bengtsson}}, \bibinfo {author} {\bibfnamefont {A.}~\bibnamefont {Bourassa}}, \bibinfo {author} {\bibfnamefont {M.}~\bibnamefont {Broughton}}, \bibinfo {author} {\bibfnamefont {B.~B.}\ \bibnamefont {Buckley}}, \bibinfo {author} {\bibfnamefont {D.~A.}\ \bibnamefont {Buell}}, \bibinfo {author} {\bibfnamefont {N.}~\bibnamefont {Bushnell}}, \bibinfo {author} {\bibfnamefont {B.}~\bibnamefont
  {Chiaro}}, \bibinfo {author} {\bibfnamefont {R.}~\bibnamefont {Collins}}, \bibinfo {author} {\bibfnamefont {W.}~\bibnamefont {Courtney}}, \bibinfo {author} {\bibfnamefont {A.~R.}\ \bibnamefont {Derk}}, \bibinfo {author} {\bibfnamefont {D.}~\bibnamefont {Eppens}}, \bibinfo {author} {\bibfnamefont {C.}~\bibnamefont {Erickson}}, \bibinfo {author} {\bibfnamefont {E.}~\bibnamefont {Farhi}}, \bibinfo {author} {\bibfnamefont {B.}~\bibnamefont {Foxen}}, \bibinfo {author} {\bibfnamefont {M.}~\bibnamefont {Giustina}}, \bibinfo {author} {\bibfnamefont {A.}~\bibnamefont {Greene}}, \bibinfo {author} {\bibfnamefont {J.~A.}\ \bibnamefont {Gross}}, \bibinfo {author} {\bibfnamefont {M.~P.}\ \bibnamefont {Harrigan}}, \bibinfo {author} {\bibfnamefont {S.~D.}\ \bibnamefont {Harrington}}, \bibinfo {author} {\bibfnamefont {J.}~\bibnamefont {Hilton}}, \bibinfo {author} {\bibfnamefont {A.}~\bibnamefont {Ho}}, \bibinfo {author} {\bibfnamefont {T.}~\bibnamefont {Huang}}, \bibinfo {author} {\bibfnamefont {W.~J.}\ \bibnamefont
  {Huggins}}, \bibinfo {author} {\bibfnamefont {L.~B.}\ \bibnamefont {Ioffe}}, \bibinfo {author} {\bibfnamefont {S.~V.}\ \bibnamefont {Isakov}}, \bibinfo {author} {\bibfnamefont {E.}~\bibnamefont {Jeffrey}}, \bibinfo {author} {\bibfnamefont {Z.}~\bibnamefont {Jiang}}, \bibinfo {author} {\bibfnamefont {K.}~\bibnamefont {Kechedzhi}}, \bibinfo {author} {\bibfnamefont {S.}~\bibnamefont {Kim}}, \bibinfo {author} {\bibfnamefont {A.}~\bibnamefont {Kitaev}}, \bibinfo {author} {\bibfnamefont {F.}~\bibnamefont {Kostritsa}}, \bibinfo {author} {\bibfnamefont {D.}~\bibnamefont {Landhuis}}, \bibinfo {author} {\bibfnamefont {P.}~\bibnamefont {Laptev}}, \bibinfo {author} {\bibfnamefont {E.}~\bibnamefont {Lucero}}, \bibinfo {author} {\bibfnamefont {O.}~\bibnamefont {Martin}}, \bibinfo {author} {\bibfnamefont {J.~R.}\ \bibnamefont {McClean}}, \bibinfo {author} {\bibfnamefont {T.}~\bibnamefont {McCourt}}, \bibinfo {author} {\bibfnamefont {X.}~\bibnamefont {Mi}}, \bibinfo {author} {\bibfnamefont {K.~C.}\ \bibnamefont {Miao}},
  \bibinfo {author} {\bibfnamefont {M.}~\bibnamefont {Mohseni}}, \bibinfo {author} {\bibfnamefont {S.}~\bibnamefont {Montazeri}}, \bibinfo {author} {\bibfnamefont {W.}~\bibnamefont {Mruczkiewicz}}, \bibinfo {author} {\bibfnamefont {J.}~\bibnamefont {Mutus}}, \bibinfo {author} {\bibfnamefont {O.}~\bibnamefont {Naaman}}, \bibinfo {author} {\bibfnamefont {M.}~\bibnamefont {Neeley}}, \bibinfo {author} {\bibfnamefont {C.}~\bibnamefont {Neill}}, \bibinfo {author} {\bibfnamefont {M.}~\bibnamefont {Newman}}, \bibinfo {author} {\bibfnamefont {M.~Y.}\ \bibnamefont {Niu}}, \bibinfo {author} {\bibfnamefont {T.~E.}\ \bibnamefont {O’Brien}}, \bibinfo {author} {\bibfnamefont {A.}~\bibnamefont {Opremcak}}, \bibinfo {author} {\bibfnamefont {E.}~\bibnamefont {Ostby}}, \bibinfo {author} {\bibfnamefont {B.}~\bibnamefont {Pató}}, \bibinfo {author} {\bibfnamefont {N.}~\bibnamefont {Redd}}, \bibinfo {author} {\bibfnamefont {P.}~\bibnamefont {Roushan}}, \bibinfo {author} {\bibfnamefont {N.~C.}\ \bibnamefont {Rubin}}, \bibinfo
  {author} {\bibfnamefont {V.}~\bibnamefont {Shvarts}}, \bibinfo {author} {\bibfnamefont {D.}~\bibnamefont {Strain}}, \bibinfo {author} {\bibfnamefont {M.}~\bibnamefont {Szalay}}, \bibinfo {author} {\bibfnamefont {M.~D.}\ \bibnamefont {Trevithick}}, \bibinfo {author} {\bibfnamefont {B.}~\bibnamefont {Villalonga}}, \bibinfo {author} {\bibfnamefont {T.}~\bibnamefont {White}}, \bibinfo {author} {\bibfnamefont {Z.~J.}\ \bibnamefont {Yao}}, \bibinfo {author} {\bibfnamefont {P.}~\bibnamefont {Yeh}}, \bibinfo {author} {\bibfnamefont {J.}~\bibnamefont {Yoo}}, \bibinfo {author} {\bibfnamefont {A.}~\bibnamefont {Zalcman}}, \bibinfo {author} {\bibfnamefont {H.}~\bibnamefont {Neven}}, \bibinfo {author} {\bibfnamefont {S.}~\bibnamefont {Boixo}}, \bibinfo {author} {\bibfnamefont {V.}~\bibnamefont {Smelyanskiy}}, \bibinfo {author} {\bibfnamefont {Y.}~\bibnamefont {Chen}}, \bibinfo {author} {\bibfnamefont {A.}~\bibnamefont {Megrant}},\ and\ \bibinfo {author} {\bibfnamefont {J.}~\bibnamefont {Kelly}},\ }\bibfield  {title}
  {\bibinfo {title} {Exponential suppression of bit or phase errors with cyclic error correction},\ }\href {https://doi.org/10.1038/s41586-021-03588-y} {\bibfield  {journal} {\bibinfo  {journal} {Nature}\ }\textbf {\bibinfo {volume} {595}},\ \bibinfo {pages} {383–387} (\bibinfo {year} {2021})}\BibitemShut {NoStop}%
\bibitem [{\citenamefont {Koh}\ \emph {et~al.}(2023)\citenamefont {Koh}, \citenamefont {Sun}, \citenamefont {Motta},\ and\ \citenamefont {Minnich}}]{koh2022experimentalrealizationof}%
  \BibitemOpen
  \bibfield  {author} {\bibinfo {author} {\bibfnamefont {J.~M.}\ \bibnamefont {Koh}}, \bibinfo {author} {\bibfnamefont {S.-N.}\ \bibnamefont {Sun}}, \bibinfo {author} {\bibfnamefont {M.}~\bibnamefont {Motta}},\ and\ \bibinfo {author} {\bibfnamefont {A.~J.}\ \bibnamefont {Minnich}},\ }\bibfield  {title} {\bibinfo {title} {Measurement-induced entanglement phase transition on a superconducting quantum processor with mid-circuit readout},\ }\href {https://www.nature.com/articles/s41567-023-02076-6} {\bibfield  {journal} {\bibinfo  {journal} {Nat. Phys.}\ }\textbf {\bibinfo {volume} {19}},\ \bibinfo {pages} {1314} (\bibinfo {year} {2023})}\BibitemShut {NoStop}%
\bibitem [{\citenamefont {Feng}\ \emph {et~al.}(2026)\citenamefont {Feng}, \citenamefont {C\^oté}, \citenamefont {Kourtis},\ and\ \citenamefont {Skinner}}]{feng2025postselectionfreeexperimentalobservationmeasurementinduced}%
  \BibitemOpen
  \bibfield  {author} {\bibinfo {author} {\bibfnamefont {X.}~\bibnamefont {Feng}}, \bibinfo {author} {\bibfnamefont {J.}~\bibnamefont {C\^oté}}, \bibinfo {author} {\bibfnamefont {S.}~\bibnamefont {Kourtis}},\ and\ \bibinfo {author} {\bibfnamefont {B.}~\bibnamefont {Skinner}},\ }\bibfield  {title} {\bibinfo {title} {Postselection-free experimental observation of the measurement-induced phase transition in circuits with universal gates},\ }\href {https://doi.org/10.1038/s42005-025-02443-0} {\bibfield  {journal} {\bibinfo  {journal} {Commun. Phys.}\ }\textbf {\bibinfo {volume} {9}},\ \bibinfo {pages} {110} (\bibinfo {year} {2026})}\BibitemShut {NoStop}%
\bibitem [{\citenamefont {Sierant}\ \emph {et~al.}(2022{\natexlab{b}})\citenamefont {Sierant}, \citenamefont {Chiriac{\`{o}}}, \citenamefont {Surace}, \citenamefont {Sharma}, \citenamefont {Turkeshi}, \citenamefont {Dalmonte}, \citenamefont {Fazio},\ and\ \citenamefont {Pagano}}]{Sierant2022dissipativefloquet}%
  \BibitemOpen
  \bibfield  {author} {\bibinfo {author} {\bibfnamefont {P.}~\bibnamefont {Sierant}}, \bibinfo {author} {\bibfnamefont {G.}~\bibnamefont {Chiriac{\`{o}}}}, \bibinfo {author} {\bibfnamefont {F.~M.}\ \bibnamefont {Surace}}, \bibinfo {author} {\bibfnamefont {S.}~\bibnamefont {Sharma}}, \bibinfo {author} {\bibfnamefont {X.}~\bibnamefont {Turkeshi}}, \bibinfo {author} {\bibfnamefont {M.}~\bibnamefont {Dalmonte}}, \bibinfo {author} {\bibfnamefont {R.}~\bibnamefont {Fazio}},\ and\ \bibinfo {author} {\bibfnamefont {G.}~\bibnamefont {Pagano}},\ }\bibfield  {title} {\bibinfo {title} {Dissipative {F}loquet {D}ynamics: from {S}teady {S}tate to {M}easurement {I}nduced {C}riticality in {T}rapped-ion {C}hains},\ }\href {https://doi.org/10.22331/q-2022-02-02-638} {\bibfield  {journal} {\bibinfo  {journal} {{Quantum}}\ }\textbf {\bibinfo {volume} {6}},\ \bibinfo {pages} {638} (\bibinfo {year} {2022}{\natexlab{b}})}\BibitemShut {NoStop}%
\bibitem [{\citenamefont {Bao}\ \emph {et~al.}(2020)\citenamefont {Bao}, \citenamefont {Choi},\ and\ \citenamefont {Altman}}]{bao2020theoryofthe}%
  \BibitemOpen
  \bibfield  {author} {\bibinfo {author} {\bibfnamefont {Y.}~\bibnamefont {Bao}}, \bibinfo {author} {\bibfnamefont {S.}~\bibnamefont {Choi}},\ and\ \bibinfo {author} {\bibfnamefont {E.}~\bibnamefont {Altman}},\ }\bibfield  {title} {\bibinfo {title} {Theory of the phase transition in random unitary circuits with measurements},\ }\href {https://doi.org/10.1103/PhysRevB.101.104301} {\bibfield  {journal} {\bibinfo  {journal} {Phys. Rev. B}\ }\textbf {\bibinfo {volume} {101}},\ \bibinfo {pages} {104301} (\bibinfo {year} {2020})}\BibitemShut {NoStop}%
\bibitem [{\citenamefont {Choi}\ \emph {et~al.}(2020)\citenamefont {Choi}, \citenamefont {Bao}, \citenamefont {Qi},\ and\ \citenamefont {Altman}}]{choi2020quantumerrorcorrection}%
  \BibitemOpen
  \bibfield  {author} {\bibinfo {author} {\bibfnamefont {S.}~\bibnamefont {Choi}}, \bibinfo {author} {\bibfnamefont {Y.}~\bibnamefont {Bao}}, \bibinfo {author} {\bibfnamefont {X.-L.}\ \bibnamefont {Qi}},\ and\ \bibinfo {author} {\bibfnamefont {E.}~\bibnamefont {Altman}},\ }\bibfield  {title} {\bibinfo {title} {Quantum error correction in scrambling dynamics and measurement-induced phase transition},\ }\href {https://doi.org/10.1103/PhysRevLett.125.030505} {\bibfield  {journal} {\bibinfo  {journal} {Phys. Rev. Lett.}\ }\textbf {\bibinfo {volume} {125}},\ \bibinfo {pages} {030505} (\bibinfo {year} {2020})}\BibitemShut {NoStop}%
\bibitem [{\citenamefont {Jian}\ \emph {et~al.}(2020)\citenamefont {Jian}, \citenamefont {You}, \citenamefont {Vasseur},\ and\ \citenamefont {Ludwig}}]{jian2020measurementinducedcriticality}%
  \BibitemOpen
  \bibfield  {author} {\bibinfo {author} {\bibfnamefont {C.-M.}\ \bibnamefont {Jian}}, \bibinfo {author} {\bibfnamefont {Y.-Z.}\ \bibnamefont {You}}, \bibinfo {author} {\bibfnamefont {R.}~\bibnamefont {Vasseur}},\ and\ \bibinfo {author} {\bibfnamefont {A.~W.~W.}\ \bibnamefont {Ludwig}},\ }\bibfield  {title} {\bibinfo {title} {Measurement-induced criticality in random quantum circuits},\ }\href {https://doi.org/10.1103/PhysRevB.101.104302} {\bibfield  {journal} {\bibinfo  {journal} {Phys. Rev. B}\ }\textbf {\bibinfo {volume} {101}},\ \bibinfo {pages} {104302} (\bibinfo {year} {2020})}\BibitemShut {NoStop}%
\bibitem [{\citenamefont {Nahum}\ \emph {et~al.}(2021)\citenamefont {Nahum}, \citenamefont {Roy}, \citenamefont {Skinner},\ and\ \citenamefont {Ruhman}}]{nahum2021measurement}%
  \BibitemOpen
  \bibfield  {author} {\bibinfo {author} {\bibfnamefont {A.}~\bibnamefont {Nahum}}, \bibinfo {author} {\bibfnamefont {S.}~\bibnamefont {Roy}}, \bibinfo {author} {\bibfnamefont {B.}~\bibnamefont {Skinner}},\ and\ \bibinfo {author} {\bibfnamefont {J.}~\bibnamefont {Ruhman}},\ }\bibfield  {title} {\bibinfo {title} {Measurement and entanglement phase transitions in all-to-all quantum circuits, on quantum trees, and in landau-ginsburg theory},\ }\href {https://doi.org/10.1103/PRXQuantum.2.010352} {\bibfield  {journal} {\bibinfo  {journal} {PRX Quantum}\ }\textbf {\bibinfo {volume} {2}},\ \bibinfo {pages} {010352} (\bibinfo {year} {2021})}\BibitemShut {NoStop}%
\bibitem [{\citenamefont {Bao}\ \emph {et~al.}(2021)\citenamefont {Bao}, \citenamefont {Choi},\ and\ \citenamefont {Altman}}]{bao2021symmetryenrichedphases}%
  \BibitemOpen
  \bibfield  {author} {\bibinfo {author} {\bibfnamefont {Y.}~\bibnamefont {Bao}}, \bibinfo {author} {\bibfnamefont {S.}~\bibnamefont {Choi}},\ and\ \bibinfo {author} {\bibfnamefont {E.}~\bibnamefont {Altman}},\ }\bibfield  {title} {\bibinfo {title} {Symmetry enriched phases of quantum circuits},\ }\href {https://doi.org/10.1016/j.aop.2021.168618} {\bibfield  {journal} {\bibinfo  {journal} {Ann. Phys.}\ }\textbf {\bibinfo {volume} {435}},\ \bibinfo {pages} {168618} (\bibinfo {year} {2021})}\BibitemShut {NoStop}%
\bibitem [{\citenamefont {Li}\ \emph {et~al.}(2021)\citenamefont {Li}, \citenamefont {Chen}, \citenamefont {Ludwig},\ and\ \citenamefont {Fisher}}]{li2021conformal}%
  \BibitemOpen
  \bibfield  {author} {\bibinfo {author} {\bibfnamefont {Y.}~\bibnamefont {Li}}, \bibinfo {author} {\bibfnamefont {X.}~\bibnamefont {Chen}}, \bibinfo {author} {\bibfnamefont {A.~W.~W.}\ \bibnamefont {Ludwig}},\ and\ \bibinfo {author} {\bibfnamefont {M.~P.~A.}\ \bibnamefont {Fisher}},\ }\bibfield  {title} {\bibinfo {title} {Conformal invariance and quantum nonlocality in critical hybrid circuits},\ }\href {https://doi.org/10.1103/PhysRevB.104.104305} {\bibfield  {journal} {\bibinfo  {journal} {Phys. Rev. B}\ }\textbf {\bibinfo {volume} {104}},\ \bibinfo {pages} {104305} (\bibinfo {year} {2021})}\BibitemShut {NoStop}%
\bibitem [{\citenamefont {Li}\ \emph {et~al.}(2023)\citenamefont {Li}, \citenamefont {Vijay},\ and\ \citenamefont {Fisher}}]{li2021entanglementdomainwalls}%
  \BibitemOpen
  \bibfield  {author} {\bibinfo {author} {\bibfnamefont {Y.}~\bibnamefont {Li}}, \bibinfo {author} {\bibfnamefont {S.}~\bibnamefont {Vijay}},\ and\ \bibinfo {author} {\bibfnamefont {M.~P.}\ \bibnamefont {Fisher}},\ }\bibfield  {title} {\bibinfo {title} {Entanglement domain walls in monitored quantum circuits and the directed polymer in a random environment},\ }\href {https://doi.org/10.1103/PRXQuantum.4.010331} {\bibfield  {journal} {\bibinfo  {journal} {PRX Quantum}\ }\textbf {\bibinfo {volume} {4}},\ \bibinfo {pages} {010331} (\bibinfo {year} {2023})}\BibitemShut {NoStop}%
\bibitem [{\citenamefont {Li}\ \emph {et~al.}(2024)\citenamefont {Li}, \citenamefont {Vasseur}, \citenamefont {Fisher},\ and\ \citenamefont {Ludwig}}]{li2021statisticalclifford}%
  \BibitemOpen
  \bibfield  {author} {\bibinfo {author} {\bibfnamefont {Y.}~\bibnamefont {Li}}, \bibinfo {author} {\bibfnamefont {R.}~\bibnamefont {Vasseur}}, \bibinfo {author} {\bibfnamefont {M.~P.~A.}\ \bibnamefont {Fisher}},\ and\ \bibinfo {author} {\bibfnamefont {A.~W.~W.}\ \bibnamefont {Ludwig}},\ }\bibfield  {title} {\bibinfo {title} {Statistical mechanics model for clifford random tensor networks and monitored quantum circuits},\ }\href {https://doi.org/10.1103/PhysRevB.109.174307} {\bibfield  {journal} {\bibinfo  {journal} {Phys. Rev. B}\ }\textbf {\bibinfo {volume} {109}},\ \bibinfo {pages} {174307} (\bibinfo {year} {2024})}\BibitemShut {NoStop}%
\bibitem [{\citenamefont {Jian}\ \emph {et~al.}(2022)\citenamefont {Jian}, \citenamefont {Bauer}, \citenamefont {Keselman},\ and\ \citenamefont {Ludwig}}]{jian2022criticality}%
  \BibitemOpen
  \bibfield  {author} {\bibinfo {author} {\bibfnamefont {C.-M.}\ \bibnamefont {Jian}}, \bibinfo {author} {\bibfnamefont {B.}~\bibnamefont {Bauer}}, \bibinfo {author} {\bibfnamefont {A.}~\bibnamefont {Keselman}},\ and\ \bibinfo {author} {\bibfnamefont {A.~W.~W.}\ \bibnamefont {Ludwig}},\ }\bibfield  {title} {\bibinfo {title} {Criticality and entanglement in nonunitary quantum circuits and tensor networks of noninteracting fermions},\ }\href {https://doi.org/10.1103/PhysRevB.106.134206} {\bibfield  {journal} {\bibinfo  {journal} {Phys. Rev. B}\ }\textbf {\bibinfo {volume} {106}},\ \bibinfo {pages} {134206} (\bibinfo {year} {2022})}\BibitemShut {NoStop}%
\bibitem [{\citenamefont {Jian}\ \emph {et~al.}(2023)\citenamefont {Jian}, \citenamefont {Shapourian}, \citenamefont {Bauer},\ and\ \citenamefont {Ludwig}}]{jian2023measurement}%
  \BibitemOpen
  \bibfield  {author} {\bibinfo {author} {\bibfnamefont {C.-M.}\ \bibnamefont {Jian}}, \bibinfo {author} {\bibfnamefont {H.}~\bibnamefont {Shapourian}}, \bibinfo {author} {\bibfnamefont {B.}~\bibnamefont {Bauer}},\ and\ \bibinfo {author} {\bibfnamefont {A.~W.}\ \bibnamefont {Ludwig}},\ }\bibfield  {title} {\bibinfo {title} {Measurement-induced entanglement transitions in quantum circuits of non-interacting fermions: Born-rule versus forced measurements},\ }\href {https://arxiv.org/abs/2302.09094} {\bibfield  {journal} {\bibinfo  {journal} {arXiv:2302.09094}\ } (\bibinfo {year} {2023})}\BibitemShut {NoStop}%
\bibitem [{\citenamefont {Nahum}\ and\ \citenamefont {Wiese}(2023)}]{nahum2023renormalizationgroupfor}%
  \BibitemOpen
  \bibfield  {author} {\bibinfo {author} {\bibfnamefont {A.}~\bibnamefont {Nahum}}\ and\ \bibinfo {author} {\bibfnamefont {K.~J.}\ \bibnamefont {Wiese}},\ }\bibfield  {title} {\bibinfo {title} {Renormalization group for measurement and entanglement phase transitions},\ }\href {https://doi.org/10.1103/PhysRevB.108.104203} {\bibfield  {journal} {\bibinfo  {journal} {Phys. Rev. B}\ }\textbf {\bibinfo {volume} {108}},\ \bibinfo {pages} {104203} (\bibinfo {year} {2023})}\BibitemShut {NoStop}%
\bibitem [{\citenamefont {Zhang}\ \emph {et~al.}(2021)\citenamefont {Zhang}, \citenamefont {Jian}, \citenamefont {Liu},\ and\ \citenamefont {Chen}}]{zhang2021emergentreplica}%
  \BibitemOpen
  \bibfield  {author} {\bibinfo {author} {\bibfnamefont {P.}~\bibnamefont {Zhang}}, \bibinfo {author} {\bibfnamefont {S.-K.}\ \bibnamefont {Jian}}, \bibinfo {author} {\bibfnamefont {C.}~\bibnamefont {Liu}},\ and\ \bibinfo {author} {\bibfnamefont {X.}~\bibnamefont {Chen}},\ }\bibfield  {title} {\bibinfo {title} {Emergent {R}eplica {C}onformal {S}ymmetry in {N}on-{H}ermitian {SYK}{$_2$} {C}hains},\ }\href {https://doi.org/10.22331/q-2021-11-16-579} {\bibfield  {journal} {\bibinfo  {journal} {{Quantum}}\ }\textbf {\bibinfo {volume} {5}},\ \bibinfo {pages} {579} (\bibinfo {year} {2021})}\BibitemShut {NoStop}%
\bibitem [{\citenamefont {Giachetti}\ and\ \citenamefont {{De Luca}}(2023)}]{giachetti2023elusive}%
  \BibitemOpen
  \bibfield  {author} {\bibinfo {author} {\bibfnamefont {G.}~\bibnamefont {Giachetti}}\ and\ \bibinfo {author} {\bibfnamefont {A.}~\bibnamefont {{De Luca}}},\ }\href@noop {} {\bibinfo {title} {Elusive phase transition in the replica limit of monitored systems}} (\bibinfo {year} {2023}),\ \Eprint {https://arxiv.org/abs/2306.12166} {arXiv:2306.12166 [cond-mat.stat-mech]} \BibitemShut {NoStop}%
\bibitem [{\citenamefont {Gullans}\ and\ \citenamefont {Huse}(2020{\natexlab{a}})}]{gullans2020dynamicalpurificationphase}%
  \BibitemOpen
  \bibfield  {author} {\bibinfo {author} {\bibfnamefont {M.~J.}\ \bibnamefont {Gullans}}\ and\ \bibinfo {author} {\bibfnamefont {D.~A.}\ \bibnamefont {Huse}},\ }\bibfield  {title} {\bibinfo {title} {Dynamical purification phase transition induced by quantum measurements},\ }\href {https://doi.org/10.1103/PhysRevX.10.041020} {\bibfield  {journal} {\bibinfo  {journal} {Phys. Rev. X}\ }\textbf {\bibinfo {volume} {10}},\ \bibinfo {pages} {041020} (\bibinfo {year} {2020}{\natexlab{a}})}\BibitemShut {NoStop}%
\bibitem [{\citenamefont {Gullans}\ and\ \citenamefont {Huse}(2020{\natexlab{b}})}]{gullans2020scalable}%
  \BibitemOpen
  \bibfield  {author} {\bibinfo {author} {\bibfnamefont {M.~J.}\ \bibnamefont {Gullans}}\ and\ \bibinfo {author} {\bibfnamefont {D.~A.}\ \bibnamefont {Huse}},\ }\bibfield  {title} {\bibinfo {title} {Scalable probes of measurement-induced criticality},\ }\href {https://doi.org/10.1103/PhysRevLett.125.070606} {\bibfield  {journal} {\bibinfo  {journal} {Phys. Rev. Lett.}\ }\textbf {\bibinfo {volume} {125}},\ \bibinfo {pages} {070606} (\bibinfo {year} {2020}{\natexlab{b}})}\BibitemShut {NoStop}%
\bibitem [{\citenamefont {Bentsen}\ \emph {et~al.}(2021)\citenamefont {Bentsen}, \citenamefont {Sahu},\ and\ \citenamefont {Swingle}}]{bentsen2021measurementinducedpurification}%
  \BibitemOpen
  \bibfield  {author} {\bibinfo {author} {\bibfnamefont {G.~S.}\ \bibnamefont {Bentsen}}, \bibinfo {author} {\bibfnamefont {S.}~\bibnamefont {Sahu}},\ and\ \bibinfo {author} {\bibfnamefont {B.}~\bibnamefont {Swingle}},\ }\bibfield  {title} {\bibinfo {title} {Measurement-induced purification in large-$n$ hybrid brownian circuits},\ }\href {https://doi.org/10.1103/PhysRevB.104.094304} {\bibfield  {journal} {\bibinfo  {journal} {Phys. Rev. B}\ }\textbf {\bibinfo {volume} {104}},\ \bibinfo {pages} {094304} (\bibinfo {year} {2021})}\BibitemShut {NoStop}%
\bibitem [{\citenamefont {Gopalakrishnan}\ and\ \citenamefont {Gullans}(2021)}]{gopalakrishnan2021entanglementandpurification}%
  \BibitemOpen
  \bibfield  {author} {\bibinfo {author} {\bibfnamefont {S.}~\bibnamefont {Gopalakrishnan}}\ and\ \bibinfo {author} {\bibfnamefont {M.~J.}\ \bibnamefont {Gullans}},\ }\bibfield  {title} {\bibinfo {title} {Entanglement and purification transitions in non-hermitian quantum mechanics},\ }\href {https://doi.org/10.1103/PhysRevLett.126.170503} {\bibfield  {journal} {\bibinfo  {journal} {Phys. Rev. Lett.}\ }\textbf {\bibinfo {volume} {126}},\ \bibinfo {pages} {170503} (\bibinfo {year} {2021})}\BibitemShut {NoStop}%
\bibitem [{\citenamefont {Ippoliti}\ and\ \citenamefont {Ho}(2023)}]{ippoliti2022dynamicalpurificationand}%
  \BibitemOpen
  \bibfield  {author} {\bibinfo {author} {\bibfnamefont {M.}~\bibnamefont {Ippoliti}}\ and\ \bibinfo {author} {\bibfnamefont {W.~W.}\ \bibnamefont {Ho}},\ }\bibfield  {title} {\bibinfo {title} {Dynamical purification and the emergence of quantum state designs from the projected ensemble},\ }\href {https://doi.org/10.1103/PRXQuantum.4.030322} {\bibfield  {journal} {\bibinfo  {journal} {PRX Quantum}\ }\textbf {\bibinfo {volume} {4}},\ \bibinfo {pages} {030322} (\bibinfo {year} {2023})}\BibitemShut {NoStop}%
\bibitem [{\citenamefont {Fidkowski}\ \emph {et~al.}(2021)\citenamefont {Fidkowski}, \citenamefont {Haah},\ and\ \citenamefont {Hastings}}]{Fidkowski_2021}%
  \BibitemOpen
  \bibfield  {author} {\bibinfo {author} {\bibfnamefont {L.}~\bibnamefont {Fidkowski}}, \bibinfo {author} {\bibfnamefont {J.}~\bibnamefont {Haah}},\ and\ \bibinfo {author} {\bibfnamefont {M.~B.}\ \bibnamefont {Hastings}},\ }\bibfield  {title} {\bibinfo {title} {How dynamical quantum memories forget},\ }\href {https://doi.org/10.22331/q-2021-01-17-382} {\bibfield  {journal} {\bibinfo  {journal} {Quantum}\ }\textbf {\bibinfo {volume} {5}},\ \bibinfo {pages} {382} (\bibinfo {year} {2021})}\BibitemShut {NoStop}%
\bibitem [{\citenamefont {Ippoliti}\ and\ \citenamefont {Khemani}(2021)}]{ippoliti2021postselection}%
  \BibitemOpen
  \bibfield  {author} {\bibinfo {author} {\bibfnamefont {M.}~\bibnamefont {Ippoliti}}\ and\ \bibinfo {author} {\bibfnamefont {V.}~\bibnamefont {Khemani}},\ }\bibfield  {title} {\bibinfo {title} {Postselection-free entanglement dynamics via spacetime duality},\ }\href {https://doi.org/10.1103/PhysRevLett.126.060501} {\bibfield  {journal} {\bibinfo  {journal} {Phys. Rev. Lett.}\ }\textbf {\bibinfo {volume} {126}},\ \bibinfo {pages} {060501} (\bibinfo {year} {2021})}\BibitemShut {NoStop}%
\bibitem [{\citenamefont {Leontica}\ and\ \citenamefont {McGinley}(2023)}]{leontica2023purification}%
  \BibitemOpen
  \bibfield  {author} {\bibinfo {author} {\bibfnamefont {S.}~\bibnamefont {Leontica}}\ and\ \bibinfo {author} {\bibfnamefont {M.}~\bibnamefont {McGinley}},\ }\bibfield  {title} {\bibinfo {title} {Purification dynamics in a continuous-time hybrid quantum circuit model},\ }\href {https://doi.org/10.1103/PhysRevB.108.174308} {\bibfield  {journal} {\bibinfo  {journal} {Phys. Rev. B}\ }\textbf {\bibinfo {volume} {108}},\ \bibinfo {pages} {174308} (\bibinfo {year} {2023})}\BibitemShut {NoStop}%
\bibitem [{\citenamefont {Anzai}\ \emph {et~al.}(2026)\citenamefont {Anzai}, \citenamefont {Matsueda},\ and\ \citenamefont {Kuno}}]{anzai2026disordered}%
  \BibitemOpen
  \bibfield  {author} {\bibinfo {author} {\bibfnamefont {K.}~\bibnamefont {Anzai}}, \bibinfo {author} {\bibfnamefont {H.}~\bibnamefont {Matsueda}},\ and\ \bibinfo {author} {\bibfnamefont {Y.}~\bibnamefont {Kuno}},\ }\bibfield  {title} {\bibinfo {title} {Disordered purification phase transition in hybrid random circuits},\ }\href {https://doi.org/10.1103/mb36-pc91} {\bibfield  {journal} {\bibinfo  {journal} {Phys. Rev. B}\ }\textbf {\bibinfo {volume} {113}},\ \bibinfo {pages} {014111} (\bibinfo {year} {2026})}\BibitemShut {NoStop}%
\bibitem [{\citenamefont {Han}\ and\ \citenamefont {Chen}(2022)}]{han2022measurement}%
  \BibitemOpen
  \bibfield  {author} {\bibinfo {author} {\bibfnamefont {Y.}~\bibnamefont {Han}}\ and\ \bibinfo {author} {\bibfnamefont {X.}~\bibnamefont {Chen}},\ }\bibfield  {title} {\bibinfo {title} {Measurement-induced criticality in ${\mathbb{z}}_{2}$-symmetric quantum automaton circuits},\ }\href {https://doi.org/10.1103/PhysRevB.105.064306} {\bibfield  {journal} {\bibinfo  {journal} {Phys. Rev. B}\ }\textbf {\bibinfo {volume} {105}},\ \bibinfo {pages} {064306} (\bibinfo {year} {2022})}\BibitemShut {NoStop}%
\bibitem [{\citenamefont {Kawabata}\ \emph {et~al.}(2023)\citenamefont {Kawabata}, \citenamefont {Numasawa},\ and\ \citenamefont {Ryu}}]{kawabata2023entanglement}%
  \BibitemOpen
  \bibfield  {author} {\bibinfo {author} {\bibfnamefont {K.}~\bibnamefont {Kawabata}}, \bibinfo {author} {\bibfnamefont {T.}~\bibnamefont {Numasawa}},\ and\ \bibinfo {author} {\bibfnamefont {S.}~\bibnamefont {Ryu}},\ }\bibfield  {title} {\bibinfo {title} {Entanglement phase transition induced by the non-hermitian skin effect},\ }\href {https://doi.org/10.1103/PhysRevX.13.021007} {\bibfield  {journal} {\bibinfo  {journal} {Phys. Rev. X}\ }\textbf {\bibinfo {volume} {13}},\ \bibinfo {pages} {021007} (\bibinfo {year} {2023})}\BibitemShut {NoStop}%
\bibitem [{\citenamefont {Claeys}\ \emph {et~al.}(2022)\citenamefont {Claeys}, \citenamefont {Henry}, \citenamefont {Vicary},\ and\ \citenamefont {Lamacraft}}]{claeys2022exact}%
  \BibitemOpen
  \bibfield  {author} {\bibinfo {author} {\bibfnamefont {P.~W.}\ \bibnamefont {Claeys}}, \bibinfo {author} {\bibfnamefont {M.}~\bibnamefont {Henry}}, \bibinfo {author} {\bibfnamefont {J.}~\bibnamefont {Vicary}},\ and\ \bibinfo {author} {\bibfnamefont {A.}~\bibnamefont {Lamacraft}},\ }\bibfield  {title} {\bibinfo {title} {Exact dynamics in dual-unitary quantum circuits with projective measurements},\ }\href {https://doi.org/10.1103/PhysRevResearch.4.043212} {\bibfield  {journal} {\bibinfo  {journal} {Phys. Rev. Res.}\ }\textbf {\bibinfo {volume} {4}},\ \bibinfo {pages} {043212} (\bibinfo {year} {2022})}\BibitemShut {NoStop}%
\bibitem [{\citenamefont {Bompais}\ \emph {et~al.}(2026)\citenamefont {Bompais}, \citenamefont {Amini}, \citenamefont {Garrahan},\ and\ \citenamefont {Guţă}}]{bompais2026rate}%
  \BibitemOpen
  \bibfield  {author} {\bibinfo {author} {\bibfnamefont {M.}~\bibnamefont {Bompais}}, \bibinfo {author} {\bibfnamefont {N.~H.}\ \bibnamefont {Amini}}, \bibinfo {author} {\bibfnamefont {J.~P.}\ \bibnamefont {Garrahan}},\ and\ \bibinfo {author} {\bibfnamefont {M.}~\bibnamefont {Guţă}},\ }\href {https://arxiv.org/abs/2601.14023} {\bibinfo {title} {The rate of purification of quantum trajectories}} (\bibinfo {year} {2026}),\ \Eprint {https://arxiv.org/abs/2601.14023} {arXiv:2601.14023 [quant-ph]} \BibitemShut {NoStop}%
\bibitem [{\citenamefont {Iaconis}\ \emph {et~al.}(2020)\citenamefont {Iaconis}, \citenamefont {Lucas},\ and\ \citenamefont {Chen}}]{PhysRevB.102.224311}%
  \BibitemOpen
  \bibfield  {author} {\bibinfo {author} {\bibfnamefont {J.}~\bibnamefont {Iaconis}}, \bibinfo {author} {\bibfnamefont {A.}~\bibnamefont {Lucas}},\ and\ \bibinfo {author} {\bibfnamefont {X.}~\bibnamefont {Chen}},\ }\bibfield  {title} {\bibinfo {title} {Measurement-induced phase transitions in quantum automaton circuits},\ }\href {https://doi.org/10.1103/PhysRevB.102.224311} {\bibfield  {journal} {\bibinfo  {journal} {Phys. Rev. B}\ }\textbf {\bibinfo {volume} {102}},\ \bibinfo {pages} {224311} (\bibinfo {year} {2020})}\BibitemShut {NoStop}%
\bibitem [{\citenamefont {Lavasani}\ \emph {et~al.}(2021{\natexlab{a}})\citenamefont {Lavasani}, \citenamefont {Alavirad},\ and\ \citenamefont {Barkeshli}}]{PhysRevLett.127.235701}%
  \BibitemOpen
  \bibfield  {author} {\bibinfo {author} {\bibfnamefont {A.}~\bibnamefont {Lavasani}}, \bibinfo {author} {\bibfnamefont {Y.}~\bibnamefont {Alavirad}},\ and\ \bibinfo {author} {\bibfnamefont {M.}~\bibnamefont {Barkeshli}},\ }\bibfield  {title} {\bibinfo {title} {Topological order and criticality in $(2+1)\mathrm{D}$ monitored random quantum circuits},\ }\href {https://doi.org/10.1103/PhysRevLett.127.235701} {\bibfield  {journal} {\bibinfo  {journal} {Phys. Rev. Lett.}\ }\textbf {\bibinfo {volume} {127}},\ \bibinfo {pages} {235701} (\bibinfo {year} {2021}{\natexlab{a}})}\BibitemShut {NoStop}%
\bibitem [{\citenamefont {Yu}\ and\ \citenamefont {Qi}(2022)}]{yu2022measurementinducedentanglementphasetransition}%
  \BibitemOpen
  \bibfield  {author} {\bibinfo {author} {\bibfnamefont {X.}~\bibnamefont {Yu}}\ and\ \bibinfo {author} {\bibfnamefont {X.-L.}\ \bibnamefont {Qi}},\ }\href {https://arxiv.org/abs/2201.12704} {\bibinfo {title} {Measurement-induced entanglement phase transition in random bilocal circuits}} (\bibinfo {year} {2022}),\ \Eprint {https://arxiv.org/abs/2201.12704} {arXiv:2201.12704 [quant-ph]} \BibitemShut {NoStop}%
\bibitem [{\citenamefont {Zabalo}\ \emph {et~al.}(2023)\citenamefont {Zabalo}, \citenamefont {Wilson}, \citenamefont {Gullans}, \citenamefont {Vasseur}, \citenamefont {Gopalakrishnan}, \citenamefont {Huse},\ and\ \citenamefont {Pixley}}]{PhysRevB.107.L220204}%
  \BibitemOpen
  \bibfield  {author} {\bibinfo {author} {\bibfnamefont {A.}~\bibnamefont {Zabalo}}, \bibinfo {author} {\bibfnamefont {J.~H.}\ \bibnamefont {Wilson}}, \bibinfo {author} {\bibfnamefont {M.~J.}\ \bibnamefont {Gullans}}, \bibinfo {author} {\bibfnamefont {R.}~\bibnamefont {Vasseur}}, \bibinfo {author} {\bibfnamefont {S.}~\bibnamefont {Gopalakrishnan}}, \bibinfo {author} {\bibfnamefont {D.~A.}\ \bibnamefont {Huse}},\ and\ \bibinfo {author} {\bibfnamefont {J.~H.}\ \bibnamefont {Pixley}},\ }\bibfield  {title} {\bibinfo {title} {Infinite-randomness criticality in monitored quantum dynamics with static disorder},\ }\href {https://doi.org/10.1103/PhysRevB.107.L220204} {\bibfield  {journal} {\bibinfo  {journal} {Phys. Rev. B}\ }\textbf {\bibinfo {volume} {107}},\ \bibinfo {pages} {L220204} (\bibinfo {year} {2023})}\BibitemShut {NoStop}%
\bibitem [{\citenamefont {Wang}\ \emph {et~al.}(2025)\citenamefont {Wang}, \citenamefont {Altman},\ and\ \citenamefont {Garratt}}]{wang2025stabilityquantumchaosweak}%
  \BibitemOpen
  \bibfield  {author} {\bibinfo {author} {\bibfnamefont {Y.-C.}\ \bibnamefont {Wang}}, \bibinfo {author} {\bibfnamefont {E.}~\bibnamefont {Altman}},\ and\ \bibinfo {author} {\bibfnamefont {S.~J.}\ \bibnamefont {Garratt}},\ }\href {https://arxiv.org/abs/2512.02934} {\bibinfo {title} {Stability of quantum chaos against weak non-unitarity}} (\bibinfo {year} {2025}),\ \Eprint {https://arxiv.org/abs/2512.02934} {arXiv:2512.02934 [quant-ph]} \BibitemShut {NoStop}%
\bibitem [{\citenamefont {Kuno}\ \emph {et~al.}(2022)\citenamefont {Kuno}, \citenamefont {Orito},\ and\ \citenamefont {Ichinose}}]{PhysRevB.106.214304}%
  \BibitemOpen
  \bibfield  {author} {\bibinfo {author} {\bibfnamefont {Y.}~\bibnamefont {Kuno}}, \bibinfo {author} {\bibfnamefont {T.}~\bibnamefont {Orito}},\ and\ \bibinfo {author} {\bibfnamefont {I.}~\bibnamefont {Ichinose}},\ }\bibfield  {title} {\bibinfo {title} {Purification and scrambling in a chaotic hamiltonian dynamics with measurements},\ }\href {https://doi.org/10.1103/PhysRevB.106.214304} {\bibfield  {journal} {\bibinfo  {journal} {Phys. Rev. B}\ }\textbf {\bibinfo {volume} {106}},\ \bibinfo {pages} {214304} (\bibinfo {year} {2022})}\BibitemShut {NoStop}%
\bibitem [{\citenamefont {De~Luca}\ \emph {et~al.}(2025)\citenamefont {De~Luca}, \citenamefont {Liu}, \citenamefont {Nahum},\ and\ \citenamefont {Zhou}}]{deluca2025universality}%
  \BibitemOpen
  \bibfield  {author} {\bibinfo {author} {\bibfnamefont {A.}~\bibnamefont {De~Luca}}, \bibinfo {author} {\bibfnamefont {C.}~\bibnamefont {Liu}}, \bibinfo {author} {\bibfnamefont {A.}~\bibnamefont {Nahum}},\ and\ \bibinfo {author} {\bibfnamefont {T.}~\bibnamefont {Zhou}},\ }\bibfield  {title} {\bibinfo {title} {Universality classes for purification in nonunitary quantum processes},\ }\href {https://doi.org/10.1103/wlj6-mkk4} {\bibfield  {journal} {\bibinfo  {journal} {Phys. Rev. X}\ }\textbf {\bibinfo {volume} {15}},\ \bibinfo {pages} {041024} (\bibinfo {year} {2025})}\BibitemShut {NoStop}%
\bibitem [{\citenamefont {Gerbino}\ \emph {et~al.}(2026)\citenamefont {Gerbino}, \citenamefont {Kim}, \citenamefont {Giachetti}, \citenamefont {{De Luca}},\ and\ \citenamefont {Turkeshi}}]{gerbino2026universal}%
  \BibitemOpen
  \bibfield  {author} {\bibinfo {author} {\bibfnamefont {F.}~\bibnamefont {Gerbino}}, \bibinfo {author} {\bibfnamefont {D.}~\bibnamefont {Kim}}, \bibinfo {author} {\bibfnamefont {G.}~\bibnamefont {Giachetti}}, \bibinfo {author} {\bibfnamefont {A.}~\bibnamefont {{De Luca}}},\ and\ \bibinfo {author} {\bibfnamefont {X.}~\bibnamefont {Turkeshi}},\ }\href {https://arxiv.org/abs/2603.10751} {\bibinfo {title} {Universal purification dynamics in real non-unitary quantum processes}} (\bibinfo {year} {2026}),\ \Eprint {https://arxiv.org/abs/2603.10751} {arXiv:2603.10751 [quant-ph]} \BibitemShut {NoStop}%
\bibitem [{\citenamefont {Xiao}\ \emph {et~al.}(2025)\citenamefont {Xiao}, \citenamefont {Ohtsuki},\ and\ \citenamefont {Kawabata}}]{PhysRevLett.134.140401}%
  \BibitemOpen
  \bibfield  {author} {\bibinfo {author} {\bibfnamefont {Z.}~\bibnamefont {Xiao}}, \bibinfo {author} {\bibfnamefont {T.}~\bibnamefont {Ohtsuki}},\ and\ \bibinfo {author} {\bibfnamefont {K.}~\bibnamefont {Kawabata}},\ }\bibfield  {title} {\bibinfo {title} {Universal stochastic equations of monitored quantum dynamics},\ }\href {https://doi.org/10.1103/PhysRevLett.134.140401} {\bibfield  {journal} {\bibinfo  {journal} {Phys. Rev. Lett.}\ }\textbf {\bibinfo {volume} {134}},\ \bibinfo {pages} {140401} (\bibinfo {year} {2025})}\BibitemShut {NoStop}%
\bibitem [{\citenamefont {Patel}\ \emph {et~al.}(2026)\citenamefont {Patel}, \citenamefont {McCarthy}, \citenamefont {Chakraborty}, \citenamefont {Huang}, \citenamefont {DiNapoli}, \citenamefont {Vasseur}, \citenamefont {Pixley},\ and\ \citenamefont {Chakram}}]{patel2026universalmonitoreddynamicsmultimode}%
  \BibitemOpen
  \bibfield  {author} {\bibinfo {author} {\bibfnamefont {S.}~\bibnamefont {Patel}}, \bibinfo {author} {\bibfnamefont {C.}~\bibnamefont {McCarthy}}, \bibinfo {author} {\bibfnamefont {A.}~\bibnamefont {Chakraborty}}, \bibinfo {author} {\bibfnamefont {J.}~\bibnamefont {Huang}}, \bibinfo {author} {\bibfnamefont {T.~J.}\ \bibnamefont {DiNapoli}}, \bibinfo {author} {\bibfnamefont {R.}~\bibnamefont {Vasseur}}, \bibinfo {author} {\bibfnamefont {J.~H.}\ \bibnamefont {Pixley}},\ and\ \bibinfo {author} {\bibfnamefont {S.}~\bibnamefont {Chakram}},\ }\href {https://arxiv.org/abs/2603.13125} {\bibinfo {title} {Universal monitored dynamics in multimode bosonic systems}} (\bibinfo {year} {2026}),\ \Eprint {https://arxiv.org/abs/2603.13125} {arXiv:2603.13125 [quant-ph]} \BibitemShut {NoStop}%
\bibitem [{\citenamefont {Bulchandani}\ \emph {et~al.}(2024)\citenamefont {Bulchandani}, \citenamefont {Sondhi},\ and\ \citenamefont {Chalker}}]{bulchandani_random-matrix_2023}%
  \BibitemOpen
  \bibfield  {author} {\bibinfo {author} {\bibfnamefont {V.~B.}\ \bibnamefont {Bulchandani}}, \bibinfo {author} {\bibfnamefont {S.~L.}\ \bibnamefont {Sondhi}},\ and\ \bibinfo {author} {\bibfnamefont {J.~T.}\ \bibnamefont {Chalker}},\ }\bibfield  {title} {\bibinfo {title} {Random-matrix models of monitored quantum circuits},\ }\href {https://doi.org/10.1007/s10955-024-03273-0} {\bibfield  {journal} {\bibinfo  {journal} {J. Stat. Phys.}\ }\textbf {\bibinfo {volume} {191}},\ \bibinfo {pages} {55} (\bibinfo {year} {2024})}\BibitemShut {NoStop}%
\bibitem [{\citenamefont {Gerbino}\ \emph {et~al.}(2024)\citenamefont {Gerbino}, \citenamefont {Le~Doussal}, \citenamefont {Giachetti},\ and\ \citenamefont {{De Luca}}}]{gerbino2024dyson}%
  \BibitemOpen
  \bibfield  {author} {\bibinfo {author} {\bibfnamefont {F.}~\bibnamefont {Gerbino}}, \bibinfo {author} {\bibfnamefont {P.}~\bibnamefont {Le~Doussal}}, \bibinfo {author} {\bibfnamefont {G.}~\bibnamefont {Giachetti}},\ and\ \bibinfo {author} {\bibfnamefont {A.}~\bibnamefont {{De Luca}}},\ }\bibfield  {title} {\bibinfo {title} {A dyson brownian motion model for weak measurements in chaotic quantum systems},\ }\href {https://www.mdpi.com/2624-960X/6/2/16} {\bibfield  {journal} {\bibinfo  {journal} {Quantum Rep.}\ }\textbf {\bibinfo {volume} {6}},\ \bibinfo {pages} {200} (\bibinfo {year} {2024})}\BibitemShut {NoStop}%
\bibitem [{\citenamefont {Mehta}(2004)}]{mehta2004random}%
  \BibitemOpen
  \bibfield  {author} {\bibinfo {author} {\bibfnamefont {M.~L.}\ \bibnamefont {Mehta}},\ }\href@noop {} {\emph {\bibinfo {title} {Random matrices}}}\ (\bibinfo  {publisher} {Elsevier},\ \bibinfo {year} {2004})\BibitemShut {NoStop}%
\bibitem [{\citenamefont {Lavasani}\ \emph {et~al.}(2021{\natexlab{b}})\citenamefont {Lavasani}, \citenamefont {Alavirad},\ and\ \citenamefont {Barkeshli}}]{lavasani2021measurementinducedtopological}%
  \BibitemOpen
  \bibfield  {author} {\bibinfo {author} {\bibfnamefont {A.}~\bibnamefont {Lavasani}}, \bibinfo {author} {\bibfnamefont {Y.}~\bibnamefont {Alavirad}},\ and\ \bibinfo {author} {\bibfnamefont {M.}~\bibnamefont {Barkeshli}},\ }\bibfield  {title} {\bibinfo {title} {Measurement-induced topological entanglement transitions in symmetric random quantum circuits},\ }\href {https://doi.org/10.1038/s41567-020-01112-z} {\bibfield  {journal} {\bibinfo  {journal} {Nature Phys.}\ }\textbf {\bibinfo {volume} {17}},\ \bibinfo {pages} {342} (\bibinfo {year} {2021}{\natexlab{b}})}\BibitemShut {NoStop}%
\bibitem [{\citenamefont {Bittel}\ \emph {et~al.}(2025)\citenamefont {Bittel}, \citenamefont {Eisert}, \citenamefont {Leone}, \citenamefont {Mele},\ and\ \citenamefont {Oliviero}}]{bittel2025completetheorycliffordcommutant}%
  \BibitemOpen
  \bibfield  {author} {\bibinfo {author} {\bibfnamefont {L.}~\bibnamefont {Bittel}}, \bibinfo {author} {\bibfnamefont {J.}~\bibnamefont {Eisert}}, \bibinfo {author} {\bibfnamefont {L.}~\bibnamefont {Leone}}, \bibinfo {author} {\bibfnamefont {A.~A.}\ \bibnamefont {Mele}},\ and\ \bibinfo {author} {\bibfnamefont {S.~F.~E.}\ \bibnamefont {Oliviero}},\ }\href {https://arxiv.org/abs/2504.12263} {\bibinfo {title} {A complete theory of the clifford commutant}} (\bibinfo {year} {2025}),\ \Eprint {https://arxiv.org/abs/2504.12263} {arXiv:2504.12263 [quant-ph]} \BibitemShut {NoStop}%
\bibitem [{\citenamefont {Magni}\ \emph {et~al.}(2026{\natexlab{a}})\citenamefont {Magni}, \citenamefont {Heinrich}, \citenamefont {Leone},\ and\ \citenamefont {Turkeshi}}]{magni2025anticoncentrationstatedesigndoped}%
  \BibitemOpen
  \bibfield  {author} {\bibinfo {author} {\bibfnamefont {B.}~\bibnamefont {Magni}}, \bibinfo {author} {\bibfnamefont {M.}~\bibnamefont {Heinrich}}, \bibinfo {author} {\bibfnamefont {L.}~\bibnamefont {Leone}},\ and\ \bibinfo {author} {\bibfnamefont {X.}~\bibnamefont {Turkeshi}},\ }\bibfield  {title} {\bibinfo {title} {Anticoncentration and state design of doped real clifford circuits and tensor networks},\ }\href {https://doi.org/10.1103/w6l4-mh6k} {\bibfield  {journal} {\bibinfo  {journal} {Phys. Rev. A}\ }\textbf {\bibinfo {volume} {113}},\ \bibinfo {pages} {062446} (\bibinfo {year} {2026}{\natexlab{a}})}\BibitemShut {NoStop}%
\bibitem [{\citenamefont {Lunt}\ \emph {et~al.}(2021)\citenamefont {Lunt}, \citenamefont {Szyniszewski},\ and\ \citenamefont {Pal}}]{PhysRevB.104.155111}%
  \BibitemOpen
  \bibfield  {author} {\bibinfo {author} {\bibfnamefont {O.}~\bibnamefont {Lunt}}, \bibinfo {author} {\bibfnamefont {M.}~\bibnamefont {Szyniszewski}},\ and\ \bibinfo {author} {\bibfnamefont {A.}~\bibnamefont {Pal}},\ }\bibfield  {title} {\bibinfo {title} {Measurement-induced criticality and entanglement clusters: A study of one-dimensional and two-dimensional clifford circuits},\ }\href {https://doi.org/10.1103/PhysRevB.104.155111} {\bibfield  {journal} {\bibinfo  {journal} {Phys. Rev. B}\ }\textbf {\bibinfo {volume} {104}},\ \bibinfo {pages} {155111} (\bibinfo {year} {2021})}\BibitemShut {NoStop}%
\bibitem [{\citenamefont {Shkolnik}\ \emph {et~al.}(2023)\citenamefont {Shkolnik}, \citenamefont {Zabalo}, \citenamefont {Vasseur}, \citenamefont {Huse}, \citenamefont {Pixley},\ and\ \citenamefont {Gazit}}]{PhysRevB.108.184204}%
  \BibitemOpen
  \bibfield  {author} {\bibinfo {author} {\bibfnamefont {G.}~\bibnamefont {Shkolnik}}, \bibinfo {author} {\bibfnamefont {A.}~\bibnamefont {Zabalo}}, \bibinfo {author} {\bibfnamefont {R.}~\bibnamefont {Vasseur}}, \bibinfo {author} {\bibfnamefont {D.~A.}\ \bibnamefont {Huse}}, \bibinfo {author} {\bibfnamefont {J.~H.}\ \bibnamefont {Pixley}},\ and\ \bibinfo {author} {\bibfnamefont {S.}~\bibnamefont {Gazit}},\ }\bibfield  {title} {\bibinfo {title} {Measurement induced criticality in quasiperiodic modulated random hybrid circuits},\ }\href {https://doi.org/10.1103/PhysRevB.108.184204} {\bibfield  {journal} {\bibinfo  {journal} {Phys. Rev. B}\ }\textbf {\bibinfo {volume} {108}},\ \bibinfo {pages} {184204} (\bibinfo {year} {2023})}\BibitemShut {NoStop}%
\bibitem [{\citenamefont {Webb}(2016)}]{webb2016clifford}%
  \BibitemOpen
  \bibfield  {author} {\bibinfo {author} {\bibfnamefont {Z.}~\bibnamefont {Webb}},\ }\bibfield  {title} {\bibinfo {title} {The {Clifford} group forms a unitary 3-design},\ }\href@noop {} {\bibfield  {journal} {\bibinfo  {journal} {Quantum Inf. Comput.}\ }\textbf {\bibinfo {volume} {16}},\ \bibinfo {pages} {1379} (\bibinfo {year} {2016})}\BibitemShut {NoStop}%
\bibitem [{\citenamefont {Zhu}(2017)}]{zhu2017multiqubit}%
  \BibitemOpen
  \bibfield  {author} {\bibinfo {author} {\bibfnamefont {H.}~\bibnamefont {Zhu}},\ }\bibfield  {title} {\bibinfo {title} {Multiqubit {Clifford} groups are unitary 3-designs},\ }\href {https://doi.org/10.1103/PhysRevA.96.062336} {\bibfield  {journal} {\bibinfo  {journal} {Phys. Rev. A}\ }\textbf {\bibinfo {volume} {96}},\ \bibinfo {pages} {062336} (\bibinfo {year} {2017})}\BibitemShut {NoStop}%
\bibitem [{\citenamefont {Gross}\ \emph {et~al.}(2021)\citenamefont {Gross}, \citenamefont {Nezami},\ and\ \citenamefont {Walter}}]{gross2021schur}%
  \BibitemOpen
  \bibfield  {author} {\bibinfo {author} {\bibfnamefont {D.}~\bibnamefont {Gross}}, \bibinfo {author} {\bibfnamefont {S.}~\bibnamefont {Nezami}},\ and\ \bibinfo {author} {\bibfnamefont {M.}~\bibnamefont {Walter}},\ }\bibfield  {title} {\bibinfo {title} {Schur--weyl duality for the clifford group with applications: Property testing, a robust hudson theorem, and de finetti representations},\ }\href {https://doi.org/10.1007/s00220-021-04118-7} {\bibfield  {journal} {\bibinfo  {journal} {Commun. Math. Phys.}\ }\textbf {\bibinfo {volume} {385}},\ \bibinfo {pages} {1325} (\bibinfo {year} {2021})}\BibitemShut {NoStop}%
\bibitem [{\citenamefont {Gottesman}(1997)}]{gottesman1997stabilizer}%
  \BibitemOpen
  \bibfield  {author} {\bibinfo {author} {\bibfnamefont {D.}~\bibnamefont {Gottesman}},\ }\emph {\bibinfo {title} {Stabilizer codes and quantum error correction}},\ \href@noop {} {Ph.D. thesis},\ \bibinfo  {school} {California Institute of Technology} (\bibinfo {year} {1997}),\ \bibinfo {note} {arXiv:quant-ph/9705052}\BibitemShut {NoStop}%
\bibitem [{\citenamefont {Aaronson}\ and\ \citenamefont {Gottesman}(2004)}]{aaronson2004improvedsimulationof}%
  \BibitemOpen
  \bibfield  {author} {\bibinfo {author} {\bibfnamefont {S.}~\bibnamefont {Aaronson}}\ and\ \bibinfo {author} {\bibfnamefont {D.}~\bibnamefont {Gottesman}},\ }\bibfield  {title} {\bibinfo {title} {Improved simulation of stabilizer circuits},\ }\href {https://doi.org/10.1103/PhysRevA.70.052328} {\bibfield  {journal} {\bibinfo  {journal} {Phys. Rev. A}\ }\textbf {\bibinfo {volume} {70}},\ \bibinfo {pages} {052328} (\bibinfo {year} {2004})}\BibitemShut {NoStop}%
\bibitem [{\citenamefont {Gidney}(2021)}]{gidney2021stimfaststabilizer}%
  \BibitemOpen
  \bibfield  {author} {\bibinfo {author} {\bibfnamefont {C.}~\bibnamefont {Gidney}},\ }\bibfield  {title} {\bibinfo {title} {Stim: a fast stabilizer circuit simulator},\ }\href {https://doi.org/10.22331/q-2021-07-06-497} {\bibfield  {journal} {\bibinfo  {journal} {{Quantum}}\ }\textbf {\bibinfo {volume} {5}},\ \bibinfo {pages} {497} (\bibinfo {year} {2021})}\BibitemShut {NoStop}%
\bibitem [{\citenamefont {Bravyi}\ and\ \citenamefont {Maslov}(2021)}]{bravyi2021hadamard}%
  \BibitemOpen
  \bibfield  {author} {\bibinfo {author} {\bibfnamefont {S.}~\bibnamefont {Bravyi}}\ and\ \bibinfo {author} {\bibfnamefont {D.}~\bibnamefont {Maslov}},\ }\bibfield  {title} {\bibinfo {title} {Hadamard-free circuits expose the structure of the {C}lifford group},\ }\href {https://doi.org/10.1109/TIT.2021.3081415} {\bibfield  {journal} {\bibinfo  {journal} {IEEE Trans. Inf. Theory}\ }\textbf {\bibinfo {volume} {67}},\ \bibinfo {pages} {4546} (\bibinfo {year} {2021})},\ \Eprint {https://arxiv.org/abs/2003.09412} {arXiv:2003.09412} \BibitemShut {NoStop}%
\bibitem [{\citenamefont {Magni}\ \emph {et~al.}(2026{\natexlab{b}})\citenamefont {Magni}, \citenamefont {Heinrich},\ and\ \citenamefont {Turkeshi}}]{turkeshi2026sampler}%
  \BibitemOpen
  \bibfield  {author} {\bibinfo {author} {\bibfnamefont {B.}~\bibnamefont {Magni}}, \bibinfo {author} {\bibfnamefont {M.}~\bibnamefont {Heinrich}},\ and\ \bibinfo {author} {\bibfnamefont {X.}~\bibnamefont {Turkeshi}},\ }\href@noop {} {\bibinfo {title} {Uniform sampling of random clifford circuits for prime-dimensional qudits}} (\bibinfo {year} {2026}{\natexlab{b}}),\ \bibinfo {note} {to appear}\BibitemShut {NoStop}%
\bibitem [{\citenamefont {Agrawal}\ \emph {et~al.}(2022)\citenamefont {Agrawal}, \citenamefont {Zabalo}, \citenamefont {Chen}, \citenamefont {Wilson}, \citenamefont {Potter}, \citenamefont {Pixley}, \citenamefont {Gopalakrishnan},\ and\ \citenamefont {Vasseur}}]{agrawal2022entanglmentandchargesharpening}%
  \BibitemOpen
  \bibfield  {author} {\bibinfo {author} {\bibfnamefont {U.}~\bibnamefont {Agrawal}}, \bibinfo {author} {\bibfnamefont {A.}~\bibnamefont {Zabalo}}, \bibinfo {author} {\bibfnamefont {K.}~\bibnamefont {Chen}}, \bibinfo {author} {\bibfnamefont {J.~H.}\ \bibnamefont {Wilson}}, \bibinfo {author} {\bibfnamefont {A.~C.}\ \bibnamefont {Potter}}, \bibinfo {author} {\bibfnamefont {J.~H.}\ \bibnamefont {Pixley}}, \bibinfo {author} {\bibfnamefont {S.}~\bibnamefont {Gopalakrishnan}},\ and\ \bibinfo {author} {\bibfnamefont {R.}~\bibnamefont {Vasseur}},\ }\bibfield  {title} {\bibinfo {title} {Entanglement and charge-sharpening transitions in u(1) symmetric monitored quantum circuits},\ }\href {https://doi.org/10.1103/PhysRevX.12.041002} {\bibfield  {journal} {\bibinfo  {journal} {Phys. Rev. X}\ }\textbf {\bibinfo {volume} {12}},\ \bibinfo {pages} {041002} (\bibinfo {year} {2022})}\BibitemShut {NoStop}%
\bibitem [{\citenamefont {Nahum}\ and\ \citenamefont {Jacobsen}(2025)}]{nahum2025bayesian}%
  \BibitemOpen
  \bibfield  {author} {\bibinfo {author} {\bibfnamefont {A.}~\bibnamefont {Nahum}}\ and\ \bibinfo {author} {\bibfnamefont {J.~L.}\ \bibnamefont {Jacobsen}},\ }\bibfield  {title} {\bibinfo {title} {Bayesian critical points in classical lattice models},\ }\href {https://doi.org/10.1103/7dpt-d4s5} {\bibfield  {journal} {\bibinfo  {journal} {Phys. Rev. B}\ }\textbf {\bibinfo {volume} {112}},\ \bibinfo {pages} {235113} (\bibinfo {year} {2025})}\BibitemShut {NoStop}%
\bibitem [{\citenamefont {Fava}\ \emph {et~al.}(2023)\citenamefont {Fava}, \citenamefont {Piroli}, \citenamefont {Swann}, \citenamefont {Bernard},\ and\ \citenamefont {Nahum}}]{fava2023nonlinear}%
  \BibitemOpen
  \bibfield  {author} {\bibinfo {author} {\bibfnamefont {M.}~\bibnamefont {Fava}}, \bibinfo {author} {\bibfnamefont {L.}~\bibnamefont {Piroli}}, \bibinfo {author} {\bibfnamefont {T.}~\bibnamefont {Swann}}, \bibinfo {author} {\bibfnamefont {D.}~\bibnamefont {Bernard}},\ and\ \bibinfo {author} {\bibfnamefont {A.}~\bibnamefont {Nahum}},\ }\bibfield  {title} {\bibinfo {title} {Nonlinear sigma models for monitored dynamics of free fermions},\ }\href {https://doi.org/10.1103/PhysRevX.13.041045} {\bibfield  {journal} {\bibinfo  {journal} {Phys. Rev. X}\ }\textbf {\bibinfo {volume} {13}},\ \bibinfo {pages} {041045} (\bibinfo {year} {2023})}\BibitemShut {NoStop}%
\bibitem [{\citenamefont {Poboiko}\ \emph {et~al.}(2024)\citenamefont {Poboiko}, \citenamefont {Gornyi},\ and\ \citenamefont {Mirlin}}]{poboiko2023measurementinduced}%
  \BibitemOpen
  \bibfield  {author} {\bibinfo {author} {\bibfnamefont {I.}~\bibnamefont {Poboiko}}, \bibinfo {author} {\bibfnamefont {I.~V.}\ \bibnamefont {Gornyi}},\ and\ \bibinfo {author} {\bibfnamefont {A.~D.}\ \bibnamefont {Mirlin}},\ }\bibfield  {title} {\bibinfo {title} {Measurement-induced phase transition for free fermions above one dimension},\ }\href {https://doi.org/10.1103/PhysRevLett.132.110403} {\bibfield  {journal} {\bibinfo  {journal} {Phys. Rev. Lett.}\ }\textbf {\bibinfo {volume} {132}},\ \bibinfo {pages} {110403} (\bibinfo {year} {2024})}\BibitemShut {NoStop}%
\bibitem [{\citenamefont {L\'oio}\ \emph {et~al.}(2023)\citenamefont {L\'oio}, \citenamefont {De~Luca}, \citenamefont {De~Nardis},\ and\ \citenamefont {Turkeshi}}]{loio2023}%
  \BibitemOpen
  \bibfield  {author} {\bibinfo {author} {\bibfnamefont {H.}~\bibnamefont {L\'oio}}, \bibinfo {author} {\bibfnamefont {A.}~\bibnamefont {De~Luca}}, \bibinfo {author} {\bibfnamefont {J.}~\bibnamefont {De~Nardis}},\ and\ \bibinfo {author} {\bibfnamefont {X.}~\bibnamefont {Turkeshi}},\ }\bibfield  {title} {\bibinfo {title} {Purification timescales in monitored fermions},\ }\href {https://doi.org/10.1103/PhysRevB.108.L020306} {\bibfield  {journal} {\bibinfo  {journal} {Phys. Rev. B}\ }\textbf {\bibinfo {volume} {108}},\ \bibinfo {pages} {L020306} (\bibinfo {year} {2023})}\BibitemShut {NoStop}%
\bibitem [{\citenamefont {Fava}\ \emph {et~al.}(2024)\citenamefont {Fava}, \citenamefont {Piroli}, \citenamefont {Bernard},\ and\ \citenamefont {Nahum}}]{fava2024monitored}%
  \BibitemOpen
  \bibfield  {author} {\bibinfo {author} {\bibfnamefont {M.}~\bibnamefont {Fava}}, \bibinfo {author} {\bibfnamefont {L.}~\bibnamefont {Piroli}}, \bibinfo {author} {\bibfnamefont {D.}~\bibnamefont {Bernard}},\ and\ \bibinfo {author} {\bibfnamefont {A.}~\bibnamefont {Nahum}},\ }\bibfield  {title} {\bibinfo {title} {Monitored fermions with conserved $u(1)$ charge},\ }\href {https://doi.org/10.1103/PhysRevResearch.6.043246} {\bibfield  {journal} {\bibinfo  {journal} {Phys. Rev. Res.}\ }\textbf {\bibinfo {volume} {6}},\ \bibinfo {pages} {043246} (\bibinfo {year} {2024})}\BibitemShut {NoStop}%
\bibitem [{\citenamefont {Buchhold}\ \emph {et~al.}(2022)\citenamefont {Buchhold}, \citenamefont {Müller},\ and\ \citenamefont {Diehl}}]{buchhold2022revealingmeasurementinduced}%
  \BibitemOpen
  \bibfield  {author} {\bibinfo {author} {\bibfnamefont {M.}~\bibnamefont {Buchhold}}, \bibinfo {author} {\bibfnamefont {T.}~\bibnamefont {Müller}},\ and\ \bibinfo {author} {\bibfnamefont {S.}~\bibnamefont {Diehl}},\ }\href {https://arxiv.org/abs/2208.10506} {\bibinfo {title} {Revealing measurement-induced phase transitions by pre-selection}} (\bibinfo {year} {2022}),\ \Eprint {https://arxiv.org/abs/2208.10506} {arXiv:2208.10506 [cond-mat.dis-nn]} \BibitemShut {NoStop}%
\bibitem [{\citenamefont {Chahine}\ and\ \citenamefont {Buchhold}(2024)}]{PhysRevB.110.054313}%
  \BibitemOpen
  \bibfield  {author} {\bibinfo {author} {\bibfnamefont {K.}~\bibnamefont {Chahine}}\ and\ \bibinfo {author} {\bibfnamefont {M.}~\bibnamefont {Buchhold}},\ }\bibfield  {title} {\bibinfo {title} {Entanglement phases, localization, and multifractality of monitored free fermions in two dimensions},\ }\href {https://doi.org/10.1103/PhysRevB.110.054313} {\bibfield  {journal} {\bibinfo  {journal} {Phys. Rev. B}\ }\textbf {\bibinfo {volume} {110}},\ \bibinfo {pages} {054313} (\bibinfo {year} {2024})}\BibitemShut {NoStop}%
\bibitem [{\citenamefont {Ladewig}\ \emph {et~al.}(2022)\citenamefont {Ladewig}, \citenamefont {Diehl},\ and\ \citenamefont {Buchhold}}]{PhysRevResearch.4.033001}%
  \BibitemOpen
  \bibfield  {author} {\bibinfo {author} {\bibfnamefont {B.}~\bibnamefont {Ladewig}}, \bibinfo {author} {\bibfnamefont {S.}~\bibnamefont {Diehl}},\ and\ \bibinfo {author} {\bibfnamefont {M.}~\bibnamefont {Buchhold}},\ }\bibfield  {title} {\bibinfo {title} {Monitored open fermion dynamics: Exploring the interplay of measurement, decoherence, and free hamiltonian evolution},\ }\href {https://doi.org/10.1103/PhysRevResearch.4.033001} {\bibfield  {journal} {\bibinfo  {journal} {Phys. Rev. Res.}\ }\textbf {\bibinfo {volume} {4}},\ \bibinfo {pages} {033001} (\bibinfo {year} {2022})}\BibitemShut {NoStop}%
\bibitem [{\citenamefont {Ha}\ \emph {et~al.}(2024)\citenamefont {Ha}, \citenamefont {Pandey}, \citenamefont {Gopalakrishnan},\ and\ \citenamefont {Huse}}]{PhysRevB.110.L140301}%
  \BibitemOpen
  \bibfield  {author} {\bibinfo {author} {\bibfnamefont {H.}~\bibnamefont {Ha}}, \bibinfo {author} {\bibfnamefont {A.}~\bibnamefont {Pandey}}, \bibinfo {author} {\bibfnamefont {S.}~\bibnamefont {Gopalakrishnan}},\ and\ \bibinfo {author} {\bibfnamefont {D.~A.}\ \bibnamefont {Huse}},\ }\bibfield  {title} {\bibinfo {title} {Measurement-induced phase transitions in systems with diffusive dynamics},\ }\href {https://doi.org/10.1103/PhysRevB.110.L140301} {\bibfield  {journal} {\bibinfo  {journal} {Phys. Rev. B}\ }\textbf {\bibinfo {volume} {110}},\ \bibinfo {pages} {L140301} (\bibinfo {year} {2024})}\BibitemShut {NoStop}%
\bibitem [{\citenamefont {Klocke}\ \emph {et~al.}(2025)\citenamefont {Klocke}, \citenamefont {Simm}, \citenamefont {Zhu}, \citenamefont {Trebst},\ and\ \citenamefont {Buchhold}}]{PhysRevB.111.224301}%
  \BibitemOpen
  \bibfield  {author} {\bibinfo {author} {\bibfnamefont {K.}~\bibnamefont {Klocke}}, \bibinfo {author} {\bibfnamefont {D.}~\bibnamefont {Simm}}, \bibinfo {author} {\bibfnamefont {G.-Y.}\ \bibnamefont {Zhu}}, \bibinfo {author} {\bibfnamefont {S.}~\bibnamefont {Trebst}},\ and\ \bibinfo {author} {\bibfnamefont {M.}~\bibnamefont {Buchhold}},\ }\bibfield  {title} {\bibinfo {title} {Entanglement dynamics in monitored kitaev circuits: Loop models, symmetry classification, and quantum lifshitz scaling},\ }\href {https://doi.org/10.1103/PhysRevB.111.224301} {\bibfield  {journal} {\bibinfo  {journal} {Phys. Rev. B}\ }\textbf {\bibinfo {volume} {111}},\ \bibinfo {pages} {224301} (\bibinfo {year} {2025})}\BibitemShut {NoStop}%
\bibitem [{\citenamefont {Sriram}\ \emph {et~al.}(2023)\citenamefont {Sriram}, \citenamefont {Rakovszky}, \citenamefont {Khemani},\ and\ \citenamefont {Ippoliti}}]{PhysRevB.108.094304}%
  \BibitemOpen
  \bibfield  {author} {\bibinfo {author} {\bibfnamefont {A.}~\bibnamefont {Sriram}}, \bibinfo {author} {\bibfnamefont {T.}~\bibnamefont {Rakovszky}}, \bibinfo {author} {\bibfnamefont {V.}~\bibnamefont {Khemani}},\ and\ \bibinfo {author} {\bibfnamefont {M.}~\bibnamefont {Ippoliti}},\ }\bibfield  {title} {\bibinfo {title} {Topology, criticality, and dynamically generated qubits in a stochastic measurement-only kitaev model},\ }\href {https://doi.org/10.1103/PhysRevB.108.094304} {\bibfield  {journal} {\bibinfo  {journal} {Phys. Rev. B}\ }\textbf {\bibinfo {volume} {108}},\ \bibinfo {pages} {094304} (\bibinfo {year} {2023})}\BibitemShut {NoStop}%
\bibitem [{\citenamefont {Zhu}\ and\ \citenamefont {Trebst}(2023)}]{zhu2023qubitfractionalizationemergentmajorana}%
  \BibitemOpen
  \bibfield  {author} {\bibinfo {author} {\bibfnamefont {G.-Y.}\ \bibnamefont {Zhu}}\ and\ \bibinfo {author} {\bibfnamefont {S.}~\bibnamefont {Trebst}},\ }\href {https://arxiv.org/abs/2311.08450} {\bibinfo {title} {Qubit fractionalization and emergent majorana liquid in the honeycomb floquet code induced by coherent errors and weak measurements}} (\bibinfo {year} {2023}),\ \Eprint {https://arxiv.org/abs/2311.08450} {arXiv:2311.08450 [quant-ph]} \BibitemShut {NoStop}%
\bibitem [{\citenamefont {Leone}\ \emph {et~al.}(2024)\citenamefont {Leone}, \citenamefont {Oliviero},\ and\ \citenamefont {Hamma}}]{Leone2024learningtdoped}%
  \BibitemOpen
  \bibfield  {author} {\bibinfo {author} {\bibfnamefont {L.}~\bibnamefont {Leone}}, \bibinfo {author} {\bibfnamefont {S.~F.~E.}\ \bibnamefont {Oliviero}},\ and\ \bibinfo {author} {\bibfnamefont {A.}~\bibnamefont {Hamma}},\ }\bibfield  {title} {\bibinfo {title} {Learning t-doped stabilizer states},\ }\href {https://doi.org/10.22331/q-2024-05-27-1361} {\bibfield  {journal} {\bibinfo  {journal} {{Quantum}}\ }\textbf {\bibinfo {volume} {8}},\ \bibinfo {pages} {1361} (\bibinfo {year} {2024})}\BibitemShut {NoStop}%
\bibitem [{\citenamefont {Bejan}\ \emph {et~al.}(2024)\citenamefont {Bejan}, \citenamefont {McLauchlan},\ and\ \citenamefont {B\'eri}}]{bejan2024dynamical}%
  \BibitemOpen
  \bibfield  {author} {\bibinfo {author} {\bibfnamefont {M.}~\bibnamefont {Bejan}}, \bibinfo {author} {\bibfnamefont {C.}~\bibnamefont {McLauchlan}},\ and\ \bibinfo {author} {\bibfnamefont {B.}~\bibnamefont {B\'eri}},\ }\bibfield  {title} {\bibinfo {title} {Dynamical magic transitions in monitored clifford+$t$ circuits},\ }\href {https://doi.org/10.1103/PRXQuantum.5.030332} {\bibfield  {journal} {\bibinfo  {journal} {PRX Quantum}\ }\textbf {\bibinfo {volume} {5}},\ \bibinfo {pages} {030332} (\bibinfo {year} {2024})}\BibitemShut {NoStop}%
\bibitem [{\citenamefont {Lami}\ \emph {et~al.}(2025)\citenamefont {Lami}, \citenamefont {Haug},\ and\ \citenamefont {De~Nardis}}]{lami2024quantum}%
  \BibitemOpen
  \bibfield  {author} {\bibinfo {author} {\bibfnamefont {G.}~\bibnamefont {Lami}}, \bibinfo {author} {\bibfnamefont {T.}~\bibnamefont {Haug}},\ and\ \bibinfo {author} {\bibfnamefont {J.}~\bibnamefont {De~Nardis}},\ }\bibfield  {title} {\bibinfo {title} {Quantum state designs with clifford-enhanced matrix product states},\ }\href {https://doi.org/10.1103/PRXQuantum.6.010345} {\bibfield  {journal} {\bibinfo  {journal} {PRX Quantum}\ }\textbf {\bibinfo {volume} {6}},\ \bibinfo {pages} {010345} (\bibinfo {year} {2025})}\BibitemShut {NoStop}%
\bibitem [{\citenamefont {Haug}\ \emph {et~al.}(2025)\citenamefont {Haug}, \citenamefont {Aolita},\ and\ \citenamefont {Kim}}]{haug2024probingquantumcomplexityuniversal}%
  \BibitemOpen
  \bibfield  {author} {\bibinfo {author} {\bibfnamefont {T.}~\bibnamefont {Haug}}, \bibinfo {author} {\bibfnamefont {L.}~\bibnamefont {Aolita}},\ and\ \bibinfo {author} {\bibfnamefont {M.}~\bibnamefont {Kim}},\ }\bibfield  {title} {\bibinfo {title} {Probing quantum complexity via universal saturation of stabilizer entropies},\ }\href {https://doi.org/10.22331/q-2025-07-21-1801} {\bibfield  {journal} {\bibinfo  {journal} {{Quantum}}\ }\textbf {\bibinfo {volume} {9}},\ \bibinfo {pages} {1801} (\bibinfo {year} {2025})}\BibitemShut {NoStop}%
\bibitem [{\citenamefont {Fux}\ \emph {et~al.}(2025)\citenamefont {Fux}, \citenamefont {B\'eri}, \citenamefont {Fazio},\ and\ \citenamefont {Tirrito}}]{fux2025disentanglingunitarydynamicsclassically}%
  \BibitemOpen
  \bibfield  {author} {\bibinfo {author} {\bibfnamefont {G.~E.}\ \bibnamefont {Fux}}, \bibinfo {author} {\bibfnamefont {B.}~\bibnamefont {B\'eri}}, \bibinfo {author} {\bibfnamefont {R.}~\bibnamefont {Fazio}},\ and\ \bibinfo {author} {\bibfnamefont {E.}~\bibnamefont {Tirrito}},\ }\bibfield  {title} {\bibinfo {title} {Disentangling magic states with classically simulable quantum circuits},\ }\href {https://doi.org/10.1103/ggp1-byj1} {\bibfield  {journal} {\bibinfo  {journal} {Phys. Rev. Lett.}\ }\textbf {\bibinfo {volume} {135}},\ \bibinfo {pages} {260605} (\bibinfo {year} {2025})}\BibitemShut {NoStop}%
\bibitem [{\citenamefont {Nakhl}\ \emph {et~al.}(2025)\citenamefont {Nakhl}, \citenamefont {Harper}, \citenamefont {West}, \citenamefont {Dowling}, \citenamefont {Sevior}, \citenamefont {Quella},\ and\ \citenamefont {Usman}}]{nakhl2025stabilizer}%
  \BibitemOpen
  \bibfield  {author} {\bibinfo {author} {\bibfnamefont {A.~C.}\ \bibnamefont {Nakhl}}, \bibinfo {author} {\bibfnamefont {B.}~\bibnamefont {Harper}}, \bibinfo {author} {\bibfnamefont {M.}~\bibnamefont {West}}, \bibinfo {author} {\bibfnamefont {N.}~\bibnamefont {Dowling}}, \bibinfo {author} {\bibfnamefont {M.}~\bibnamefont {Sevior}}, \bibinfo {author} {\bibfnamefont {T.}~\bibnamefont {Quella}},\ and\ \bibinfo {author} {\bibfnamefont {M.}~\bibnamefont {Usman}},\ }\bibfield  {title} {\bibinfo {title} {Stabilizer tensor networks with magic state injection},\ }\href {https://doi.org/10.1103/PhysRevLett.134.190602} {\bibfield  {journal} {\bibinfo  {journal} {Phys. Rev. Lett.}\ }\textbf {\bibinfo {volume} {134}},\ \bibinfo {pages} {190602} (\bibinfo {year} {2025})}\BibitemShut {NoStop}%
\bibitem [{\citenamefont {Liu}\ and\ \citenamefont {Clark}(2026)}]{liu2024classicalsimulabilitycliffordtcircuits}%
  \BibitemOpen
  \bibfield  {author} {\bibinfo {author} {\bibfnamefont {Z.}~\bibnamefont {Liu}}\ and\ \bibinfo {author} {\bibfnamefont {B.~K.}\ \bibnamefont {Clark}},\ }\bibfield  {title} {\bibinfo {title} {Classical simulability of $\mathrm{Clifford}+t$ circuits with clifford-augmented matrix product states},\ }\href {https://doi.org/10.1103/ybnf-rjw8} {\bibfield  {journal} {\bibinfo  {journal} {Phys. Rev. Res.}\ }\textbf {\bibinfo {volume} {8}},\ \bibinfo {pages} {023116} (\bibinfo {year} {2026})}\BibitemShut {NoStop}%
\bibitem [{\citenamefont {Leone}\ \emph {et~al.}(2026)\citenamefont {Leone}, \citenamefont {Oliviero}, \citenamefont {Hamma}, \citenamefont {Eisert},\ and\ \citenamefont {Bittel}}]{Leone_2026}%
  \BibitemOpen
  \bibfield  {author} {\bibinfo {author} {\bibfnamefont {L.}~\bibnamefont {Leone}}, \bibinfo {author} {\bibfnamefont {S.~F.}\ \bibnamefont {Oliviero}}, \bibinfo {author} {\bibfnamefont {A.}~\bibnamefont {Hamma}}, \bibinfo {author} {\bibfnamefont {J.}~\bibnamefont {Eisert}},\ and\ \bibinfo {author} {\bibfnamefont {L.}~\bibnamefont {Bittel}},\ }\bibfield  {title} {\bibinfo {title} {Non-clifford cost of random unitaries},\ }\href {https://doi.org/10.1103/25v1-my1x} {\bibfield  {journal} {\bibinfo  {journal} {PRX Quantum}\ }\textbf {\bibinfo {volume} {7}},\ \bibinfo {pages} {020321} (\bibinfo {year} {2026})}\BibitemShut {NoStop}%
\bibitem [{\citenamefont {Turkeshi}\ \emph {et~al.}(2025)\citenamefont {Turkeshi}, \citenamefont {Tirrito},\ and\ \citenamefont {Sierant}}]{turkeshi2025magicspreadingrandomquantum}%
  \BibitemOpen
  \bibfield  {author} {\bibinfo {author} {\bibfnamefont {X.}~\bibnamefont {Turkeshi}}, \bibinfo {author} {\bibfnamefont {E.}~\bibnamefont {Tirrito}},\ and\ \bibinfo {author} {\bibfnamefont {P.}~\bibnamefont {Sierant}},\ }\bibfield  {title} {\bibinfo {title} {Magic spreading in random quantum circuits},\ }\href {https://doi.org/10.1038/s41467-025-57704-x} {\bibfield  {journal} {\bibinfo  {journal} {Nat. Commun.}\ }\textbf {\bibinfo {volume} {16}},\ \bibinfo {pages} {2575} (\bibinfo {year} {2025})}\BibitemShut {NoStop}%
\bibitem [{\citenamefont {Aditya}\ \emph {et~al.}(2025{\natexlab{a}})\citenamefont {Aditya}, \citenamefont {Summer}, \citenamefont {Sierant},\ and\ \citenamefont {Turkeshi}}]{aditya2025mpembaeffectsquantumcomplexity}%
  \BibitemOpen
  \bibfield  {author} {\bibinfo {author} {\bibfnamefont {S.}~\bibnamefont {Aditya}}, \bibinfo {author} {\bibfnamefont {A.}~\bibnamefont {Summer}}, \bibinfo {author} {\bibfnamefont {P.}~\bibnamefont {Sierant}},\ and\ \bibinfo {author} {\bibfnamefont {X.}~\bibnamefont {Turkeshi}},\ }\href {https://arxiv.org/abs/2509.22176} {\bibinfo {title} {Mpemba effects in quantum complexity}} (\bibinfo {year} {2025}{\natexlab{a}}),\ \Eprint {https://arxiv.org/abs/2509.22176} {arXiv:2509.22176 [quant-ph]} \BibitemShut {NoStop}%
\bibitem [{\citenamefont {Paviglianiti}\ \emph {et~al.}(2026)\citenamefont {Paviglianiti}, \citenamefont {Lumia}, \citenamefont {Tirrito}, \citenamefont {Silva}, \citenamefont {Collura}, \citenamefont {Turkeshi},\ and\ \citenamefont {Lami}}]{paviglianiti2025emergencegenericentanglementstructure}%
  \BibitemOpen
  \bibfield  {author} {\bibinfo {author} {\bibfnamefont {A.}~\bibnamefont {Paviglianiti}}, \bibinfo {author} {\bibfnamefont {L.}~\bibnamefont {Lumia}}, \bibinfo {author} {\bibfnamefont {E.}~\bibnamefont {Tirrito}}, \bibinfo {author} {\bibfnamefont {A.}~\bibnamefont {Silva}}, \bibinfo {author} {\bibfnamefont {M.}~\bibnamefont {Collura}}, \bibinfo {author} {\bibfnamefont {X.}~\bibnamefont {Turkeshi}},\ and\ \bibinfo {author} {\bibfnamefont {G.}~\bibnamefont {Lami}},\ }\bibfield  {title} {\bibinfo {title} {Emergence of generic entanglement structure in doped matchgate circuits},\ }\href {https://doi.org/10.1103/w97w-7zny} {\bibfield  {journal} {\bibinfo  {journal} {Phys. Rev. Lett.}\ }\textbf {\bibinfo {volume} {136}},\ \bibinfo {pages} {020403} (\bibinfo {year} {2026})}\BibitemShut {NoStop}%
\bibitem [{\citenamefont {Lóio}\ \emph {et~al.}(2025)\citenamefont {Lóio}, \citenamefont {Lami}, \citenamefont {Leone}, \citenamefont {McGinley}, \citenamefont {Turkeshi},\ and\ \citenamefont {Nardis}}]{loio2025quantumstatedesignsmagic}%
  \BibitemOpen
  \bibfield  {author} {\bibinfo {author} {\bibfnamefont {H.}~\bibnamefont {Lóio}}, \bibinfo {author} {\bibfnamefont {G.}~\bibnamefont {Lami}}, \bibinfo {author} {\bibfnamefont {L.}~\bibnamefont {Leone}}, \bibinfo {author} {\bibfnamefont {M.}~\bibnamefont {McGinley}}, \bibinfo {author} {\bibfnamefont {X.}~\bibnamefont {Turkeshi}},\ and\ \bibinfo {author} {\bibfnamefont {J.~D.}\ \bibnamefont {Nardis}},\ }\href {https://arxiv.org/abs/2510.13950} {\bibinfo {title} {Quantum state designs via magic teleportation}} (\bibinfo {year} {2025}),\ \Eprint {https://arxiv.org/abs/2510.13950} {arXiv:2510.13950 [quant-ph]} \BibitemShut {NoStop}%
\bibitem [{\citenamefont {Aditya}\ \emph {et~al.}(2025{\natexlab{b}})\citenamefont {Aditya}, \citenamefont {Turkeshi},\ and\ \citenamefont {Sierant}}]{aditya2025growthspreadingquantumresources}%
  \BibitemOpen
  \bibfield  {author} {\bibinfo {author} {\bibfnamefont {S.}~\bibnamefont {Aditya}}, \bibinfo {author} {\bibfnamefont {X.}~\bibnamefont {Turkeshi}},\ and\ \bibinfo {author} {\bibfnamefont {P.}~\bibnamefont {Sierant}},\ }\href {https://arxiv.org/abs/2512.14827} {\bibinfo {title} {Growth and spreading of quantum resources under random circuit dynamics}} (\bibinfo {year} {2025}{\natexlab{b}}),\ \Eprint {https://arxiv.org/abs/2512.14827} {arXiv:2512.14827 [quant-ph]} \BibitemShut {NoStop}%
\bibitem [{\citenamefont {Varikuti}\ \emph {et~al.}(2026)\citenamefont {Varikuti}, \citenamefont {Bandyopadhyay},\ and\ \citenamefont {Hauke}}]{varikuti2025impactcliffordoperationsnonstabilizing}%
  \BibitemOpen
  \bibfield  {author} {\bibinfo {author} {\bibfnamefont {N.~D.}\ \bibnamefont {Varikuti}}, \bibinfo {author} {\bibfnamefont {S.}~\bibnamefont {Bandyopadhyay}},\ and\ \bibinfo {author} {\bibfnamefont {P.}~\bibnamefont {Hauke}},\ }\bibfield  {title} {\bibinfo {title} {Impact of {C}lifford operations on non-stabilizing power and quantum chaos},\ }\href {https://doi.org/10.22331/q-2026-03-10-2017} {\bibfield  {journal} {\bibinfo  {journal} {{Quantum}}\ }\textbf {\bibinfo {volume} {10}},\ \bibinfo {pages} {2017} (\bibinfo {year} {2026})}\BibitemShut {NoStop}%
\bibitem [{\citenamefont {Haferkamp}\ \emph {et~al.}(2022)\citenamefont {Haferkamp}, \citenamefont {Montealegre-Mora}, \citenamefont {Heinrich}, \citenamefont {Eisert}, \citenamefont {Gross},\ and\ \citenamefont {Roth}}]{haferkamp2022}%
  \BibitemOpen
  \bibfield  {author} {\bibinfo {author} {\bibfnamefont {J.}~\bibnamefont {Haferkamp}}, \bibinfo {author} {\bibfnamefont {F.}~\bibnamefont {Montealegre-Mora}}, \bibinfo {author} {\bibfnamefont {M.}~\bibnamefont {Heinrich}}, \bibinfo {author} {\bibfnamefont {J.}~\bibnamefont {Eisert}}, \bibinfo {author} {\bibfnamefont {D.}~\bibnamefont {Gross}},\ and\ \bibinfo {author} {\bibfnamefont {I.}~\bibnamefont {Roth}},\ }\bibfield  {title} {\bibinfo {title} {Efficient unitary designs with a system-size independent number of non-clifford gates},\ }\href {https://doi.org/10.1007/s00220-022-04507-6} {\bibfield  {journal} {\bibinfo  {journal} {Commun. Math. Phys.}\ }\textbf {\bibinfo {volume} {397}},\ \bibinfo {pages} {995–1041} (\bibinfo {year} {2022})}\BibitemShut {NoStop}%
\bibitem [{\citenamefont {Heinrich}\ and\ \citenamefont {Gross}(2019)}]{Heinrich2019robustnessofmagic}%
  \BibitemOpen
  \bibfield  {author} {\bibinfo {author} {\bibfnamefont {M.}~\bibnamefont {Heinrich}}\ and\ \bibinfo {author} {\bibfnamefont {D.}~\bibnamefont {Gross}},\ }\bibfield  {title} {\bibinfo {title} {Robustness of {M}agic and {S}ymmetries of the {S}tabiliser {P}olytope},\ }\href {https://doi.org/10.22331/q-2019-04-08-132} {\bibfield  {journal} {\bibinfo  {journal} {{Quantum}}\ }\textbf {\bibinfo {volume} {3}},\ \bibinfo {pages} {132} (\bibinfo {year} {2019})}\BibitemShut {NoStop}%
\bibitem [{\citenamefont {True}\ and\ \citenamefont {Hamma}(2022)}]{True2022transitionsin}%
  \BibitemOpen
  \bibfield  {author} {\bibinfo {author} {\bibfnamefont {S.}~\bibnamefont {True}}\ and\ \bibinfo {author} {\bibfnamefont {A.}~\bibnamefont {Hamma}},\ }\bibfield  {title} {\bibinfo {title} {Transitions in {E}ntanglement {C}omplexity in {R}andom {C}ircuits},\ }\href {https://doi.org/10.22331/q-2022-09-22-818} {\bibfield  {journal} {\bibinfo  {journal} {{Quantum}}\ }\textbf {\bibinfo {volume} {6}},\ \bibinfo {pages} {818} (\bibinfo {year} {2022})}\BibitemShut {NoStop}%
\bibitem [{\citenamefont {Magni}\ and\ \citenamefont {Turkeshi}(2025)}]{Magni_2025}%
  \BibitemOpen
  \bibfield  {author} {\bibinfo {author} {\bibfnamefont {B.}~\bibnamefont {Magni}}\ and\ \bibinfo {author} {\bibfnamefont {X.}~\bibnamefont {Turkeshi}},\ }\bibfield  {title} {\bibinfo {title} {Quantum complexity and chaos in many-qudit doped clifford circuits},\ }\href {https://doi.org/10.22331/q-2025-12-24-1956} {\bibfield  {journal} {\bibinfo  {journal} {Quantum}\ }\textbf {\bibinfo {volume} {9}},\ \bibinfo {pages} {1956} (\bibinfo {year} {2025})}\BibitemShut {NoStop}%
\bibitem [{\citenamefont {Magni}\ \emph {et~al.}(2025)\citenamefont {Magni}, \citenamefont {Christopoulos}, \citenamefont {De~Luca},\ and\ \citenamefont {Turkeshi}}]{p8dn-glcw}%
  \BibitemOpen
  \bibfield  {author} {\bibinfo {author} {\bibfnamefont {B.}~\bibnamefont {Magni}}, \bibinfo {author} {\bibfnamefont {A.}~\bibnamefont {Christopoulos}}, \bibinfo {author} {\bibfnamefont {A.}~\bibnamefont {De~Luca}},\ and\ \bibinfo {author} {\bibfnamefont {X.}~\bibnamefont {Turkeshi}},\ }\bibfield  {title} {\bibinfo {title} {Anticoncentration in clifford circuits and beyond: From random tensor networks to pseudomagic states},\ }\href {https://doi.org/10.1103/p8dn-glcw} {\bibfield  {journal} {\bibinfo  {journal} {Phys. Rev. X}\ }\textbf {\bibinfo {volume} {15}},\ \bibinfo {pages} {031071} (\bibinfo {year} {2025})}\BibitemShut {NoStop}%
\bibitem [{\citenamefont {Zhu}\ \emph {et~al.}(2016)\citenamefont {Zhu}, \citenamefont {Kueng}, \citenamefont {Grassl},\ and\ \citenamefont {Gross}}]{zhu2016cliffordgroupfailsgracefully}%
  \BibitemOpen
  \bibfield  {author} {\bibinfo {author} {\bibfnamefont {H.}~\bibnamefont {Zhu}}, \bibinfo {author} {\bibfnamefont {R.}~\bibnamefont {Kueng}}, \bibinfo {author} {\bibfnamefont {M.}~\bibnamefont {Grassl}},\ and\ \bibinfo {author} {\bibfnamefont {D.}~\bibnamefont {Gross}},\ }\href {https://arxiv.org/abs/1609.08172} {\bibinfo {title} {The clifford group fails gracefully to be a unitary 4-design}} (\bibinfo {year} {2016}),\ \Eprint {https://arxiv.org/abs/1609.08172} {arXiv:1609.08172 [quant-ph]} \BibitemShut {NoStop}%
\end{thebibliography}%

\clearpage
\appendix

\section{End Matter}

\textit{A. Born weights, outcome resummation, and the rank process.}---Consider the trajectory average of an observable $F$ that depends on the state only through its spectrum -- hence, for stabilizer states, on the rank alone, $F(\rho)=f(r)$. Its exact expression is
\be
\average{F}=\exv_{\rm hist}\sum_{\rm out} p_{\rm out}\,F(\rho_{\rm out})\,,
\label{eq:bornavg}
\ee
where $\exv_{\rm hist}$ averages over the circuit realizations, the sum runs over the measurement outcomes, and $p_{\rm out}$ is the Born probability of the corresponding trajectory. For the projective measurement of a Pauli string $P$ on a stabilizer state, the Born weights are degenerate and follow from the three classes of the main text: each measurement contributes a multiplicative factor to $p_{\rm out}$, equal to $1$ in class (i), where the outcome is deterministic, and to $1/q$ in classes (ii) and (iii), where the $q$ outcomes are equiprobable. Moreover, the post-measurement rank is fixed by the class of $P$ alone: different outcomes yield different states (different $\pmb{\varphi}$) but the same $r$. The outcome sum in \eqref{eq:bornavg} then resums exactly at each step, $\sum_{\rm out}p_{\rm out}\,F(\rho_{\rm out})=f(r')$ with $r'$ the new rank: for the Born-rule average, all weights cancel, and the rank remains the \emph{only} dynamical variable, evolving stochastically with the class label. Consequently, the residual randomness is only that of the circuit history, and the Born average of any rank observable reduces to an expectation over the induced stochastic process of the rank, $\average{F}=\exv[f(R)]$: the process expectation $\exv$ of the main text is nothing but $\exv_{\rm hist}$ restricted to rank observables.
In the notation of the main text, a uniformly random global Clifford maps $\Wsp$ to a uniformly random isotropic subspace of the same dimension, so that the probabilities of the three classes depend on $r$ alone. The total $q^{2L}-1$ nontrivial Pauli strings split in three sets with probabilities
\be
\begin{aligned}
&P\in\Wsp\setminus\{\id\}: && q_1(r)=\tfrac{q^{L-r}-1}{q^{2L}-1},\\
&P\notin\Wsp^\perp: && q_2(r)=\tfrac{q^{2L}-q^{L+r}}{q^{2L}-1},\\
&P\in\Wsp^\perp\!\setminus\Wsp: && q_3(r)=\tfrac{q^{L+r}-q^{L-r}}{q^{2L}-1},
\end{aligned}
\label{eq:counting}
\ee
with $q_1+q_2+q_3=1$; only the third class lowers the rank, $q_3(r)=p_L(r)$ of \eqref{eq:pL}. At fixed $r$ and large $L$, $q_1\sim q^{-L}q^{-r}$, $q_3\sim q^{-L}\gamma_r$ and $q_2 \sim 1-q^{-L}q^{r}$, so on the rescaled time $x=q^{-L}t$ the rank-reducing events occur at rate $\gamma_r=q^r-q^{-r}$, giving the master equation \eqref{eq:master}.

\textit{B. Death process from infinity.}---
With the waiting and entrance times $\tau_r$, $\Tpass_r=\sum_{k>r}\tau_k$ of the main text, we can define the probability distribution for the variable $r$ as
\begin{equation}
    P(r; x)= \mathbb{E}_{\boldsymbol{\tau}_r}[\mathbf{1}(\Tpass_r\le x<\Tpass_{r-1})]\,,
\end{equation}
where $\mathbf{1}(\ldots)$ is the indicator function of a set. Since $\mathbf{1}(\Tpass_r\le x<\Tpass_{r-1}) = \mathbf{1}(\tau_r > x - \Tpass_r) \mathbf{1}(x \geq \Tpass_{r})$, and $\tau_r$ is independent from $\Tpass_r$,
we can first average on $\tau_r$, obtaining
\begin{equation}
\label{eq:prx1}
    P(r; x) = \mathbb{E}_{\boldsymbol{\tau}_{r-1}}[e^{-\gamma_r(x - \Tpass_r)} \mathbf{1}(x \geq \Tpass_{r})]\,,
\end{equation}
and then Laplace transform in $x$ which, using the independence of the $\tau_k$, yields \eqref{eq:laplace}.
The simple poles of the transform $\hat P(r,u)$ sit at $u=-\gamma_k$, $k\ge r$. For $r=1$ the pole closest to the origin is $u=-\gamma_1=-(q-q^{-1})$, with residue $A_q = \prod_{k\ge2}\gamma_k/(\gamma_k-\gamma_1)$, which governs the late-time behavior of $\average{S}$. The value of the residue can be simplified further: since $\gamma_k-\gamma_1=(q^{k-1}-1)(q+q^{-k})$ and $\gamma_k=q^{-k}(q^k-1)(q^k+1)$, each ratio equals $\frac{(q^k-1)(q^k+1)}{(q^{k-1}-1)(q^{k+1}+1)}$ and the product simplifies to $A_q=\frac{q^2+1}{q(q-1)}$, appearing in Eq.~\eqref{eq:Slarge}.
% Since $\overline S\simeq P(1,x)$ at late times, this yields \eqref{eq:Slarge} with $A_q=\prod_{k\ge2}a_k/(a_k-a_1)$.

\textit{C. Small $x$ expansion.}---The non-periodic part of the entropy follows from the integer moments $\average{S(x)}=-(\ln q)^{-1}\partial_n m_n|_{n=0}$, and since, from the hierarchy~\eqref{eq:momhier}, $\partial_s[x^s a_0(s)]|_{0}=\ln x+ \sum_k a_k'(0)x^{2k}$, the expansion results in
\be
\average{S(x)}=\log_q\tfrac1x-\frac{a_0'(0)}{\ln q} -\frac{a_1'(0)}{\ln q}x^2 + O(x^4),
\ee
with the universal coefficient $C_q=-a_0'(0)/\ln q$ and $c_2(q) = -\frac{a_1'(0)}{\ln q}$. Logarithmic differentiation of \eqref{eq:a0} gives $a_0'(0)/\ln q=\tfrac12+\gamma_{\rm E}/\ln q-\sum_{k\ge1}(q^k-1)^{-1}$, which reproduces \eqref{eq:C}.
Similarly, we obtain $a_1(s)$ from the recursion at $k=1$, finding $a_1(s) = - a_0(s) b_1(s)/[(s+2)(s+1)]$
with
\begin{equation}
    b_1(s)= \frac{q \left(q^{2 s+2}-(q+1)^2 q^s+q (s+2)+1\right)-s}{q^2-1}.
\label{eq:b1}
\end{equation}
After differentiation, this leads to $c_2(q)$ in Eq.~\eqref{eq:c2}.

\textit{D. Replica weights as a Feynman--Kac tilt.}--- Here we show that this dynamics admits a replica description which is common in the treatment of non-unitary purification processes.
We follow the unnormalized state $\urho$, whose trace coincides with the Born probability of the trajectory introduced in Sec.~A, $p_{\rm out}=\Tr\urho$. For an observable $F_k$ homogeneous of degree $k$ (with $F_k(\rho)=f_k(r)$ on a flat rank-$q^r$ state), the exact average \eqref{eq:bornavg} reads $\average{F_k(\rho)}=\exv_{\rm hist}\sum_{\rm out}p_{\rm out}\,F_k(\urho/p_{\rm out})$.
Since $F_k$ is homogeneous,
the replica trick allows
avoiding the denominator by introducing an integer index $N\ge k$, and defining the $N$-replica average
\begin{equation}
    \average{F_k(\rho)}^{(N)} = \exv_{\rm hist}\sum_{\rm out}p_{\rm out}^{\,N-k}\,F_k(\urho) \,,
\end{equation}
which reproduces the Born average, $\average{F_k(\rho)}\equiv\average{F_k(\rho)}^{(1)}$, in the replica limit $N\to 1$.
The outcome resummation that cancels the Born weights (Sec.~A) is special to $N=1$: for $N\neq1$ each branch retains a residual factor $p_{\rm out}^{\,N-1}$, which is sensitive also to the class-(ii) measurements -- $q$ outcomes each -- that change the state without changing $r$. This residual weight is the origin of the tilt derived below.

In our setting, we start from the normalized maximally entangled state $\urho_{0}=\id/q^{L}$. 
As established in Sec.~A, each measurement of class $q_2$ (rank-preserving) or $q_3$ (rank-reducing) divides $p_{\rm out}$ by $q$, whereas the rare class-$q_1$ events leave it unchanged. Hence, after $t=q^Lx$ steps the Born weight is divided by $q$ exactly $t-I(x)$ times, where $I(x)$ counts the rare type-$q_1$ events: $p_{\rm out}(x)=q^{-\,t+I(x)}=q^{-q^Lx+I(x)}$.
Conditioned on the rank trajectory $R(\cdot)$, the type-$q_1$ events form an inhomogeneous Poisson process. Indeed, while the rank sits at a given value $r$ — i.e., between two consecutive rank-reducing events — every step is of type $q_1$ or $q_2$, and the former occurs with conditional probability $q_1/(q_1+q_2)\sim q^{-L}q^{-r}$; since a unit of rescaled time $x$ corresponds to $q^L$ steps, type-$q_1$ events accumulate at rate $q^{-r}$ per unit $x$. Therefore, conditioned on $R(\cdot)$, the counter $I(x)$ is a Poisson random variable of mean $\Lambda(x)=\int_0^x q^{-R(u)}\dd u$, i.e.
\be
\operatorname{Prob}\big[I(x)=j\,\big|\,R(\cdot)\big]=\e^{-\Lambda(x)}\,\frac{\Lambda(x)^j}{j!}\,.
\label{eq:poisson}
\ee

Because all branches with nonzero weight are equiprobable and are in total $p_{\rm out}^{-1}$ for a specific history, the outcome sum gives $\average{F_k(\rho)}^{(N)}=\exv_{\rm hist}[f_k(R(x))\,p_{\rm out}(x)^{N-1}]$.
Inserting $p_{\rm out}(x)$ and using the generating function of \eqref{eq:poisson}, $\exv[q^{(N-1)I(x)}\,|\,R(\cdot)]=\e^{\mu_N\Lambda(x)}$ with $\mu_N=q^{N-1}-1$, we obtain
\be
\average{F_k(\rho)}^{(N)}=q^{-(N-1)q^Lx}
\exv\!\Big[f_k(R(x))\,\e^{\,\mu_N\Lambda(x)}\Big].
\label{eq:FK}
\ee
Aside from the prefactor, the average is the Feynman--Kac tilt by $\Lambda(x)$. For $N\to1$, $\mu_N\to0$ and \eqref{eq:FK} collapses to the bare marginal of $R(x)$ used in the main text.

\textit{E. Tilted moment hierarchy.}---Promoting $N$ to a continuous index $\alpha$, we set $\mu_\alpha=q^{\alpha-1}-1$ and define the tilted moments $M_s^{(\alpha)}(x)=\exv[q^{-sR(x)}\,\e^{\,\mu_\alpha\Lambda(x)}]$. The death-process generator acts as $(\mathcal L g)(r)=\gamma_r[g(r-1)-g(r)]$, and the Feynman--Kac identity $\tfrac{\dd}{\dd x}\exv[g\,\e^{\mu_\alpha\Lambda}]=\exv[(\mathcal L g+\mu_\alpha q^{-r}g)\e^{\mu_\alpha\Lambda}]$ with $g=q^{-sr}$ gives the closed hierarchy
\be
\frac{\dd M_s^{(\alpha)}}{\dd x}=(q^s-1)\big(M_{s-1}^{(\alpha)}-M_{s+1}^{(\alpha)}\big)+\mu_\alpha M_{s+1}^{(\alpha)} .
\label{eq:tilthier}
\ee
With the same ansatz, $M_s^{(\alpha)}=x^s\sum_k a_k^{(\alpha)}(s)x^{2k}$, matching powers of $x$ yields
\be
\begin{aligned}
(s+2k)\,a_k^{(\alpha)}(s)=\;&(q^s-1)\,a_k^{(\alpha)}(s-1)\\
&+\big[\mu_\alpha-(q^s-1)\big]\,a_{k-1}^{(\alpha)}(s+1),
\end{aligned}
\label{eq:tiltrec}
\ee
with $a_{-1}^{(\alpha)}\equiv0$, reducing to \eqref{eq:akrec} at $\alpha=1$. At $k=0$, we get exactly $a_0^{(\alpha)}(s)=a_0(s)$.
Writing $a_1^{(\alpha)}(s)=-a_0(s)\,b_1^{(\alpha)}(s)/[(s+1)(s+2)]$
the recursion at $k=1$ simplifies,
and is solved with boundary value $b_1^{(\alpha)}(0)=-\mu_\alpha(q-1)$ (read off from \eqref{eq:tiltrec} at $s=0$) by
\be
\begin{aligned}
b_1^{(\alpha)}(s) =&\frac{q^3}{q^2-1}\,q^{2s}-\frac{q\,(q+1+\mu_\alpha q)}{q-1}\,q^{s} \\
& +(1+\mu_\alpha)\,s
+\frac{q\,(2q+1)}{q^2-1}+\mu_\alpha\,\frac{2q-1}{q-1}.
\end{aligned}
\ee
For $\mu_\alpha=0$ this is the untilted $b_1$ Eq.~\eqref{eq:b1} entering \eqref{eq:Ssmall} through $c_2(q)$, which evaluates to \eqref{eq:c2}.

\begin{figure}[t]
\centering
\includegraphics[width=0.7\columnwidth]{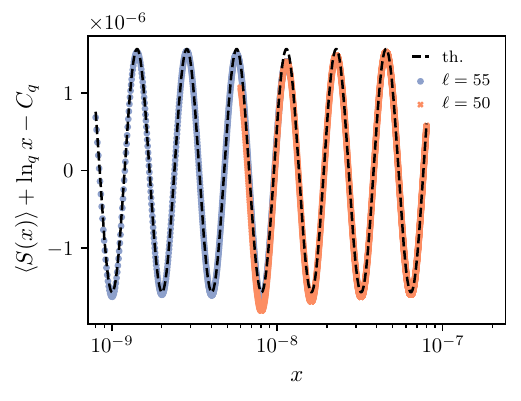}
\caption{Log-periodic modulation of $\average{S(x)}$ for $q=2$. The dashed black represents Eq.~\eqref{eq:logperiodicS} while markers display finite-$\ell$ data for the discrete-time Markov process.}
\label{fig:logper}
\end{figure}

\textit{F. Match with the Clifford commutant.}---For integer $N$, the same $\average{F_k(\rho)}^{(N)}$ can be computed directly from the $N$-replica channel, averaging $N$ copies of the global Clifford and contracting with the measurement insertion and $\rho_0$. The Clifford average is fixed by the $N$-th order Clifford commutant, Refs.~\cite{gross2021schur,bittel2025completetheorycliffordcommutant}. For small $N$ and $F_k(\rho) = \Tr\rho^k$, we verify the exact matching between a full replica computation and the Feynman-Kac tilt method in the small $x$ expansion. The results are also consistent with the design properties of the Clifford group, which is a unitary 3-design for $q=2$ and a 2-design for prime $q\geq3$~\cite{zhu2016cliffordgroupfailsgracefully, zhu2017multiqubit}.

 \textit{G. Log-periodic corrections.}---We now explain the appearance of the small periodic modulation in $\log_q(1/x)$, which starkly distinguishes Clifford from generic monitored purification. The reason is a discrete scale invariance special to the Clifford problem. The stabilizer rank is quantized in powers of $q$, so the rates grow by a factor $q$ from one level to the next, $\gamma_{r+1}\simeq q\,\gamma_r$: dividing the observation time $x$ by $q$ is nearly equivalent to advancing the front by one rank, $r\to r+1$, and the dynamics is invariant not under arbitrary rescalings of $x$ but only under the discrete subgroup $x\to x/q$.

To make this precise, zoom in on the moving front: write $\log_q(1/x)=\ell+\theta$, with $\ell$ integer and $\theta\in[0,1)$ the fractional scale, and shift the rank by the front position, $J=R-\ell$. The scaling limit must be taken at fixed $\theta$, i.e.\ along the sequence $x=q^{-\ell-\theta}$ with $\ell\to\infty$: as we now show, the limit law of $J$ retains a periodic dependence on $\theta$, an effect we refer to as \emph{rank quantization}. For $\ell \gg 1$,
$J$ performs a death process on $j\in\mathbb Z$, with rate $q^{j}$ for the transition $j\to j-1$, observed at the single time $x q^{\ell}=q^{-\theta}\in(q^{-1},1]$. Its probability distribution function (pdf) has a well-defined limit shape $P(r = \ell+j,x = q^{-\ell-\theta}) \stackrel{\ell \gg 1}{\simeq} h(q^{j-\theta})$. This can be understood from the Laplace transform Eq.~\eqref{eq:laplace}, where one can scale $u\to q^\ell u$, as the rates $\gamma_{\ell+j}\simeq q^{\ell+j}$ for $\ell\gg1$. The function $h$ is then given by the inverse Laplace transform
\begin{equation}
\begin{split}
    & h(q^{j-\theta}) = \mathcal L^{-1}[H](q^{j-\theta}) \,, \\
    & H(u) = \prod_{k=0}^{\infty}\left( 1+uq^{-k} \right)^{-1} \,.
\end{split}
\end{equation}
Notice that, with these definitions any averaged function 
\begin{multline}
\label{eq:favefront}
\left.\mathbb{E}[f(R(x) - \log_q (1/x))]\right|_{x = q^{-\ell-\theta}} = \\
   = \sum_{j \geq -\ell} f(j-\theta) P(j+\ell,q^{-\ell-\theta})  \stackrel{\ell\to\infty}{\simeq}
    \sum_{j \in \mathbb Z} f(j-\theta) h(q^{j-\theta})
\end{multline}
becomes periodic of unit period in $\theta$, since a shift $\theta \to \theta + T$ with $T \in \mathbb{Z}$ can be re-absorbed in the dummy variable $j$. Therefore, for the average rank and moments one has the short-time expressions
\begin{equation}
\begin{split}
    & \mathbb E[R(x)] - \log_q \frac 1 x \stackrel{x \to0}{\simeq}
    \sum_{j\in \mathbb Z} (j-\theta)h(q^{j-\theta}) \,,\\
    & \mathbb E[q^{-s(R(x) - \log_q 1/x)}]\stackrel{x \to0}{\simeq}
    \sum_{j\in \mathbb Z} q^{-s(j-\theta)} \ h(q^{j-\theta}) \,,
\end{split}
\end{equation}
which are unit-periodic. 
For the integer moments $s = n\in \mathbb{N}$, one obtains the closed formula
\begin{equation}
\label{eq:intmomEM}
    \lim_{x \to 0} \mathbb{E}[q^{-n (R(x) - \log_q 1/x)}] = \frac{1}{n!}\,
q^{n(n+1)/2}(q^{-1};q^{-1})_n\,;
\end{equation}
which is independent of $\theta$: the rank quantization is invisible at the level of integer moments, by the same mechanism that hides the modulation $\phi$ from the expansion \eqref{eq:akrec}. As a consequence, the continuation of \eqref{eq:intmomEM} to $s\notin\mathbb N$ is not unique  and the subtleties involved point to different protocols regarding the choice of $\theta$ as $\ell \to \infty$, leading to distinct laws $Y_\theta \stackrel{\ell \to \infty}{=} \left.R(x)-\log_q(1/x)\right|_{x = q^{-\ell - \theta}}$. A physically meaningful
definition is considering the law $\bar Y$, defined as the limit law of $Y_\theta$ when the fractional part $\theta$ is drawn uniformly on $[0,1)$ -- as happens when $x$ is sampled log-uniformly within one period. Its second cumulant gives the saturation value of the variance,
\begin{equation}
\label{eq:varR}
\sigma^2_q := \operatorname{Var}(\bar Y)= \frac{\pi^2}{6\ln^2 q}+\frac{1}{12}-\sum_{k= 1}^\infty \frac{q^k}{(q^k-1)^2}\,.
\end{equation}
Numerically, $\sigma_q^2=0.7630,\ 0.4968,\ 0.3523$ for $q=2,3,5$, in agreement with the exact solution of \eqref{eq:master}; at these small $q$ the periodic modulation of the variance itself is negligible, so $\sigma^2_q$ coincides with the fixed-$\theta$ variance $\operatorname{Var}(Y_\theta | \theta)$. At larger $q$ the two notions part ways, as made explicit by the law of total variance,
\be
\label{eq:totvarEM}
\sigma_q^2 = \mathbb E_\theta\big[\operatorname{Var}(Y_\theta|\theta)\big] + \operatorname{Var}_\theta\big(\mathbb E[Y_\theta|\theta]\big)\,.
\ee
The first term originates from the intrinsic variance of the $\theta$-dependent limit, while the second is related to the flat average over $\theta$. 
At fixed $\theta$ and large $q$, the entrance times $\Tpass_r = \sum_{k>r}\tau_k$ are dominated by their last term, $\log_q \Tpass_r = -(r+1) + O(1/\ln q)$, so that $R(x)\stackrel{q\gg1}{\to}\lfloor\log_q(1/x)\rfloor$ becomes \emph{deterministic}: the dynamical term $\operatorname{Var}(Y_\theta|\theta)\to0$ (slowly, as $\sim1/\ln q$), while the conditional mean approaches $\mathbb E[Y_\theta|\theta]\to-\theta$. The surviving variance is that of the deterministic offset $-\theta$ under the flat average over $\theta$, $\operatorname{Var}_\theta(\theta)=\tfrac1{12}$: the limit $\sigma_q^2\to\tfrac1{12}$ as $q\to\infty$ in Eq.~\eqref{eq:varR} is thus due to the flat average over the fractional part $\theta$. However, even at relatively large values of $q$, Eq.~\eqref{eq:varR} provides a good estimate of the dynamical variance, which dominates the decomposition, e.g.\ at $q=100$, $\sigma_q^2\simeq0.1506=0.1446+0.0060$ in \eqref{eq:totvarEM}. 

To fully characterize the periodicity, it is convenient to consider the Fourier expansion of these periodic functions,
\begin{equation}
    g(\theta) = \sum_{m \in \mathbb Z} \mathcal G_m e^{2\pi i m \theta} ,\,\,
    \mathcal G_m=\int_0^1 d\theta\  e^{-2\pi i m\theta}g(\theta)\,.
\end{equation}
The computation of the Fourier modes is described in the Supplemental Material. For the short-time entropy
\begin{equation}
\begin{split}
    \average{S(x)} & -  \log_q\frac 1 x = C_q + \\
    & \frac{2}{\ln q}  \sum_{m \geq 1} \operatorname{Re}\left[\Gamma\left(\frac{2\pi i m}{\ln q}\right)\ e^{2\pi i m\theta} \right] \,,
\end{split}
\label{eq:logperiodicS}
\end{equation}
where the zero-mode $m=0$ is the universal constant $C_q$ Eq.~\eqref{eq:C}.
This oscillatory behavior is amplified as $q$ grows, with a peak-to-peak amplitude $\approx3.1\times10^{-6}$ at $q=2$ (dominated by $m=\pm1$, Fig.~\ref{fig:logper}) rising to nearly the percent level already at $q=5$.

%%%%%%%%%%%%%%%%%%%%%%%%%%%%%%%%%%%%%%%%%%%%%%%%%%%%%%%%%%%%%%%%%%%%%%%%%%%%%%%%%%%%%%%%%%%%%%

\clearpage
\onecolumngrid

\setcounter{page}{1}
\setcounter{equation}{0}
\setcounter{figure}{0}
\setcounter{table}{0}
\setcounter{section}{0}

\renewcommand{\theequation}{S\arabic{equation}}
\renewcommand{\thefigure}{S\arabic{figure}}
\renewcommand{\thetable}{S\arabic{table}}
\renewcommand{\thesection}{S\arabic{section}}
\renewcommand{\theHequation}{S\arabic{equation}}
\renewcommand{\theHfigure}{S\arabic{figure}}
\renewcommand{\theHtable}{S\arabic{table}}
\renewcommand{\theHsection}{S\arabic{section}}

\begin{center}
    {\Large \bfseries Universal purification dynamics of monitored Clifford circuits\\[0.4em]
    Supplemental Material\par}
    \vspace{1em}
    {\normalsize
    Beatrice Magni,$^{1}$ Federico Gerbino,$^{2}$, Xhek Turkeshi$^{1}$, and Andrea De Luca$^{3}$\par}
    \vspace{0.5em}
    {\small
    $^{1}$Institut f\"ur Theoretische Physik, Universit\"at zu K\"oln, Z\"ulpicher Strasse 77, 50937 K\"oln, Germany\\
    $^{2}$Laboratoire de Physique Th\'eorique et Mod\`eles Statistiques, Universit\'e Paris-Saclay, CNRS, 91405 Orsay, France\\ 
    $^{3}$Laboratoire de Physique de l'\'Ecole Normale Sup\'erieure, CNRS, ENS \& PSL University, Sorbonne Universit\'e, Universit\'e Paris Cit\'e, 75005 Paris, France\par}

    \vspace{1em}
\end{center}
\title{Universal purification dynamics of monitored Clifford circuits\\  Supplemental Material}

This Supplemental Material reports the derivation of the Fourier Coefficients for the periodic parts of expectation values at small $x$. We focus on the average entropy and on the moments. We then clarify in which sense the limit $x\to0$ has to be taken, the distinction between the fixed-$\theta$ and $\theta$-averaged limit laws of $R(x)-\log_q(1/x)$, with $\theta$ the fractional part of $\log_q(1/x)$, the associated ambiguity in the analytic continuation of the integer moments, and the deterministic behavior emerging as $q\to\infty$. 

Let us start from the Laplace transform of $P(r,x)$ Eq.~(10) of the main text. By scaling $x = q^{-\ell-\theta}$, with $\ell=\lfloor -\log_q x \rfloor \in \mathbb N$ integer and $\theta = \{-\log_qx\} \in [0,1)$, shifting $r=\ell+j$ to the front position, and assuming $\ell \gg 1$, we have
\begin{equation}
    P(\ell+j,q^{-\ell-\theta}) = \int_{\mathcal B} \frac{du}{2\pi i}e^{u q^{-\ell-\theta}} \left[\frac1{\gamma_{\ell+j}}\prod_{k\ge \ell+j}\Big(1+\frac{u}{\gamma_k}\Big)^{-1}\right] \stackrel{\ell \to \infty}{\simeq}  \int_{\mathcal B} \frac{du}{2\pi i}e^{u q^{-\ell-\theta}} \left[\frac1{q^{\ell+j}}\prod_{k\ge \ell+j}\Big(1+\frac{u}{q^{k}}\Big)^{-1}\right] \,,
\end{equation}
where $\mathcal B$ indicates the Bromwich contour.
This suggests setting $u \to q^{\ell+j} u$ and $k \to \ell+j+k$, yielding a well-defined limit shape for the probability distribution function
\begin{equation}
    P(\ell+j,q^{-\ell-\theta}) \stackrel{\ell \to \infty}{\simeq} \int_{\mathcal B} \frac{du}{2\pi i}e^{u q^{j-\theta}}  \prod_{k \geq 0} \Big(1+u q^{-k}\Big)^{-1} =
    \int_{\mathcal B} \frac{du}{2\pi i}e^{u q^{j-\theta}}  H(u)
    %=: g(j,\theta)\,.
\end{equation}
where we wrote the product in terms of the function
\begin{equation}
    H(u) :=  \prod_{k \geq 0} \Big(1+u q^{-k}\Big)^{-1} = \frac{1}{(-u;q^{-1})_\infty} \;, \quad (a; Q)_\infty = \prod_{k=0}^{\infty}(1 - a Q^k)
\end{equation}
Introducing the inverse Laplace transform 
\begin{equation}
h(y)=\mathcal{L}^{-1}[H](y)
= \int_{\mathcal B} \frac{du}{2\pi i}e^{u y} H(u)
\end{equation}
we arrive at the relation presented in the End matter
\begin{equation}
    \lim_{\ell \to \infty} P(\ell+j,q^{-\ell-\theta}) = h(q^{j-\theta})
\end{equation}
Note that the limit $x\to0$ is taken here along the sequence $x = q^{-\ell-\theta}$ at fixed fractional part $\theta$: since $\theta=\{-\log_q x\}$ does not converge as $x\to0$ along real values, this is the only way the limit distribution exists, and it retains a dependence on $\theta$. This indicates that any expectation value
\begin{equation}
    \lim_{\ell\to\infty} \left. \mathbb{E}\big[f\big(R(x) - \log_q (1/x)\big)\big]\right|_{x = q^{-\ell - \theta}} = \sum_{j \in \mathbb{Z}} f(j - \theta) h(q^{j - \theta})
\end{equation}
which is thus a periodic function of $\theta$, with unit period. Specifically, in the same scaling for small $x$, we have for the entropy and moments
\begin{align}
&\left.\lim_{\ell\to\infty} \mathbb{E}[S(x)] - \log_q\frac 1 x \right|_{x = q^{-\ell -\theta}} = \sum_{j\in\mathbb{Z}}
(j-\theta)h\left(q^{j-\theta}\right) \,,
\label{eq:fourentr}
\\
&\left.\lim_{\ell\to\infty} x^{-s}\mathbb{E}[q^{-s R(x)}] \right|_{x = q^{-\ell -\theta}}
    = \sum_{j\in \mathbb Z} q^{-s(j-\theta)} \ h(q^{j-\theta})
    \label{eq:fourmom}
\end{align}
Being periodic functions of $\theta \in [0,1)$, it is convenient to look at the Fourier expansion
\begin{equation}
    g(\theta) = \sum_{m \in \mathbb Z} \mathcal{G}_m e^{2\pi i m \theta} \,,\quad
    \mathcal G_m=\int_0^1 d\theta\  e^{-2\pi i m\theta}g(\theta) \,.
\end{equation}
Setting $w = j-\theta$, one can express the Fourier modes $\mathcal{G}_m$ by turning the discrete sum over $j$ and the $\theta$-integral into a single real-valued integral
\begin{equation}
\mathcal G_m := \int_0^1 d\theta \ e^{-2\pi i m \theta} \sum_{j\in \mathbb Z}f(j-\theta)\ h(q^{j-\theta}) =
\int_{-\infty}^{\infty}dw\ 
e^{2\pi i m w}\,f(w)\,h(q^w) \,,
\end{equation}
In particular, for the Fourier modes of the moments in Eq.~\eqref{eq:fourmom}, setting $y=q^w$ and introducing $z_m=\frac{2\pi i m}{\ln q}$, we find
\begin{equation}
 \mathcal{M}_m(s) = \frac{1}{\ln q}
\int_0^\infty dy\ y^{z_m-1-s}\ h(y)
= \frac{1}{\ln q} M_h\left(z_m-s\right) \,,
\end{equation}
where $M_h$
is the Mellin transform of the function $h$
\begin{equation}
    M_h(z)=\int_0^\infty y^{z-1}h(y)\,dy \,.
\end{equation}
Clearly, the relation between the coefficients for the entropy and for the moments is $\mathcal S_m = -\frac{1}{\ln q}\partial_s \mathcal{M}_m(s)|_{s=0}$, implying for the Fourier modes of the entropy in Eq.~\eqref{eq:fourentr}
\begin{equation}
 \mathcal S_m = \frac{1}{(\ln q)^2}
\int_0^\infty dy \ y^{z_m-1}\ln y\ h(y)
= \frac{1}{(\ln q)^2} \partial_z M_h(z)|_{z=z_m}  \,, 
\end{equation}
Therefore, the fundamental ingredient is the Mellin transform $M_h$.  Since $H$ is the Laplace transform of $h$, we can relate $M_h$ to the Mellin transform $M_H$ of $H$: 
\begin{equation}
    M_H(1-z) = \int_0^\infty du \ u^{-z} H(u) = \int_0^\infty du\ u^{-z} \int_0^\infty dy \ e^{-uy} h(y) = \Gamma(1-z) M_h(z) \,,
\end{equation}
where one exchanges the $u$- and $y$-integrals, for
$0<\operatorname{Re}z<1$. Also, using the Euler reflection formula,
\begin{equation}
M_h(z) =
\Gamma(z)\frac{\sin(\pi z)}{\pi}
\int_0^\infty du \ u^{-z}H(u) \,.
\end{equation}
The latter integral can be computed in terms of Ramanujan's beta integral which in general reads
% \cite{ramanujan2015collected}
\begin{equation}
\int_0^\infty
t^{\nu-1}
\frac{(-at;Q)_\infty}{(-t;Q)_\infty}\,dt =
\frac{\pi}{\sin(\pi \nu)}\frac{(Q^{1-\nu};Q)_\infty\,(a;Q)_\infty}
     {(Q;Q)_\infty\,(aQ^{-\nu};Q)_\infty} \;.
%\frac{(Q^{1-\nu},a;Q)_\infty}{(Q,aQ^{-\nu};Q)_\infty} \,.%\quad \int_0^\infty\frac{t^{\nu-1}}{(-t;Q)_\infty}\,dt = \frac{(Q^{1-\nu};Q)_\infty}{(Q;Q)_\infty} \,.
\end{equation}
Choosing $\nu=1-z$, $Q=q^{-1}$ and $a = 0$, and using that $(0; Q)_\infty = 1$,
this yields
\begin{equation}
M_h(z) =
\Gamma(z) A(z) \,, 
\quad 
A(z) = \frac{(q^{-z};q^{-1})_\infty}{(q^{-1};q^{-1})_\infty} = (1-q^{-z})
\prod_{n=1}^{\infty}
\frac{1-q^{-n-z}}{1-q^{-n}}
\,.
\end{equation}
This expression provides the analytic continuation required to evaluate $M_h$
and $\partial_z M_h(z)$ at points $z=z_m$.

The computation is easier for the coefficients $\mathcal{M}_m$ at integer positive $s = n \in \mathbb{N}$. Indeed, at $z = z_m - n$ the function $A(z)$ contains the factor $(q^{-z_m+n}; q^{-1})_\infty = \prod_{k=0}^\infty(1-q^{n-k})$ (we used $q^{-z_m}=1$), which vanishes for every integer $n\geq 0$ because of the term $k=n$. Thus, the coefficients $\mathcal{M}_{m\neq 0}(n)$ vanish, i.e., the integer moments are independent of $\theta$.
For the zero mode, $z_0=0$, and at positive integer $s = n$ the pole of
$\left.\Gamma(z)\right|_{z = -n}$ cancels the simple zero of
$(q^{n};q^{-1})_\infty$. Taking the limit $z\to -n$, one finds
\begin{equation}
M_h(-n) =
\frac{\ln q}{n!}\prod_{r=1}^{n}(q^r-1)
= \frac{\ln q}{n!}\,
q^{n(n+1)/2}(q^{-1};q^{-1})_n \,,
\quad n\in\mathbb N.
\end{equation}
This already implies that 
\begin{equation}
\label{eq:Ythetadef}
\lim_{\ell \to \infty} \left[ R(x) - \log_q (1/x) \right]_{x = q^{-\ell - \theta}} \stackrel{\text{in law}}{=} Y_\theta
\end{equation}
where the random variable $Y_{\theta}$
has a $\theta$-dependent distribution and $O(1)$ fluctuations; we denote as $\mathbb{E}[\ldots|\theta]$ the expectation conditioned to the chosen value of $\theta$. The integer moments are independent of $\theta$
\begin{equation}
\label{eq:intmom}
     \mathbb{E}[q^{-n Y_\theta}|\theta] = \frac{1}{n!}\,
q^{n(n+1)/2}(q^{-1};q^{-1})_n = \lim_{x \to 0} \mathbb{E}[q^{-n (R(x) - \log_q 1/x)}]
\end{equation}
where the last equality holds because for integer $n$, the $x \to 0$ limit is well-defined.
A naive analytical continuation of this expression for non-integer $n$ hides some subtleties related to the dependence on $\theta$, which we discuss in the next section, particularly in relation to the calculation of the variance for small $x$. 

Let us conclude the computation by writing explicitly the coefficients $\mathcal S_m$ for the entropy.
For $m\neq0$, one has $
q^{-z_m}=e^{-2\pi i m}=1$, 
therefore, $A(z_m)=0$, $A'(z_m)=\ln q$. 
Since $\Gamma(z_m)$ is finite for $m\neq0$,
\begin{equation}
M_h'(z_m)
= \Gamma(z_m)A'(z_m)
= \ln q \ \Gamma(z_m) \,,
\end{equation}
therefore,
\begin{equation}
\mathcal S_m = \frac{1}{\ln q}
\Gamma\left(\frac{2\pi i m}{\ln q}\right), \qquad m\neq0.
\end{equation}
The mode $m=0$ is different because $\Gamma(z)$ has a pole at $z=0$,
while $(q^{-z};q^{-1})_\infty$ has a compensating zero. Near $z=0$, indeed, 
\begin{equation}
A(z) = 
z \ln q \left[
1+z \ln q \left(\sum_{n=1}^{\infty}\frac{1}{q^n-1}-\frac12\right)+O(z^2)
\right]\,,\quad 
\Gamma(z)=\frac{1}{z}-\gamma_{\mathrm E}+O(z).
\end{equation}
Hence,
\begin{equation}
M_h(z) = \ln q + z\left[
(\ln q)^2\left(\sum_{n=1}^{\infty}\frac{1}{q^n-1}-\frac12\right)
-\gamma_{\mathrm E} \ln q
\right]
+O(z^2),
\end{equation}
and thus
\begin{equation}
\mathcal S_0 = C_q= \sum_{n=1}^{\infty}\frac{1}{q^n-1} -\frac12 -\frac{\gamma_{\mathrm E}}{\ln q} \,.
\end{equation}

Finally, because $\mathbb{E}[ S(x)]$ is real, $\mathcal S_{-m}=\mathcal S_m^*$, and its Fourier expansion may be written as
\begin{equation}
\mathbb{E}[S(x)] - \log_q \frac 1 x
\stackrel{x\to 0}{\simeq} C_q + \frac{2}{\ln q} \sum_{m\geq1} 
\operatorname{Re}\left[\Gamma\left(\frac{2\pi i m}{\ln q}\right)e^{2\pi im\theta}  \right].
\end{equation}

\section*{Fixed-$\theta$ vs.\ $\theta$-averaged limit, analytic continuation of the moments, and the $q\to\infty$ behavior}

We now discuss in more detail the residual role of the fractional part $\theta$ in the $x\to0$ limit, which becomes crucial at large $q$. Let $Y_\theta$ be the random variable defined in Eq.~\eqref{eq:Ythetadef} by the fixed-$\theta$ limit 
\begin{equation}
    \operatorname{Prob}[Y_\theta = j-\theta] = h(q^{j-\theta})\,, \quad j \in \mathbb Z\,.
\end{equation}
By Eq.~\eqref{eq:fourmom}, its generating function is the periodic function of $\theta$
\begin{equation}
G_\theta(s) := \mathbb E\big[q^{-s Y_\theta}|\theta\big] = \sum_{m\in\mathbb Z}\mathcal M_m(s)\, e^{2\pi i m \theta}
= \frac{1}{\ln q}\sum_{m\in\mathbb Z} M_h(z_m - s)\, e^{2\pi i m \theta}\,.
\label{eq:Gtheta}
\end{equation}
Averaging flatly over $\theta$ defines the law $\bar Y$, that of $Y_\theta$ with $\theta$ uniform on $[0,1)$ -- equivalently, the law obtained when $x$ is sampled log-uniformly within one period, as is effectively the case when data are collected over many periods of $\log_q(1/x)$. Its generating function is the zero mode of \eqref{eq:Gtheta},
\begin{equation}
\mathbb E_{\rm flat}\big[q^{-s\bar Y}\big] := \mathbb{E}_\theta[ \mathbb{E}[q^{-s Y_\theta}|\theta]] = \int_0^1 \dd\theta\, G_\theta(s) = \mathcal M_0(s) = \frac{M_h(-s)}{\ln q}
= \frac{\Gamma(-s)}{\ln q}\, \frac{(q^{s};q^{-1})_\infty}{(q^{-1};q^{-1})_\infty}\,.
\label{eq:annealedGF}
\end{equation}
For clarity, in Eq.~\eqref{eq:totvar}, $\mathbb{E}[\ldots|\theta]$ represents the expectation over the law of $Y_\theta$, conditioned to a given value of $\theta$,  while $\mathbb{E}_{\theta}[\ldots]$ is the expectation over the uniform choice of $\theta \in [0,1)$.

\emph{Analytic continuation off the integers.} Since $\mathcal M_{m\neq0}(n)=0$ at integer $s=n$, 
Eq.~\eqref{eq:annealedGF} coincides with Eq.~\eqref{eq:intmom}. However, the continuation of the integer moments to $s\notin\mathbb N$ is not unique and signals different protocols regarding the choice of $\theta$ as $\ell \to \infty$: adding any smooth combination of terms vanishing at the integers, e.g.\ built from $e^{2\pi i m s}-1$, produces another interpolation with the same integer values. For instance, the elementary continuation of Eq.~\eqref{eq:intmom} through $(a;Q)_s := (a;Q)_\infty/(a Q^{s};Q)_\infty$ would lead to the candidate generating function
\begin{equation}
G_{\rm ele}(s) = \frac{q^{s(s+1)/2}}{\Gamma(s+1)}\, \frac{(q^{-1};q^{-1})_\infty}{(q^{-s-1};q^{-1})_\infty}\,,
\label{eq:naive}
\end{equation}
that agrees with \eqref{eq:annealedGF} at all $s=n\in\mathbb N$, but differs off the integers: the ratio of the two is a periodic function of $s$, equal to one at the integers, as follows from the reflection formula $\Gamma(-s)\Gamma(1+s)=-\pi/\sin(\pi s)$ together with the Jacobi triple product applied to $(q^{s};q^{-1})_\infty\, (q^{-s-1};q^{-1})_\infty$. The discrepancy is governed by the same log-periodic amplitudes responsible for the modulations of the entropy: it is invisible at small $q$ (relative size $\sim 10^{-24}$ at $q=2$, $\sim 10^{-10}$ at $q=5$), but reaches $\sim 10^{-3}$ at $q=100$ and becomes $O(1)$ in the scaled regime $s = O(1/\ln q)$ relevant below. 
Note that the analytic continuation \eqref{eq:naive} does not correspond to any sensible protocol for the average over $\theta$ at finite $q$, while at $q \to \infty$, it reduces to the choice $\theta = 1/2$ deterministically.

\emph{Variance decomposition.} By the law of total variance, the variance of $\bar Y$ splits into a dynamical and a quantization contribution,
\begin{equation}
\sigma_q^2 = \operatorname{Var}_{\rm flat}(\bar Y) = \underbrace{\mathbb E_\theta\big[\operatorname{Var}(Y_\theta|\theta)\big]}_{\rm dynamical} + \underbrace{\operatorname{Var}_\theta\big(\mathbb E[Y_\theta|\theta]\big)}_{\rm quantization} = \frac{
\partial_s^2 \log \mathcal M_0(s)\big|_{s=0}
}{(\ln q)^2} = \frac{\pi^2}{6\ln^2 q}+\frac{1}{12}-\sum_{n=1}^\infty \frac{q^n}{(q^n-1)^2}\,,
\label{eq:totvar}
\end{equation}
where the first term measures the trajectory-to-trajectory fluctuations at fixed observation time and the second the periodic drift of the conditional mean $\mathbb E[Y_\theta|\theta] = C_q + \phi(\theta)$ with $\theta$.

By Parseval's theorem, the quantization term is fully determined by the log-periodic amplitudes $\mathcal S_m$ computed above,
\begin{equation}
\operatorname{Var}_\theta\big(\mathbb E[Y_\theta | \theta]\big) = \sum_{m\neq0}|\mathcal S_m|^2 = \frac{2}{\ln^2 q}\sum_{m\geq1}\Big|\Gamma\Big(\frac{2\pi i m}{\ln q}\Big)\Big|^2 = \frac{2}{\ln^2 q}\sum_{m\geq1}\frac{\pi}{y_m \sinh(\pi y_m)}\,, \qquad y_m = \frac{2\pi m}{\ln q}\,,
\label{eq:parseval}
\end{equation}
which grows from $\simeq 1.2\times10^{-12}$ at $q=2$ and $\simeq 5.9\times10^{-6}$ at $q=5$ to $\tfrac1{12}$ as $q\to\infty$: term by term, $|\Gamma(i y_m)|^2 = \pi/(y_m\sinh \pi y_m) \to 1/y_m^2$, so that $\eqref{eq:parseval}\to\sum_{m\neq0} 1/(2\pi m)^2 = \tfrac1{12}$, reproducing the Fourier variance of a uniform $\theta$. Correspondingly, $\sigma_2^2\simeq 0.7630$ is entirely dynamical, at $q=100$ one finds $\sigma_q^2 \simeq 0.1506 = 0.1446+0.0060$, and as $q\to\infty$, as we now show, the dynamical term vanishes and the quantization term saturates the total.

\emph{Deterministic $q\to\infty$ limit at fixed $\theta$.} The natural variable for the $q\to\infty$ limit is $\lambda = s\ln q$, so that $q^{-sY}=e^{-\lambda Y}$. Setting $s=\lambda/\ln q$ in \eqref{eq:Gtheta}, the arguments $w_m := z_m - s = (2\pi i m - \lambda)/\ln q \to 0$, while $q^{-w_m} = e^{\lambda - 2\pi i m} = e^{\lambda}$; hence $A(w_m)\to 1-e^{\lambda}$, $\Gamma(w_m)\simeq 1/w_m = \ln q/(2\pi i m - \lambda)$, and each Fourier coefficient has the finite limit
\begin{equation}
\mathcal M_m\!\Big(\frac{\lambda}{\ln q}\Big) = \frac{M_h(w_m)}{\ln q} \;\xrightarrow{\;q\to\infty\;}\; \frac{e^{\lambda}-1}{\lambda-2\pi i m} = \int_0^1 \dd\theta\; e^{\lambda\theta}\, e^{-2\pi i m\theta}\,.
\label{eq:modelimit}
\end{equation}
Resumming the Fourier series, the fixed-$\theta$ generating function becomes
\begin{equation}
\mathbb E\big[e^{-\lambda Y_\theta}|\theta\big] = G_\theta\!\Big(\frac{\lambda}{\ln q}\Big) \;\xrightarrow{\;q\to\infty\;}\; e^{\lambda \theta}\,,
\qquad\text{i.e.}\qquad Y_\theta \xrightarrow{\;q\to\infty\;} -\theta \quad\text{deterministically}\,.
\label{eq:quenchdet}
\end{equation}
This matches the direct probabilistic argument: the entrance times $\Tpass_r = \sum_{k>r}\tau_k$ are dominated by their last term, $\log_q \Tpass_r = -(r+1) + \log_q\varepsilon_{r+1}$ with $\varepsilon_{r+1}$ an exponential variable of unit mean, and $\log_q \varepsilon \to 0$ as $q\to\infty$; hence at fixed $\theta\in(0,1)$ the rank freezes onto the front, $R(x)\to\lfloor \log_q (1/x)\rfloor$, with probability one. Thus, $Y_\theta = R - \log_q (1/x) \to -\theta$ and the $\theta$-fixed fluctuations disappear, $\operatorname{Var}(Y_\theta)\to0$ (the decay is slow, $\sim 1/\ln q$, due to the boundary layers at $\theta\to0,1$). For a uniform fractional part $\theta \in [0,1)$, only the zero mode of \eqref{eq:modelimit} survives the $\theta$-average and from Eq.~\eqref{eq:annealedGF}, one has
\begin{equation}
\mathbb E_{\rm flat}\big[e^{-\lambda \bar Y}\big] = \mathbb{E}_\theta[ \mathbb{E}[e^{-\lambda Y_\theta}|\theta]] = \mathcal M_0\!\Big(\frac{\lambda}{\ln q}\Big) \;\xrightarrow{\;q\to\infty\;}\; \frac{e^{\lambda}-1}{\lambda} = \int_0^1 e^{\lambda\theta}\,\dd\theta\,,
\label{eq:annlimit}
\end{equation}
the generating function of $-\bar Y = \theta$ uniform on $[0,1)$. The entire $q\to\infty$ variance $\sigma_\infty^2 = \tfrac1{12}$ in Eq.~\eqref{eq:totvar} is therefore the quantization variance of the deterministic sawtooth offset $-\theta$ under the flat average over $\theta$, and not a dynamical fluctuation. Note that the naive continuation \eqref{eq:naive} yields yet another limit, $G_{\rm ele}(\lambda/\ln q)\to e^{\lambda/2}$: the generating function of a variable deterministically pinned at the flat-average mean $-1/2$. It correctly tracks the mean but suppresses the fluctuations of $\theta$ entirely.

\end{document}